\documentclass[superscriptaddress,twocolumn,pre,nofootinbib]{revtex4}

\usepackage{amsmath}
\usepackage{graphicx,color}
\usepackage{amssymb}
\usepackage{amsfonts}
\usepackage{bm}

\newcommand{\be}{\begin{equation}}
\newcommand{\ee}{\end{equation}}
\newcommand{\bea}{\begin{eqnarray}}
\newcommand{\eea}{\end{eqnarray}}

\begin{document}
\sloppy


\title{Jeans mass-radius relation of self-gravitating Bose-Einstein condensates
and\\
typical parameters of the dark matter particle}

\author{Pierre-Henri Chavanis}
\affiliation{Laboratoire de Physique Th\'eorique, Universit\'e de Toulouse,
CNRS, UPS, France}

\begin{abstract}

We study the Jeans mass-radius relation of Bose-Einstein
condensate dark matter in Newtonian gravity. We show at a general level that it
is similar to the
mass-radius relation of Bose-Einstein condensate dark matter halos [P.H.
Chavanis, Phys. Rev. D {\bf 84}, 043531 (2011)]. Bosons with a repulsive
self-interaction generically evolve from the Thomas-Fermi regime to the
noninteracting regime as the Universe expands. In the Thomas-Fermi regime, the
Jeans radius remains approximately constant while the Jeans mass decreases. In
the noninteracting regime, the Jeans radius increases while the Jeans mass
decreases. Bosons with
an attractive self-interaction generically evolve from the nongravitational
regime to the noninteracting regime as the Universe expands. In the
nongravitational regime, the Jeans radius and the Jeans mass increase. In the
noninteracting regime, the Jeans radius increases while the Jeans mass
decreases. The transition occurs at a maximum Jeans mass which is similar to the
maximum
mass of Bose-Einstein condensate dark matter halos with an attractive
self-interaction. We use the mass-radius
relation of dark matter halos and the observational evidence of a ``minimum
halo'' (with typical radius $R\sim 1\, {\rm kpc}$ and typical mass $M\sim 10^8\,
M_{\odot}$) to constrain the mass $m$ and the scattering length $a_s$ of the
dark matter
particle.  For noninteracting bosons, $m$ is of
the
order of $2.92\times 10^{-22}\, {\rm eV}/c^2$. The mass of bosons with an
attractive self-interaction can only be
slightly smaller ($2.19\times 10^{-22}\, {\rm eV}/c^2<m<2.92\times 10^{-22}\,
{\rm eV}/c^2$ and $-1.11\times
10^{-62}\, {\rm fm}\le a_s\le 0$) otherwise the minimum halo would be unstable.
Constraints from particle physics and cosmology imply $m=2.92\times
10^{-22}\, {\rm eV}/c^2$ and $a_s=-3.18\times
10^{-68}\, {\rm fm}$ for ultralight axions and it is then found that attractive
self-interactions can
be neglected both in the linear and nonlinear regimes of structure formation.
The mass of bosons with a repulsive self-interaction can be
larger by $18$ orders of magnitude ($2.92\times 10^{-22}\, {\rm
eV}/c^2<m<1.10\times
10^{-3}\, {\rm eV}/c^2$ and $0\le a_s\le 4.41\times 10^{-6}\,
{\rm fm}$). The maximum allowed mass ($m=1.10\times
10^{-3}\, {\rm eV}/c^2$ and $a_s=4.41\times 10^{-6}\,
{\rm fm}$) is determined by the Bullet Cluster
constraint
while the transition between the noninteracting limit and the Thomas-Fermi
limit corresponds to $m=2.92\times 10^{-22}\, {\rm eV}/c^2$ and
$a_s=8.13\times
10^{-62}\, {\rm fm}$. For each of these models, we
calculate the Jeans length and the Jeans mass at the epoch of radiation-matter
equality and at the present epoch.

\end{abstract}

\maketitle


\section{Introduction}

Even after  $100$ years of research, the nature of dark matter (DM) is still elusive.
The standard cold dark matter (CDM) model, in which DM is represented by a
classical pressureless fluid at zero temperature ($T=0$) or by a collisionless
$N$-body system of classical particles, works extremely well at large
(cosmological) scales and can account for precise measurements of the cosmic
microwave background (CMB) from WMAP and Planck missions
\cite{planck2013,planck2016}.
However,  in addition to the lack
of evidence for any CDM particle such as a weakly
interacting massive particle (WIMP) with a mass in the
GeV-TeV range, the CDM model faces
serious problems at small (galactic) scales that are known as the ``core-cusp''
problem \cite{moore}, the ``missing satellites'' problem
\cite{satellites1,satellites2,satellites3}, and the ``too big to fail'' problem
\cite{boylan}. This ``small-scale crisis of CDM''  \cite{crisis} is somehow
related to the assumption that DM is pressureless implying that gravitational
collapse takes place at all scales. A possibility to solve these problems is to 
consider self-interacting dark matter
(SIDM) \cite{spergelsteinhardt}, warm dark matter (WDM)  \cite{wdm}, or the
feedback of
baryons that can transform cusps into cores \cite{romano1,romano2,romano3}.
Another
promising possibility
to solve the CDM crisis is to
take into account the quantum (or wave) nature of the DM particle. Indeed, in
quantum mechanics, an effective pressure is present even at $T=0$. This quantum
pressure may balance the gravitational attraction at small scales and solve the
CDM crisis.

In this paper, we shall consider the possibility that the DM particle is a
boson, e.g., an ultralight axion (ULA)
\cite{marshrevue}.\footnote{Some
authors have considered the case where the DM particle is a fermion like a
massive
neutrino \cite{markov,cmc1,cmc2,tg,r,gao,gao2,kallman,cgr,fjr,zcls,bgr,stella,
cls,cl,ir,
gmr,merafina,imrs,
vlt,vtt,viollierseul,bvn,tv,bmv,csmnras,chavmnras,robert,bvcqg,bvr,paolis,bmtv,
pt,dark,ispolatov,btv,rieutord,ptdimd,nm,ijmpb,mb,narain,ren,kupi,dvs1,dvs2,vss,
ar,arf,sar,rar,du,clm1,clm2,vsedd,vs2,rsu,vsbh,krut}. In this model,
gravitational collapse is prevented
by the quantum pressure arising from the Pauli exclusion principle.} At $T=0$,
bosons
form Bose-Einstein condensates (BECs) and they are described by a single
wavefunction $\psi({\bf r},t)$. They can therefore be interpreted as a scalar
field (SF). The bosons may be noninteracting or may have a repulsive 
or an attractive self-interaction (for example, the QCD axion has an attractive
self-interaction). On astrophysical scales, one must generally take into account
gravitational interactions between the bosons.  The evolution of the wave
function of self-gravitating BECs is then governed by the Schr\"odinger-Poisson
(SP) equations when the bosons are noninteracting or by the
Gross-Pitaevskii-Poisson (GPP) equations when the bosons are self-interacting.
BECDM halos can thus be viewed as gigantic bosonic atoms where the
bosonic particles are condensed in a single
macroscopic quantum state. The wave properties of
the SF are negligible at large (cosmological)
scales where the SF behaves as CDM, but they become important at small (galactic)
scales where they can prevent gravitational collapse, providing
halo cores and suppressing small-scale
structures. This model has been given several names such as wave DM, fuzzy dark
matter
(FDM), BECDM, $\psi$DM, or
SFDM
\cite{baldeschi,khlopov,membrado,maps,bianchi,sin,jisin,leekoh,schunckpreprint,
matosguzman,sahni,
guzmanmatos,hu,peebles,goodman,mu,arbey1,silverman1,matosall,silverman,
lesgourgues,arbey,fm1,bohmer,fm2,bmn,fm3,sikivie,mvm,lee09,ch1,lee,prd1,prd2,
prd3,briscese,
harkocosmo,harko,abrilMNRAS,aacosmo,velten,pires,park,rmbec,rindler,lora2,
abrilJCAP,mhh,lensing,glgr1,ch2,ch3,shapiro,bettoni,lora,mlbec,madarassy,marsh,
abrilph,playa,stiff,souza,freitas,alexandre,schroven,pop,cembranos,
schwabe,fan,calabrese,bectcoll,chavmatos,hui,abrilphas,
ggpp,shapironew,mocz,zhang,suarezchavanisprd3,veltmaat,moczSV,phi6,bbbs,
cmnjv,psgkk,ekhe,matosbh,tkachevprl2,epjpbh,ag,lhb,nmibv,zlc,bm,modeldm,bft,
bblp,dn,bbes,bvc,moczamin,mabc,gga,moczprl,mcmh,harkoj,mcmhbh,reig,dm,adgltt,
moczmnras,verma,braxbh,lancaster,tunnel}
(see the
Introduction of \cite{prd1} and Ref. \cite{leerevue} for an early history of
this model and Refs. \cite{srm,rds,chavanisbook,marshrevue,niemeyer,ferreira}
for reviews). Here, we shall
use the name BECDM. In the BECDM model, gravitational
collapse is prevented by the quantum pressure arising from the Heisenberg
uncertainty principle or from the scattering of the bosons 
(when the self-interaction is repulsive).\footnote{A
repulsive self-interaction ($a_s>0$) stabilizes the quantum
core. By
contrast, an attractive  self-interaction (like for the axion) destabilizes the
quantum core
above a maximum mass $M_{\rm max}=1.012\, \hbar/\sqrt{Gm|a_s|}$ first
identified in \cite{prd1} (see Refs.
\cite{prd1,prd2,bectcoll,phi6,epjpbh,mcmh,mcmhbh,tunnel,bb,ebyinfrared,guth,
ebybosonstars,braaten,braatenEFT,davidson,ebycollapse,bbb,ebylifetime,cotner,
ebycollisions,ebychiral,tkachevprl,helfer,kling,svw,visinelli,moss,ebyexpansion,
ebybh,ebydecay,namjoo,ebyapprox,nsh,nhs,chs,croon,ebyclass,elssw,guerra}  for
recent
works on axion stars \cite{tkachev,tkachevrt,kt} and Ref. \cite{braatenrevue}
for a review). This maximum mass has a nonrelativistic origin. It is physically
different from the maximum mass of fermion stars \cite{ov} and boson stars
\cite{kaup,rb,colpi,chavharko} that is due to general
relativity.}
Therefore, quantum mechanics (or a repulsive
self-interaction) 
may solve the small-scale problems of the CDM model mentioned above.

It is usually considered that large-scale structures such as galaxies or dark
matter halos form in a homogeneous universe by Jeans instability
\cite{jeans1902}. For
a cold
classical gas, the Jeans length vanishes or is extremely small 
($\lambda_J\simeq 0$) implying that structures can form at all scales. This is
not what we observe and this leads to the CDM crisis. By contrast,
when
quantum mechanics (or a repulsive self-interaction) is taken into account, the
Jeans length is non-zero implying the absence of structures below a minimum
scale
in agreement with the observations. The Jeans instability of a self-gravitating
BEC with repulsive or 
attractive self-interaction was first considered by Khlopov {\it et al.}
\cite{khlopov} and Bianchi {\it et al.} \cite{bianchi} in a general relativistic
framework based on the Klein-Gordon-Einstein equations. The Jeans instability of
a noninteracting self-gravitating BEC in Newtonian gravity described by the SP
equations was studied by Hu
{\it et al.} \cite{hu} and Sikivie and Yang \cite{sikivie}. Finally, 
the Jeans instability of a Newtonian self-gravitating BEC  with repulsive or
attractive self-interactions described by the GPP equations was studied by
Chavanis \cite{prd1}. 
These results were extended in general relativity by Su\'arez and Chavanis
\cite{suarezchavanisprd3} going beyond some of the approximations made by
Khlopov {\it et al.} \cite{khlopov} (see footnote
7 of \cite{suarezchavanisprd3}). Recently, Harko \cite{harkoj}
considered the Jeans instability of rotating Newtonian BECs  in the TF limit. 
In these
different studies, the authors
determined the Jeans length and the Jeans mass of the BECs and used them to
obtain an estimate of the minimum size and minimum length of
BECDM halos.\footnote{These studies were performed in a static Universe. The
Jeans instability of an infinite homogeneous self-gravitating BEC
in an expanding universe has been
studied by Bianchi {\it et
al.} \cite{bianchi}, Su\'arez and Matos \cite{abrilMNRAS} and Su\'arez and
Chavanis \cite{abrilph}  in general relativity and by Sikivie and Yang
\cite{sikivie} and Chavanis  \cite{aacosmo} in Newtonian gravity. These studies
are valid for a complex SF describing the wave function of a BEC. They rely
on a hydrodynamical representation of the wave equation. The Jeans
instability of a real SF has been studied by numerous
authors in Refs.
\cite{sasaki1,sasaki2,mukhanov,ratrapeebles,nambu,ratra,mukhanovrevue,hwang,
jetzer,hucosmo,joyce,ma,pb,brax,matos,hn1,spintessence,jk,hn2,malik,axiverse,mf,
easther,park,mmss,nph,nhp,hlozek,alcubierre1,cembranos,ug,marshrevue,fm2,fm3}.}

The Jeans instability study is only valid in the linear regime of structure
formation. It describes the initiation of the large-scale structures of the
Universe. The Jeans instability leads to a growth of the perturbations and the
formation of condensations (clumps). When the
density contrast reaches a sufficiently large value,  the condensations
experience a free fall, followed by a complicated process of gravitational
cooling and violent relaxation. They can also undergo merging and accretion.
This corresponds to the nonlinear
regime of structure formation leading to DM halos. BECDM halos result from the
balance between the gravitational attraction and the quantum pressure due to the
Heisenberg principle and the self-interaction of the bosons.
Observations reveal that, contrary to the prediction of the CDM
model, there are no halos with a  mass smaller than $M\sim 10^8\, M_{\odot}$ and
with a size smaller than $R\sim 1\, {\rm kpc}$. 
These ultracompact DM halos correspond typically to dwarf spheroidal galaxies
(dSphs) like Fornax. To be specific, we shall
assume
that Fornax
is the smallest  halo observed in the Universe. In the BECDM model, this
``minimum halo'' is interpreted as the ground state of the self-gravitating
BEC at $T=0$. Bigger halos have a core-halo structure with a quantum core
(ground state) surrounded by an approximately isothermal atmosphere which
results from quantum
interferences. This core-halo structure is observed in numerical simulations of
BECDM
\cite{ch2,ch3,schwabe,mocz,moczSV,veltmaat,moczprl,moczmnras}. The quantum
core may solve the
small-scale problems of the CDM model such as the cusp problem and the missing
satellite problem. The approximately isothermal atmosphere is consistent with
the classical NFW profile and accounts for the flat rotation curves of the
galaxies at large distances. The mass-radius relation of BECDM halos at $T=0$
(ground state) representing the minimum halo or the quantum core of larger halos
has been determined in Refs. \cite{prd1,prd2} for bosons with vanishing,
repulsive, or
attractive self-interaction. It can be obtained either numerically \cite{prd2}
by solving the GPP equations or analytically \cite{prd1} by using a gaussian
ansatz for the wavefunction. The quantum core mass -- halo
mass relation $M_c(M_h)$ was first obtained by Schive {\it et al.} \cite{ch3},
in
the case of noninteracting bosons, from direct numerical simulations and
heuristic arguments. This relation was later derived in Refs.
\cite{modeldm,mcmh,mcmhbh} from an effective thermodynamic approach by
maximizing the entropy at fixed mass and energy. It was also extended to the
case of self-interacting bosons (with a repulsive of an attractive
self-interaction) and fermions \cite{modeldm,mcmh,mcmhbh}.

It was noticed in Ref. \cite{prd1} that the Jeans mass-radius relation obtained
from the dispersion relation of self-gravitating homogeneous BECs is similar to
the mass-radius relation of BECDM halos obtained by solving the equation of
quantum hydrostatic equilibrium with a Gaussian ansatz. This
agreement is surprising because the two relations apply to very different
regimes of structure formation: linear versus nonlinear. It
results, 
however, essentially from dimensional analysis. The aim of the present paper is
to further develop this analogy and study its consequences. In
Sec. \ref{sec_sgbec} we
recall the basic equations describing self-gravitating BECs. Using the Madelung
\cite{madelung}
transformation, we write the GPP equations in the form of hydrodynamic
equations. We then consider spatially inhomogeneous solutions of
these equations
representing BECDM halos. They correspond to stationary solutions of the GPP
equations or to stationary solutions of the quantum Euler-Poisson equations
satisfying the condition of hydrostatic equilibrium. Stable equilibrium states
follow a minimum energy principle. We also consider the Jeans instability of an
infinite homogeneous self-gravitating BEC. We recall the general dispersion
relation and
the Jeans wavenumber obtained in Ref. \cite{prd1} from which we can
obtain the Jeans length and the Jeans mass. We briefly discuss the Jeans
instability in an expanding universe. In Sec. \ref{sec_pa} we
derive the analytical
mass-radius relation of BECDM halos from a general ansatz on the wavefunction
($f$-ansatz). We determine the parameters of this relation by comparing its
asymptotic limits with the exact results obtained by solving the GPP equations
numerically \cite{prd2}. In this manner, the analytical mass-radius relation
that we obtain 
provides a very good agreement with the exact numerical mass-radius relation. In
Sec. \ref{sec_mas}, we use the fact that this mass-radius relation applies to
the minimum
halo (with $R\sim 1\, {\rm kpc}$ and $M\sim 10^8\, M_{\odot}$) to obtain the
dark matter particle mass-scattering length relation. This is a constraint that
the parameters of the DM particle must satisfy in order to reproduce the
characteristics of the minimum halo (assumed to correspond to the ground state
of the BECDM model). Using additional constraints such as the Bullet Cluster
constraint, or constraints from particle physics and
cosmology, we can put
some bounds on the possible values of $m$ and $a_s$. We consider specific models
of physical interest that we call BECNI, BECTF, BECt, BECcrit and BECth. Once
the
values of $m$ and $a_s$ have
been determined from the previous considerations, we study in Sec.
\ref{sec_jmr} the
evolution of the Jeans radius and Jeans mass as a function of the cosmic
density, as
the Universe expands. We confirm that the Jeans
mass-radius relation is similar to the mass-radius relation of DM halos, the
density of the universe playing in this analogy the role of the average density
of DM halos. We characterize different regimes (noninteracting, TF,
nongravitational) for bosons with repulsive or attractive self-interaction.
Finally, we explain how our results can be extended to more general forms of
self-interaction.

\section{Self-gravitating Bose-Einstein condensates}
\label{sec_sgbec}

\subsection{Gross-Pitaevskii-Poisson equations}
\label{sec_gpp}

We assume that DM is made of bosons (like the axion) in the form of BECs at $T=0$.  We use a nonrelativistic approach based on Newtonian gravity.  The evolution
of the wave function $\psi({\bf r},t)$ of a self-gravitating BEC is governed by
the
Gross-Pitaevskii-Poisson (GPP) equations (see, e.g., \cite{prd1})
\begin{eqnarray}
\label{gpp1}
i\hbar \frac{\partial\psi}{\partial
t}=-\frac{\hbar^2}{2m}\Delta\psi+m\frac{dV}{d|\psi|^2}\psi+m\Phi\psi,
\end{eqnarray}
\begin{equation}
\label{gpp2}
\Delta\Phi=4\pi G |\psi|^2,
\end{equation}
where $\Phi({\bf r},t)$ is the gravitational potential and $m$ is the mass of the
bosons. The first term in Eq. (\ref{gpp1}) is the
kinetic term which accounts for the Heisenberg uncertainty
principle. The second term takes into account the self-interaction of the
bosons via a potential $V(|\psi|^2)$. The third term accounts for the
self-gravity of the BEC. The mass density of the BEC is $\rho({\bf
r},t)=|\psi|^2$.

For the standard BEC, we have
\begin{equation}
\label{gpp3}
V(|\psi|^2)=\frac{2\pi a_s\hbar^2}{m^3}|\psi|^{4},
\end{equation}
where  $a_s$ is the scattering
length of the bosons. The interaction between the bosons is repulsive when $a_s>0$ and
attractive when $a_s<0$. This potential is valid provided that the gas is sufficiently dilute. It corresponds to a power-law potential of the form
\begin{equation}
\label{gpp4}
V(|\psi|^2)=\frac{K}{\gamma-1}|\psi|^{2\gamma}
\end{equation}
with
\begin{equation}
\label{gpp5}
\gamma=2 \quad {\rm and} \quad K=\frac{2\pi a_s\hbar^2}{m^3}.
\end{equation}

The GPP equations conserve the mass
\begin{eqnarray}
\label{gpp6}
M=\int |\psi|^2\, d{\bf r}
\end{eqnarray}
and the energy
\begin{equation}
\label{gpp7}
E_{\rm tot}=\frac{\hbar^2}{2m^2}\int |\nabla\psi|^2\, d{\bf r}+\int V(|\psi|^2)\, d{\bf r}+\frac{1}{2}\int |\psi|^2 \Phi\, d{\bf r},
\end{equation}
which is the sum of the kinetic energy $\Theta$, the internal energy $U$, and the gravitational energy $W$ (i.e. $E_{\rm tot}=\Theta+U+W$).

{\it Remark:} the GPP equations (\ref{gpp1}) and (\ref{gpp2}) may be obtained
from the KGE equations in the nonrelativistic limit $c\rightarrow +\infty$  for
a SF interacting via a potential $V_R(\varphi)$ (see, e.g.,
\cite{playa,chavmatos} for a complex SF and \cite{phi6,tunnel} for a
real SF).

\subsection{The Madelung transformation}
\label{sec_mad}

Writing the wave function as
\begin{equation}
\label{mad1}
\psi({\bf r},t)=\sqrt{{\rho({\bf r},t)}} e^{iS({\bf r},t)/\hbar},
\end{equation}
where $\rho({\bf r},t)$ is the mass density  and $S({\bf r},t)$ is the action,
and making the Madelung \cite{madelung} transformation
\begin{equation}
\label{mad2}
\rho({\bf
r},t)=|\psi|^2\qquad {\rm and} \qquad {\bf u}=\frac{\nabla S}{m},
\end{equation}
where ${\bf u}({\bf r},t)$ is the velocity field, the GPP
equations (\ref{gpp1}) and  (\ref{gpp2}) can be written under the form of
hydrodynamic equations
\begin{equation}
\label{mad3}
\frac{\partial\rho}{\partial t}+\nabla\cdot (\rho {\bf u})=0,
\end{equation}
\begin{equation}
\label{mad3b}
\frac{\partial S}{\partial t}+\frac{(\nabla S)^2}{2m}+m\left\lbrack
\Phi+V'(\rho)\right\rbrack+Q=0,
\end{equation}
\begin{equation}
\label{mad4}
\frac{\partial {\bf u}}{\partial t}+({\bf u}\cdot
\nabla){\bf
u}=-\frac{1}{m}\nabla
Q-\frac{1}{\rho}\nabla P-\nabla\Phi,
\end{equation}
\begin{equation}
\label{mad5}
\Delta\Phi=4\pi G \rho,
\end{equation}
where
\begin{equation}
\label{mad6}
Q=-\frac{\hbar^2}{2m}\frac{\Delta
\sqrt{\rho}}{\sqrt{\rho}}=-\frac{\hbar^2}{4m}\left\lbrack
\frac{\Delta\rho}{\rho}-\frac{1}{2}\frac{(\nabla\rho)^2}{\rho^2}\right\rbrack
\end{equation}
is the quantum potential which takes into account the Heisenberg uncertainty
principle. On the other hand, the pressure $P$ is
a function $P({\bf r},t)=P\lbrack \rho({\bf
r},t)\rbrack$ of the density (the fluid is barotropic) which is determined by the potential $V(\rho)$
through the relation\footnote{This relation is consistent with
the first principle of
thermodynamics for a barotropic gas at $T=0$ (see
Appendix \ref{sec_ti}). It shows that $V(\rho)$ represents the density of
internal energy ($u=V$). Then, the enthalpy is given by $h=(P+V)/\rho=V'(\rho)$
and it satisfies the identity $h'(\rho)=P'(\rho)/\rho$. This allows us, for
example, to replace
$(1/\rho)\nabla P$ by $\nabla h$ in Eq. (\ref{mad4}).} 
\begin{equation}
\label{mad7}
P(\rho)=\rho
V'(\rho)-V(\rho)=\rho^2\left\lbrack
\frac{V(\rho)}{\rho}\right\rbrack'
\end{equation}
equivalent to
\begin{equation}
\label{mad8}
P'(\rho)=\rho V''(\rho).
\end{equation}
The speed of sound $c_s$ is
given by
\begin{equation}
\label{mad8b}
c_s^2=P'(\rho)=\rho V''(\rho).
\end{equation}
For a power-law potential of the form of Eq. (\ref{gpp4}), we get a polytropic equation of state
\begin{eqnarray}
\label{mad9}
V(\rho)=\frac{K}{\gamma-1}\rho^{\gamma}  \quad &\Rightarrow& \quad P=K\rho^{\gamma}\nonumber\\
 &\Rightarrow& \quad c_s^2=K\gamma\rho^{\gamma-1}.
\end{eqnarray}
In particular, for the standard BEC, we obtain
\begin{eqnarray}
\label{mad10}
V(\rho)=\frac{2\pi a_s\hbar^2}{m^3}\rho^{2}  \quad &\Rightarrow& \quad P=\frac{2\pi a_s\hbar^2}{m^3}\rho^{2}\nonumber\\
&\Rightarrow& \quad c_s^2=\frac{4\pi a_s\hbar^2}{m^3}\rho.
\end{eqnarray}
The equation of state is quadratic. The hydrodynamic equations
(\ref{mad3})-(\ref{mad5}) are called the
quantum Euler-Poisson equations \cite{prd1}. Equation (\ref{mad3b}) is the
quantum Hamilton-Jacobi (or Bernoulli) equation. In the TF limit  where the
quantum potential
can be neglected (formally $\hbar=0$), they become equivalent to the classical
Euler-Poisson equations for a barotropic gas \cite{bt}.\footnote{In the
classical limit $\hbar=0$ and for $P=0$,  the
quantum Euler-Poisson equations (\ref{mad3})-(\ref{mad5}) reduce to the 
pressureless hydrodynamic equations of the CDM model.}

The quantum Euler equations conserve the mass
\begin{eqnarray}
\label{mad11}
M=\int \rho\, d{\bf r}
\end{eqnarray}
and the energy (see, e.g., \cite{prd1})
\begin{eqnarray}
\label{mad12}
E_{\rm tot}=\Theta_c+\Theta_Q+U+W,
\end{eqnarray}
which is the sum of the classical kinetic energy
\begin{eqnarray}
\label{mad12a}
\Theta_c=\int \rho\frac{{\bf u}^2}{2}\, d{\bf r},
\end{eqnarray}
the quantum kinetic energy
\begin{eqnarray}
\label{mad12b}
\Theta_Q=\frac{\hbar^2}{8m^2}\int \frac{(\nabla\rho)^2}{\rho}\, d{\bf r}=\frac{1}{m}\int\rho Q\, d{\bf r},
\end{eqnarray}
the internal energy
\begin{eqnarray}
\label{mad12c}
U=\int V(\rho)\, d{\bf r}=\int \rho\int^{\rho}\frac{P(\rho')}{{\rho'}^2}\,
d\rho'\, d{\bf r},
\end{eqnarray}
and the gravitational energy
\begin{eqnarray}
\label{mad12d}
W=\frac{1}{2}\int \rho \Phi\, d{\bf r}.
\end{eqnarray}
At equilibrium, the classical (macroscopic) kinetic energy vanishes and we get
\begin{eqnarray}
\label{mad12e}
E_{\rm tot}=\Theta_Q+U+W.
\end{eqnarray}

The quantum virial theorem writes (see, e.g., \cite{prd1,ggpp})
\begin{eqnarray}
\label{mad13}
\frac{1}{2}\ddot I=2(\Theta_c+\Theta_Q)+3\int P\, d{\bf r}+W,
\end{eqnarray}
where
\begin{eqnarray}
\label{mad13b}
I=\int\rho r^2\, d{\bf r}
\end{eqnarray}
is the moment of inertia. At equilibrium, it reduces to
\begin{eqnarray}
\label{mad14}
2\Theta_Q+3\int P\, d{\bf r}+W=0.
\end{eqnarray}
For a polytropic equation of state $P=K\rho^{\gamma}$, we have $\int P\, d{\bf r}=(\gamma-1)U$ and the equilibrium virial theorem may be written as $2\Theta_Q+3(\gamma-1)U+W=0$. In particular, for the standard BEC for which $\gamma=2$, we have $\int P\, d{\bf r}=U$ and the equilibrium virial theorem reduces to $2\Theta_Q+3U+W=0$.

By using the Madelung transformation, the GPP
equations (\ref{gpp1}) and (\ref{gpp2}) have be written in the form of
hydrodynamic equations involving a quantum
potential taking into account the Heisenberg uncertainty principle and a
pressure force arising from the self-interaction of the bosons. This
transformation allows us to treat the BEC as a quantum fluid
(superfluid) and to apply standard methods developed in astrophysics as
discussed below.

{\it Remark:} The GPP equations  (\ref{gpp1}) and (\ref{gpp2}) and the
quantum Euler-Poisson equations  (\ref{mad3})-(\ref{mad5})  can be written in
terms of the functional derivative of the total energy $E_{\rm tot}$ (see Sec.
3.6 of \cite{ggpp}). They can also be obtained from a least
action principle and a Lagrangian (see Appendix B of \cite{bectcoll} and
Appendix F of \cite{ggpp}).

\subsection{Spatially inhomogeneous equilibrium states in the nonlinear regime: BECDM halos}
\label{sec_dm}

We first apply the  GPP equations (\ref{gpp1}) and  (\ref{gpp2}), or 
equivalently the quantum Euler-Poisson equations (\ref{mad3})-(\ref{mad5}), to
BECDM halos that appear in the nonlinear regime of structure formation in
cosmology.

A stationary solution of GPP equations is of the form
\begin{eqnarray}
\label{dm1}
\psi({\bf r},t)=\phi({\bf r})e^{-iEt/\hbar},
\end{eqnarray}
where $\phi({\bf r})=\sqrt{\rho({\bf r})}$ and $E$ are real. 
Substituting Eq. (\ref{dm1}) into Eqs.  (\ref{gpp1}) and  (\ref{gpp2}), we
obtain the eigenvalue problem
\begin{eqnarray}
\label{dm2}
-\frac{\hbar^2}{2m}\Delta\phi+m\frac{d V}{d\phi^2}\phi+m\Phi\phi=E\phi,
\end{eqnarray}
\begin{equation}
\label{dm3}
\Delta\phi=4\pi G \phi^2,
\end{equation}
determining the eigenfunctions $\phi({\bf r})$ and the eigenvalues $E$. For the fundamental mode (the one with the lowest energy) the  wavefunction $\phi(r)$ is spherically symmetric and has no node so that the density profile decreases monotonically with the radial distance.  Dividing Eq. (\ref{dm2}) by $\phi$ and using $\rho=\phi^2$, we obtain the identity
\begin{eqnarray}
\label{dm4}
Q+mV'(\rho)+m\Phi=E,
\end{eqnarray}
which can also be obtained from the quantum Hamilton-Jacobi (or Bernoulli) equation (\ref{mad3b}) with $S=-Et$. Multiplying Eq. (\ref{dm4}) by $\rho/m$ and integrating over the system we get
\begin{eqnarray}
\label{dm4b}
NE=\Theta_Q+\int \rho V'(\rho)\, d{\bf r}+2W.
\end{eqnarray}
For a polytropic equation of state $P=K\rho^{\gamma}$, we have $\int \rho V'(\rho)\, d{\bf r}=\gamma U$ and Eq. (\ref{dm4b}) may be written as $NE=\Theta_Q+\gamma U+2W$. In particular, for the standard BEC for which $\gamma=2$, we get $\int \rho V'(\rho)\, d{\bf r}=2 U$ and Eq. (\ref{dm4b}) reduces to $NE=\Theta_Q+2U+2W$.

Equivalent results can be obtained from the hydrodynamic equations 
(\ref{mad3})-(\ref{mad5}). 
Indeed, the condition of quantum
hydrostatic equilibrium,  corresponding to a
steady state of the quantum Euler equation (\ref{mad4}), writes
\begin{eqnarray}
\label{dm5}
\frac{\rho}{m}\nabla
Q+\nabla P+\rho\nabla\Phi={\bf 0}.
\end{eqnarray}
Dividing Eq. (\ref{dm5}) by $\rho$ and integrating the resulting expression with the help of Eq. (\ref{mad8}), we recover Eq. (\ref{dm4}) where $E$ appears as a constant of integration. On the other hand, combining Eq. (\ref{dm5}) with the Poisson equation (\ref{mad5}), we obtain the
fundamental differential
equation of quantum
hydrostatic equilibrium
\begin{equation}
\label{dm6}
\frac{\hbar^2}{2m^2}\Delta
\left (\frac{\Delta\sqrt{\rho}}{\sqrt{\rho}}\right )-\nabla\cdot \left
(\frac{\nabla P}{\rho}\right )=4\pi G\rho.
\end{equation}
This equation describes the balance between the quantum potential taking into
account the Heisenberg uncertainty principle, the pressure due to the
self-interaction of the bosons,  and the self-gravity.
For the standard BEC, using Eq. (\ref{mad10}), it takes the form
\begin{eqnarray}
\label{dm7}
\frac{\hbar^2}{
2m^2}\Delta
\left (\frac{\Delta\sqrt{\rho}}{\sqrt{\rho}}\right
)-\frac{4\pi a_s\hbar^2}{m^3}\Delta\rho=4\pi
G\rho.
\end{eqnarray}

These results can also be obtained from an energy principle (see
Appendix \ref{sec_alt}). Indeed, one can show that an equilibrium state of the
GPP equations is an extremum of energy $E_{\rm tot}$ at fixed mass $M$ and that
an equilibrium state is stable if, and only if, it is a
minimum of energy at fixed mass. We are led therefore to considering the
minimization problem
\begin{eqnarray}
\label{dm8}
\min\quad \lbrace E_{\rm tot}\quad | \quad M\quad {\rm fixed}\rbrace.
\end{eqnarray}
Writing the variational problem for the first variations (extremization
problem) as
\begin{eqnarray}
\label{dm9}
\delta E_{\rm tot}-\frac{\mu}{m}\delta M=0,
\end{eqnarray}
where $\mu$ (global chemical potential) is a Lagrange multiplier taking into
account the mass constraint, we obtain ${\bf u}={\bf 0}$ and
\begin{eqnarray}
\label{dm10}
Q+mV'(\rho)+m\Phi=\mu.
\end{eqnarray}
This relation is equivalent to Eq. (\ref{dm4}) provided that we make the
identification
$E=\mu$.\footnote{Using the results of Appendix \ref{sec_ti},
Eq. (\ref{dm10}) can
be interpreted as a quantum Gibbs condition $Q+m h+m\Phi=\mu$ expressing
the fact that the quantum
potential $Q/m$ $+$
the enthalpy $h=V'(\rho)$ (equal to the local chemical potential
$\mu_{\rm loc}(\rho)/m$)
$+$ the gravitational potential $\Phi$ is a constant equal to the global
chemical potential $\mu/m$.}
Therefore, the eigenenergy $E$ coincides with the global chemical
potential $\mu$. Eq.
(\ref{dm10}) is also equivalent to the condition of quantum hydrostatic
equilibrium (\ref{dm5}). Therefore,  an extremum of energy at fixed mass is an
equilibrium state of the
GPP equations. Furthermore, as shown in Appendix \ref{sec_alt}, among all
possible equilibria, only minima of energy
at
fixed mass are dynamically stable with respect to the GPP equations (maxima or
saddle points are linearly unstable). The
stability of an equilibrium state can be settled by studying the sign
of $\delta^2E_{\rm tot}$ or, equivalently, by linearizing the equations of
motion about the equilibrium state and investigating the sign of
the square pulsation $\omega^2$ (see Appendix \ref{sec_alt}). In each case,
these methods require to solve
a rather complicated eigenvalue equation. Alternatively, the
stability of an equilibrium state can be settled
more
directly by plotting the series of equilibria and using the Poincar\'e-Katz
\cite{poincare,katzpoincare} turning point criterion applied on the curve
$\mu(M)$ or the Wheeler
\cite{htww} theorem applied on the curve  $M(R)$ (see \cite{prd1,prd2,phi6} for
a specific application of these methods to the case of axion
stars). It may also be useful to plot the curve $E_{\rm
tot}(M)$ in order to compare the energy of different equilibrium states
with the same mass $M$. Since $\delta M=0 \Leftrightarrow \delta E_{\rm
tot}=0$ according to Eq. (\ref{dm9}), the extrema of mass coincide with the
extrema of energy in the series of equilibria. As a result, the curve  $E_{\rm
tot}(M)$ presents cusps at these critical points (see, e.g., Fig. 11 of
\cite{prd2} for illustration).

The fundamental equation of hydrostatic equilibrium of BECDM halos, 
Eq. (\ref{dm7}), has been solved analytically (using a Gaussian ansatz) in
\cite{prd1}, and numerically in \cite{prd2}, for an arbitrary self-interaction
(repulsive or attractive). It describes a compact quantum object (soliton/BEC). 
Because of quantum effects, the central density is finite instead of diverging
as in the CDM model. Therefore, quantum mechanics is able to solve the cusp-core
problem.

For noninteracting bosons ($a_s=0$), Eq. (\ref{dm7}) reduces to
\begin{eqnarray}
\label{dm11}
\frac{\hbar^2}{
2m^2}\Delta
\left (\frac{\Delta\sqrt{\rho}}{\sqrt{\rho}}\right
)=4\pi
G\rho.
\end{eqnarray}
The density profile is plotted in Fig. 2 of \cite{mcmh} using the results of
\cite{prd2}. The density has not a
compact support: the 
density decreases to zero at infinity exponentially
rapidly.
The mass-radius relation is given by \cite{membrado,prd2}
\begin{eqnarray}
\label{dm12}
M=9.95\frac{\hbar^2}{Gm^2R_{99}},
\end{eqnarray}
where $R_{99}$ represents the radius containing $99\%$ of the mass. 
The mass decreases as the radius increases. The equilibrium states are all
stable.

For bosons with a repulsive self-interaction ($a_s>0$), some density 
profiles are plotted in \cite{prd2}. The density has not a compact support
except in the TF limit (see below). The mass-radius relation is plotted in Fig.
4 of \cite{prd2} (see also Fig. \ref{M-R-chi-pos-part1} below). There is a
minimum radius (see Eq. (\ref{dm15}) below) reached for $M\rightarrow +\infty$.
The mass decreases as the radius increases. The equilibrium states are all
stable. In the TF limit, Eq. (\ref{dm7}) reduces to
\begin{eqnarray}
\label{dm13}
\frac{4\pi a_s\hbar^2}{m^3}\Delta\rho+4\pi G\rho=0.
\end{eqnarray}
This equation is equivalent to the Lane-Emden equation for a polytrope of index
$n=1$ \cite{chandra}. It has a simple
analytical solution\footnote{The Helmholtz-type equation (\ref{dm13})
and its solution (\ref{dm14}) have a long history. As mentioned by Chandrasekhar
\cite{chandra}, 
the analytical solution (\ref{dm14}) was first given by Ritter \cite{ritter} in
the context of self-gravitating polytropic spheres. Actually, this solution was
already familiar to Laplace \cite{laplace}. It corresponds indeed to Laplace's
celebrated law of density for the earth interior ($\sin (nr)/r$) which he
suggested as a consequence of supposing the earth to be a liquid globe, having
pressure increasing from the surface inward in proportion to the augmentation of
the square of the density.}
\begin{eqnarray}
\label{dm14}
\rho=\frac{\rho_0 R}{\pi r}\sin \left (\frac{\pi r}{R_{\rm TF}}\right ).
\end{eqnarray}
The density profile is plotted in Fig. 3 of \cite{mcmh}. In the TF limit, the
density has a compact support: the density vanishes at a finite radius $R_{\rm
TF}$. 
The equilibrium states have a unique radius given
by \cite{tkachev,maps,leekoh,goodman,arbey,bohmer,prd1}
\begin{eqnarray}
\label{dm15}
R_{\rm TF}=\pi\left (\frac{a_s\hbar^2}{Gm^3}\right )^{1/2},
\end{eqnarray}
independent of their mass $M$. In the noninteracting (NI) limit
$R\gg R_{\rm TF}$, we recover Eq. (\ref{dm12}).

\begin{figure}[!h]
\begin{center}
\includegraphics[clip,scale=0.3]{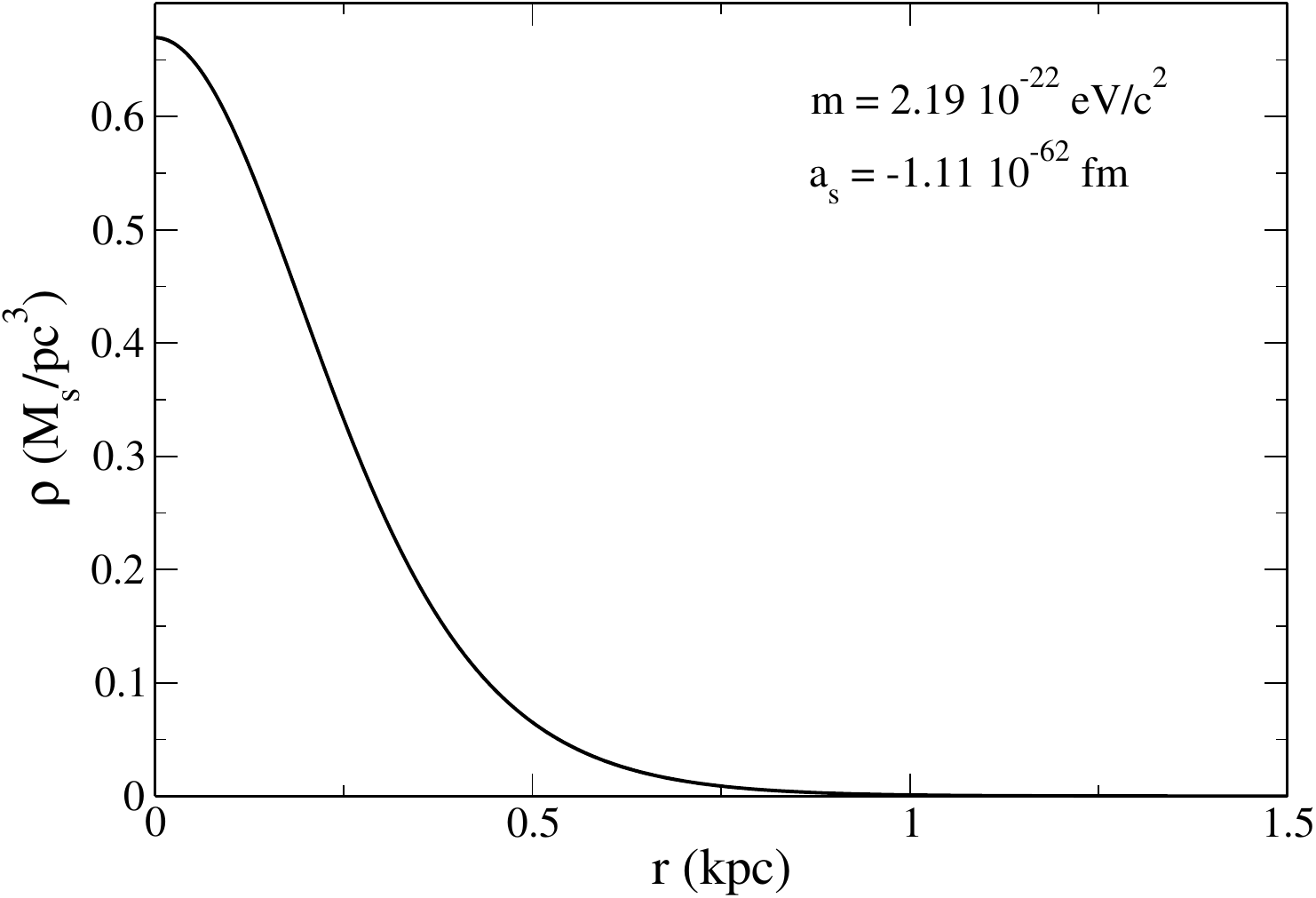}
\caption{Density profile of a self-gravitating BEC with an attractive
self-interaction at the maximum mass $M_{\rm max}$. We have adopted the
values $m=2.19\times 10^{-22}\, {\rm eV}/c^2$ and
$a_s=-1.11\times
10^{-62}\, {\rm fm}$ (see Sec. \ref{sec_paab}) corresponding to a DM halo of
mass $M=10^8\, M_{\odot}$ and radius $R=1\, {\rm kpc}$ (minimum halo).}
\label{profilecrit}
\end{center}
\end{figure}

For bosons with an attractive  self-interaction ($a_s<0$), some density profiles
are plotted in \cite{prd2}. The density has not a compact support. The
mass-radius relation is plotted in Fig. 6 of \cite{prd2} (see also Fig.
\ref{M-R-chi-neg-part1} below). 
There is a maximum mass \cite{prd1,prd2} 
\begin{eqnarray}
\label{dm16}
M_{\rm max}=1.012\, \frac{\hbar}{\sqrt{Gm|a_s|}}
\end{eqnarray}
at
\begin{eqnarray}
\label{dm17}
R_{99}^*=5.5\, \left (\frac{|a_s|\hbar^2}{Gm^3}\right )^{1/2}.
\end{eqnarray}
The density profile at the maximum mass is plotted in Fig. \ref{profilecrit}
using the results of \cite{prd2}. 
There is no equilibrium state with $M>M_{\rm max}$. In that case, the BEC is
expected to collapse \cite{bectcoll}. The outcome of the collapse (dense
axion star, black hole, bosenova...) is discussed in
\cite{braaten,cotner,bectcoll,ebycollapse,tkachevprl,helfer,phi6,visinelli,
moss}. For $M<M_{\rm max}$ the equilibrium
states with $R>R_{99}^*$ are stable and 
the equilibrium states with $R<R_{99}^*$ are
unstable.\footnote{This can be shown by using the Poincar\'e criterion, the
Wheeler theorem, or by computing the square pulsation \cite{prd1,prd2,phi6}.}
In the nongravitational (NG) limit $R\ll
R_{99}^*$, Eq. (\ref{dm7}) can be written as 
\begin{eqnarray}
\label{dm18}
-\frac{\hbar^2}{
2m}\frac{\Delta\sqrt{\rho}}{\sqrt{\rho}}+\frac{4\pi
a_s\hbar^2}{m^2}\rho=E.
\end{eqnarray}
It is equivalent to the standard stationary GP equation. The mass-radius
relation is given by (see, e.g., \cite{prd2})
\begin{eqnarray}
\label{dm19}
M=0.275\, \frac{m R_{99}}{|a_s|}.
\end{eqnarray}
These equilibrium states are unstable. In the NI limit $R\gg R_{99}^*$, we
recover Eqs. (\ref{dm11}) and (\ref{dm12}). These equilibrium states are stable.

{\it Remark:} We have seen that self-gravitating
BECs with an attractive self-interaction ($a_s<0$) can be at
equilibrium only below a maximum mass given
by Eq. (\ref{dm16}). Conversely, a self-gravitating BEC of mass
$M$ can be at equilibrium only if the scattering length of the bosons is above
a minimum negative value \cite{prd1,prd2}
\begin{eqnarray}
(a_s)_{\rm min}=-1.024\frac{\hbar^2}{GmM^2}.
\label{dm19b}
\end{eqnarray}

\subsection{Infinite homogeneous BEC in the linear regime: quantum Jeans problem}
\label{sec_j}

We now apply the  GPP equations (\ref{gpp1}) and  (\ref{gpp2}),
or equivalently the quantum Euler-Poisson equations (\ref{mad3})-(\ref{mad5}),
to the universe as a whole in order to study the initiation of
structure formation. Specifically, following \cite{prd1},
we study the linear dynamical stability of an infinite homogeneous
self-gravitating BEC with density $\rho$ and velocity ${\bf u}={\bf 0}$
described by the quantum Euler-Poisson equations
(\ref{mad3})-(\ref{mad5}). This is a generalization of the classical Jeans
problem \cite{jeans1902} to a quantum fluid.

Considering a small perturbation about an infinite homogeneous equilibrium
state, making the Jeans swindle \cite{bt,chavanisjeans}, 
and linearizing the hydrodynamic equations
(\ref{mad3})-(\ref{mad5}), we
obtain\footnote{See Refs. \cite{prd1,suarezchavanisprd3,jeansuniverse} for a
more detailed discussion and some comments about the Jeans swindle.}
\begin{equation}
\label{j1}
\frac{\partial\delta}{\partial t}+\nabla\cdot {\bf u}=0,
\end{equation}
\begin{equation}
\label{j2}
\frac{\partial{\bf u}}{\partial t}=-c_s^2\nabla
\delta-\nabla\delta\Phi+\frac{\hbar^2}{4m^2}\nabla (\Delta\delta),
\end{equation}
\begin{equation}
\label{j3}
\Delta\delta\Phi=4\pi G \rho\delta,
\end{equation}
where $c_s^2=P'(\rho)$ is the square of the speed of sound and
$\delta({\bf r},t)=\delta\rho({\bf r},t)/\rho$ the density contrast. Taking the
time derivative of Eq. (\ref{j1}) and the divergence of
Eq. (\ref{j2}), and using the Poisson equation (\ref{j3}), we obtain a single
equation for the density contrast
\begin{eqnarray}
\label{j4}
\frac{\partial^2\delta}{\partial t^2}=-\frac{\hbar^2}{4m^2}\Delta^2\delta+c_s^2\Delta\delta+4\pi G\rho\delta.
\end{eqnarray}
Expanding the solutions of this equation into plane waves of the form
$\delta({\bf
r},t)\propto {\rm exp}\lbrack i({\bf k}\cdot{\bf r}-\omega t)\rbrack$, we
obtain the general dispersion relation \cite{prd1}
\begin{eqnarray}
\label{j5}
\omega^2=\frac{\hbar^2k^4}{4m^2}+c_s^2k^2-4\pi G\rho.
\end{eqnarray}
This quantum dispersion relation may also be obtained from the
gravitational Bogoliubov equations (see Appendix D of Ref.
\cite{jeansuniverse}). For $\hbar=0$, we recover the classical Jeans dispersion
relation. The dispersion relation (\ref{j5}) is studied in detail in
\cite{prd1,suarezchavanisprd3,jeansuniverse}.  The
generalized Jeans
wavenumber $k_J$, corresponding to $\omega=0$, is
determined by the quadratic equation
\begin{eqnarray}
\label{j6}
\frac{\hbar^2k_J^4}{4m^2}+c_s^2k_J^2-4\pi G\rho=0.
\end{eqnarray}
It is given
by \cite{prd1}
\begin{eqnarray}
\label{j7}
k_J^2=\frac{2m^2}{\hbar^2}\left (-c_s^2+\sqrt{c_s^4+\frac{4\pi
G\rho\hbar^2}{m^2}}\right).
\end{eqnarray}
The Jeans length is $\lambda_J=2\pi/k_J$. The Jeans radius and the Jeans mass
are defined by
\begin{equation}
\label{j8}
R_J=\frac{\lambda_J}{2}=\frac{\pi}{k_J},\qquad M_J=\frac{4}{3}\pi\rho R_J^3.
\end{equation}
They represent the
minimum radius and the minimum mass of a fluctuation
that can collapse at a given
epoch. They are therefore expected to provide an order of magnitude 
of the minimum size and minimum mass of DM halos interpreted as
self-gravitating BECs.\footnote{This is only an order of magnitude because
the true mass and the true size of the structures is determined by  the complex
evolution of the system in
the nonlinear regime.}

Extending the Jeans instability study for a self-gravitating BEC in an expanding
universe, using the equations of Appendix \ref{sec_eu}, we find that the
evolution of the density contrast $\delta_{\bf k}(t)$ is determined by the
equation \cite{aacosmo}
\begin{eqnarray}
\label{lin5}
\ddot\delta+2\frac{\dot a}{a}\dot\delta+\left
(\frac{\hbar^2k^4}{4m^2a^4}+\frac{c_s^2k^2}{a^2}-4\pi G\rho_b\right )\delta=0.
\end{eqnarray}
This equation extends the classical Bonnor equation to a quantum gas. A
detailed study of this equation  has been performed in
Refs. \cite{aacosmo,suarezchavanisprd3}. In a static universe
($a=1$), writing
$\delta\propto e^{-i\omega t}$, we recover the dispersion relation
(\ref{j5}). The comoving Jeans length is $\lambda^c_J=\lambda_J/a$.

{\it Remark:} for the CDM model for which $\hbar=0$ and $c_s\simeq 0$, we find
that  $\lambda_J\simeq 0$. This implies that structures can form at all scales.
This is not what is observed and this is why the BECDM model has been
introduced. In that case, there is a nonzero Jeans length even at $T=0$ because
of quantum effects.

\section{Mass-radius relation of BECDM halos from the $f$-ansatz}
\label{sec_pa}

In Ref. \cite{prd1}, using a Gaussian ansatz for the wave function, we have
obtained an approximate analytical expression of the mass-radius relation of
self-gravitating BECs. In Appendix \ref{sec_mra}, we show that the form of this
relation is independent of the ansatz. Indeed, it is always given by
\begin{equation}
\label{pa1}
M=\frac{a\frac{\hbar^2}{Gm^2R}}{1-b^2\frac{a_s\hbar^2}{Gm^3R^2}},
\end{equation}
where only the value of the coefficients $a$ and $b$ depends on the ansatz. 
Here, we shall determine the coefficients $a$ and $b$ so as to recover the exact
mass-radius relation in some particular limits. Once the mass-radius relation is
known, we can compute the average density of the DM halo by
\begin{equation}
\label{pa2}
\rho= \frac{3M}{4\pi R^3}.
\end{equation}

{\it Remark:} With the Gaussian ansatz, we get $a_G^*=2\sigma_G/\nu_G=3.76$ and
$b_G^*=(6\pi\zeta_G/\nu_G)^{1/2}=1.73$, where we have used Eq. (\ref{mrga})
with $\sigma_G=3/4$,  $\zeta_G=1/(2\pi)^{3/2}$ and $\nu_G=1/\sqrt{2\pi}$.
However, below, we shall identify the radius $R$
with $R_{99}$, not with the radius $R$ of the $f$-ansatz defined in Eq.
(\ref{mra1}). Since
$R_{99}=2.38167\, R$ for the Gaussian ansatz, we obtain $a_G=2.38167\,
a_G^*=8.96$ and $b_G=2.38167\,
b_G^*=4.12$ to be compared with the more exact values of $a$ and
$b$ found below. 

\subsection{Noninteracting bosons}
\label{sec_panb}

For noninteracting bosons ($a_s=0$), the mass-radius relation from Eq. (\ref{pa1}) reduces to
\begin{equation}
\label{pa3}
M=a\frac{\hbar^2}{Gm^2R}.
\end{equation}
The mass decreases as the radius increases. 
If we identify $R$ with the radius $R_{99}$ containing $99\%$ of the mass and
compare Eq. (\ref{pa3}) with the exact mass-radius relation of noninteracting
self-gravitating BECs from Eq. (\ref{dm12}), we get $a=9.946$.

The average density is given by 
\begin{equation}
\label{pa4}
\rho=\frac{3a}{4\pi}\frac{\hbar^2}{Gm^2R^4}=
\frac{3}{4\pi a^3}\frac{G^3m^6M^4}{\hbar^6}.
\end{equation}
The density decreases along the series of equilibria going 
from small radii to large radii. The equilibrium states are all stable.

\subsection{Repulsive self-interaction}
\label{sec_parb}

\begin{figure}[!h]
\begin{center}
\includegraphics[clip,scale=0.3]{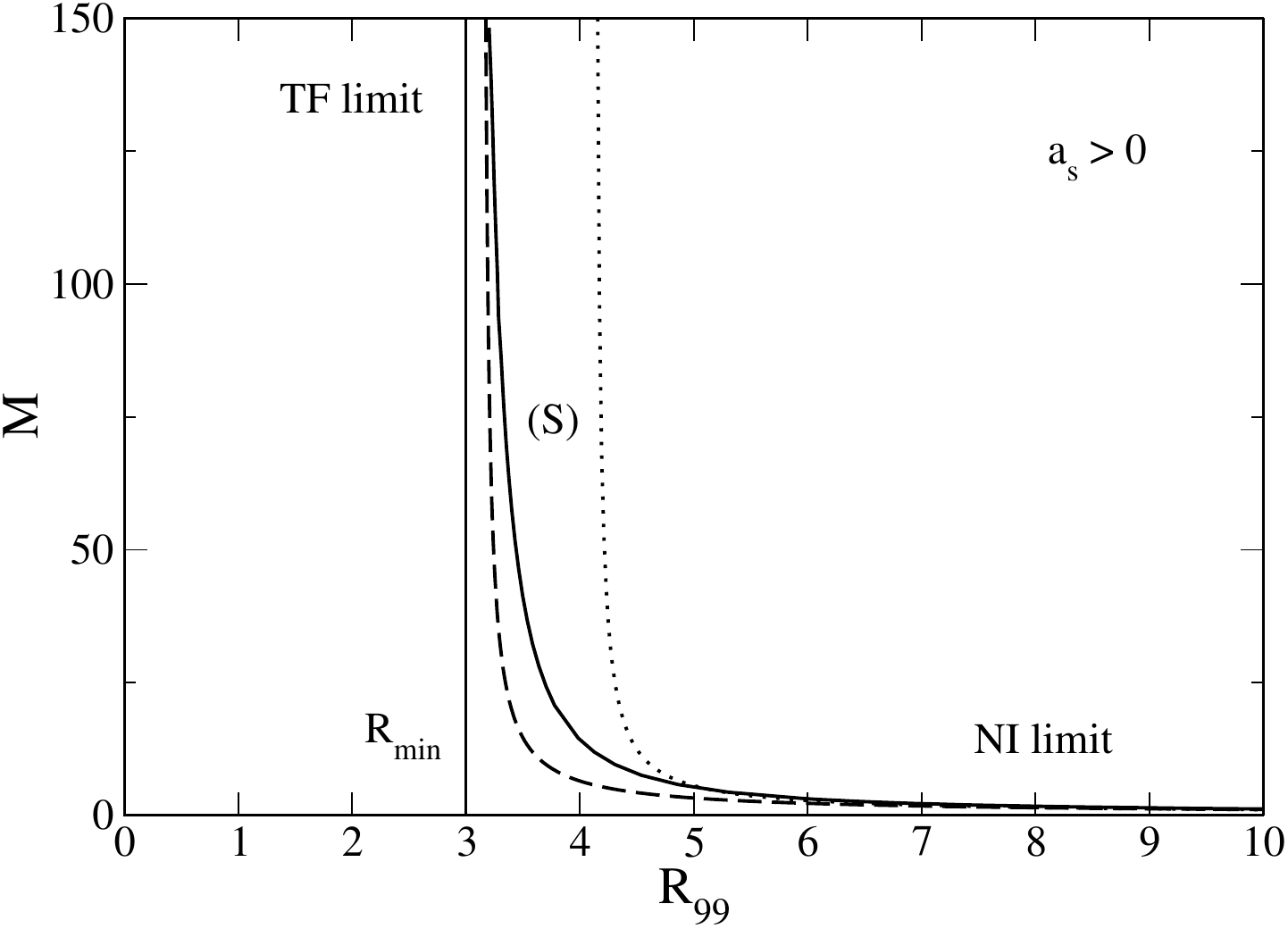}
\caption{Mass-radius relation of self-gravitating BECs with $a_s>0$ (full line:
exact \cite{prd2}; dotted line: Gaussian ansatz \cite{prd1}; dashed line: fit
from Eq. (\ref{pa1}) with $a=9.946$ and $b=\pi$). The mass is normalized by
$M_a=\hbar/\sqrt{Gma_s}$ and the radius by $R_a=(a_s\hbar^2/Gm^3)^{1/2}$.}
\label{M-R-chi-pos-part1}
\end{center}
\end{figure}

For bosons with a repulsive self-interaction ($a_s>0$), 
the exact mass-radius relation is represented in Fig. \ref{M-R-chi-pos-part1}.
The mass decreases as the radius increases. In the TF limit ($\hbar\rightarrow
0$ with $a_s\hbar^2\neq 0$), the mass-radius
relation from Eq. (\ref{pa1}) reduces to
\begin{equation}
\label{pa5}
R=b\left (\frac{a_s\hbar^2}{Gm^3}\right )^{1/2}.
\end{equation}
The radius is independent of the mass. If we identify $R$ with 
the radius at which the density vanishes and compare Eq. (\ref{pa5}) with the
exact radius of self-gravitating BECs in the TF limit from Eq. (\ref{dm15}), we
get $b=\pi$. On the other
hand, in the NI limit, we recover the
result from Eq. (\ref{pa3}) leading to $a=9.946$. We shall adopt these values of
$a$ and $b$ in the repulsive case (see Fig. \ref{M-R-chi-pos-part1} for a
comparison with the exact result).

The average density decreases along the series of equilibria going from small 
radii to large radii, i.e., from the TF regime to the NI regime. In the TF
regime, the average density is given by
\begin{equation}
\label{pa6}
\rho=\frac{3M}{4\pi b^3}\left (\frac{Gm^3}{a_s\hbar^2}\right )^{3/2}.
\end{equation}
In the NI regime it is given by Eq. (\ref{pa4}). 
The equilibrium states are all stable.

The
transition between the TF regime and the NI regime (obtained by
equating Eqs. (\ref{pa3}) and (\ref{pa5})) typically occurs at
\begin{equation}
\label{pa7}
M_{t}=\frac{a}{b} \frac{\hbar}{\sqrt{Gm a_s}},\quad R_t=b \left
(\frac{a_s\hbar^2}{Gm^3}\right )^{1/2},\quad \rho_{t}=\frac{3a}{4\pi b^4}
\frac{Gm^4}{a_s^2\hbar^2}.
\end{equation}
We are in the TF regime when $M\gg M_t$ and $R\sim R_t$. We are in the NI
regime when $M\ll M_t$ and $R\gg R_t$. Note that $R_t$
corresponds to the minimum radius $R_{\rm min}$ (i.e., the radius in the TF
regime).

\subsection{Attractive self-interaction}
\label{sec_paab}

\begin{figure}[!h]
\begin{center}
\includegraphics[clip,scale=0.3]{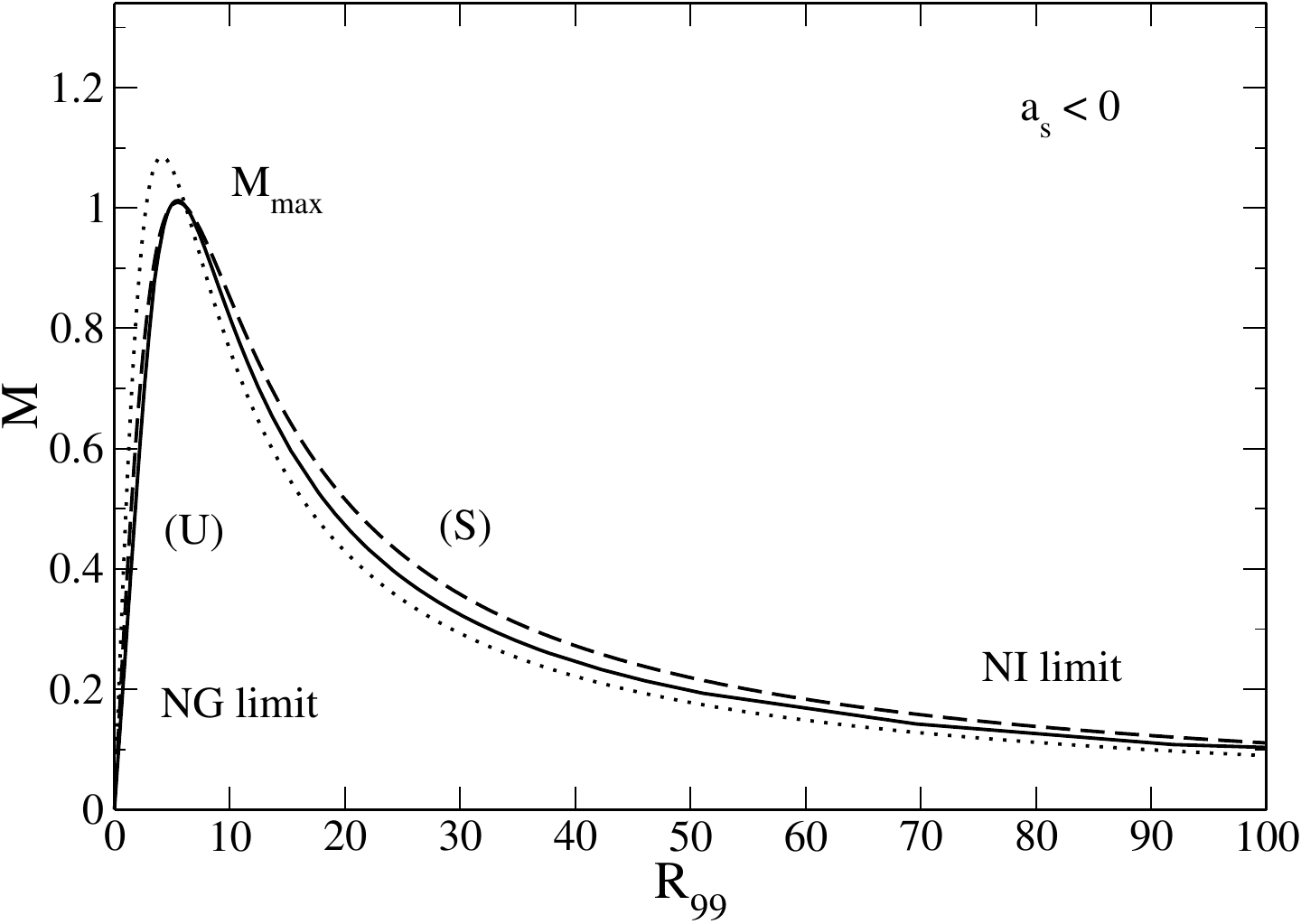}
\caption{Mass-radius relation of self-gravitating BECs with $a_s<0$ (full line:
exact \cite{prd2}; dotted line: Gaussian ansatz \cite{prd1}; dashed line: fit
from Eq. (\ref{pa1}) with $a=11.1$ and $b=5.5$). The mass is normalized by
$M_a=\hbar/\sqrt{Gm|a_s|}$ and the radius by
$R_a=(|a_s|\hbar^2/Gm^3)^{1/2}$.}
\label{M-R-chi-neg-part1}
\end{center}
\end{figure}

For bosons with an attractive self-interaction ($a_s<0$), 
the exact mass-radius relation is represented in
Fig. \ref{M-R-chi-neg-part1}. The mass increases as the radius increases,
reaches a maximum value
\begin{equation}
\label{pa8}
M_{\rm max}=\frac{a}{2b} \frac{\hbar}{\sqrt{Gm|a_s|}}\qquad {\rm at}\qquad R_*=b\left (\frac{|a_s|\hbar^2}{Gm^3}\right )^{1/2},
\end{equation}
and decreases.  If we identify $R_*$ with the radius $(R_*)_{99}$ 
containing $99\%$ of the mass and compare Eq. (\ref{pa8}) with the exact values
of the maximum mass and of the corresponding  radius from Eqs. (\ref{dm16}) and
(\ref{dm17}), we get $b=5.5$ and $a/2b=1.012$, leading to $a=11.1$. We shall
adopt these values in the attractive case (see Fig. \ref{M-R-chi-neg-part1} for
a comparison with the exact result). We note that the value  $a=11.1$
obtained from the maximum mass is relatively close to  the value $a=9.946$
obtained from the NI limit (see Sec. \ref{sec_panb}). In the NG
limit, the mass-radius relation from Eq. (\ref{pa1}) reduces to
\begin{equation}
\label{pa9}
M=\frac{a}{b^2} \frac{mR}{|a_s|}.
\end{equation}
The value $a/b^2=0.367$ obtained from the maximum mass is relatively close to
the exact value $0.275$ from Eq.
(\ref{dm19}). This is a consistency check.
The density decreases along the series of equilibria going from small radii to
large radii, i.e., from the NG regime to the NI regime. In the NG regime, the
average density is given by 
\begin{equation}
\label{pa10}
\rho=\frac{3a}{4\pi b^2} \frac{m}{|a_s|R^2}=\frac{3a^3}{4\pi
b^6}\frac{m^3}{|a_s|^3M^2}.
\end{equation}
In the NI regime it is given by Eq. (\ref{pa4}). The average 
density at the maximum mass is 
\begin{equation}
\label{pa11}
\rho_{*}=\frac{3a}{8\pi b^4} \frac{Gm^4}{a_s^2\hbar^2}.
\end{equation}
The equilibrium states are unstable before the turning point of mass
($R<R_*$) and  stable after the turning point of mass ($R>R_*$).

The transition between the NG regime and the NI regime (obtained by
equating Eqs. (\ref{pa3}) and (\ref{pa9})) typically occurs at
\begin{equation}
\label{pa7w}
M_{t}=\frac{a}{b} \frac{\hbar}{\sqrt{Gm |a_s|}},\, R_t=b \left
(\frac{|a_s|\hbar^2}{Gm^3}\right )^{1/2},\, \rho_{t}=\frac{3a}{4\pi b^4}
\frac{Gm^4}{a_s^2\hbar^2}.
\end{equation}
These
scales are similar to those corresponding to the maximum mass (we have
$\rho_t=2\rho_*$, $M_{t}=2 M_{\rm max}$ and
$R_t=R_*$).
We are in the NG regime when $M\ll M_t$ and $R\ll R_t$ but these equilibrium
states are unstable. We
are in the NI regime when $M\ll M_t$ and $R\gg R_t$. There is no equilibrium
state of mass $M>M_{\rm max}$. 

{\it Remark:} the scales (\ref{pa7w}) determining the transition between the NG
regime and the  NI regime in the attractive case are similar to the scales
(\ref{pa7}) determining the transition between the TF regime and the NI regime
in the
repulsive case provided that $a_s$ is replaced by $|a_s|$.

\section{Dark matter particle mass - scattering length relation}
\label{sec_mas}

\begin{figure}[!h]
\begin{center}
\includegraphics[clip,scale=0.3]{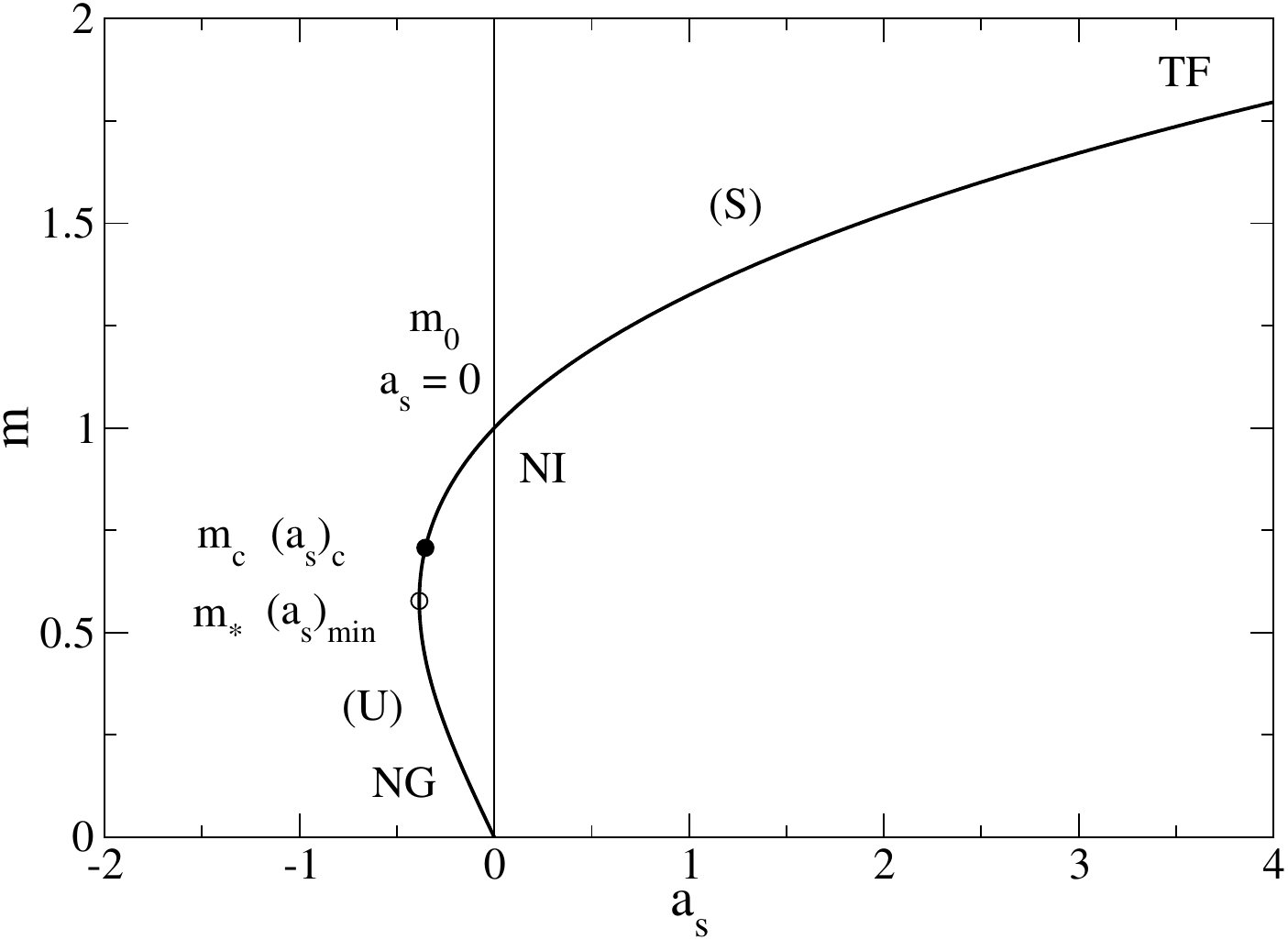}
\caption{Mass $m$ of the DM particle as a function of the scattering
length $a_s$ in order to match the characteristics of the minimum halo. The
mass is normalized by
$m_0$ and the scattering length by $a'_*$. The stable part of
the curve starts at the critical minimum halo point ($(a_s)_c,m_{c}$). It
differs from the minimum of the curve $a_s(m)$.}
\label{abism}
\end{center}
\end{figure}

As explained previously, in the BECDM model, the mass-radius relation
(\ref{pa1}) of a self-gravitating BEC at $T=0$ (ground state) describes the
smallest halos observed in the Universe.\footnote{It also describes the
quantum core of larger DM halos \cite{mcmh}.} They correspond to dSphs like
Fornax.
From the observations, these ultracompact DM halos have a typical radius
$\sim1\, {\rm kpc}$ and a typical mass $\sim 10^8\, M_{\odot}$. To fix the
ideas, we shall consider a ``minimum halo'' of radius and
mass\footnote{If more accurate values of $R$ and $M$ are adopted,
the numerical applications of our paper would slightly change. However, the main
ideas and the main results would remain substantially the same.}
\begin{equation}
\label{mas1}
R=1\, {\rm kpc},\qquad M=10^8\, M_{\odot}\qquad ({\rm Fornax}).
\end{equation}
Its average density is 
\begin{equation}
\label{mas1b}
\rho=\frac{3M}{4\pi R^3}=1.62\times 10^{-18}\, {\rm g/m^3} \qquad
({\rm Fornax}).
\end{equation}
Since $R$ and $M$ are prescribed by
Eq. (\ref{mas1}), then Eq. (\ref{pa1}) provides a relation
\begin{equation}
\label{asm1}
a_s=\frac{amR}{b^2M}\left (\frac{GMm^2R}{a\hbar^2}-1\right )
\end{equation}
between the mass $m$
and the scattering length $a_s$ of the bosonic DM particle. Such a relation is
necessary to obtain a minimum halo consistent with the observations. The DM
particle
mass-scattering length relation (\ref{asm1}) may be written as
\begin{equation}
\label{mas2}
\frac{a_s}{a'_*}=\left (\frac{m}{m_0}\right )^3-\frac{m}{m_0},
\end{equation}
where we have
introduced the
scales
\begin{equation}
\label{mas3}
m_0=\left (\frac{a\hbar^2}{GMR}\right )^{1/2}
\end{equation}
and
\begin{equation}
\label{mas4}
a'_*=\frac{a^{3/2}}{b^2}\left (\frac{\hbar^2 R}{GM^3}\right
)^{1/2}.
\end{equation}
The relation $m(a_s)$ is plotted in Fig. \ref{abism}. Taking
$a=9.946$
and $b=\pi$ (see Secs. \ref{sec_panb} and \ref{sec_parb}) adapted to bosons
with a repulsive self-interaction (or no interaction), we get $m_0=2.92\times
10^{-22}\, {\rm eV}/c^2$ and $a'_*=8.13\times 10^{-62}\, {\rm fm}$. Taking
$a=11.1$ and $b=5.5$ (see Sec. \ref{sec_paab}) adapted to bosons with an
attractive
self-interaction, we get $m_0=3.08\times 10^{-22}\,
{\rm eV}/c^2$ and $a'_*=3.12\times 10^{-62}\, {\rm fm}$.

\subsection{Noninteracting bosons}

For noninteracting bosons ($a_s=0$), we get
\begin{equation}
\label{mas5}
m=m_{0}=2.92\times
10^{-22}\, {\rm eV}/c^2 \qquad ({\rm BECNI})
\end{equation}
This is the typical mass considered in the literature when the bosons are
assumed to be noninteracting.

{\it Remark:} In the NI regime, the
mass from Eq. (\ref{mas5}) can be written as
\begin{equation}
\label{mas7qw}
m=\left (\frac{a\hbar^2}{GMR}\right )^{1/2},
\end{equation}
which is equivalent to Eq. (\ref{pa3}).

\subsection{Repulsive self-interaction}
\label{sec_rsi}

\begin{figure}[!h]
\begin{center}
\includegraphics[clip,scale=0.3]{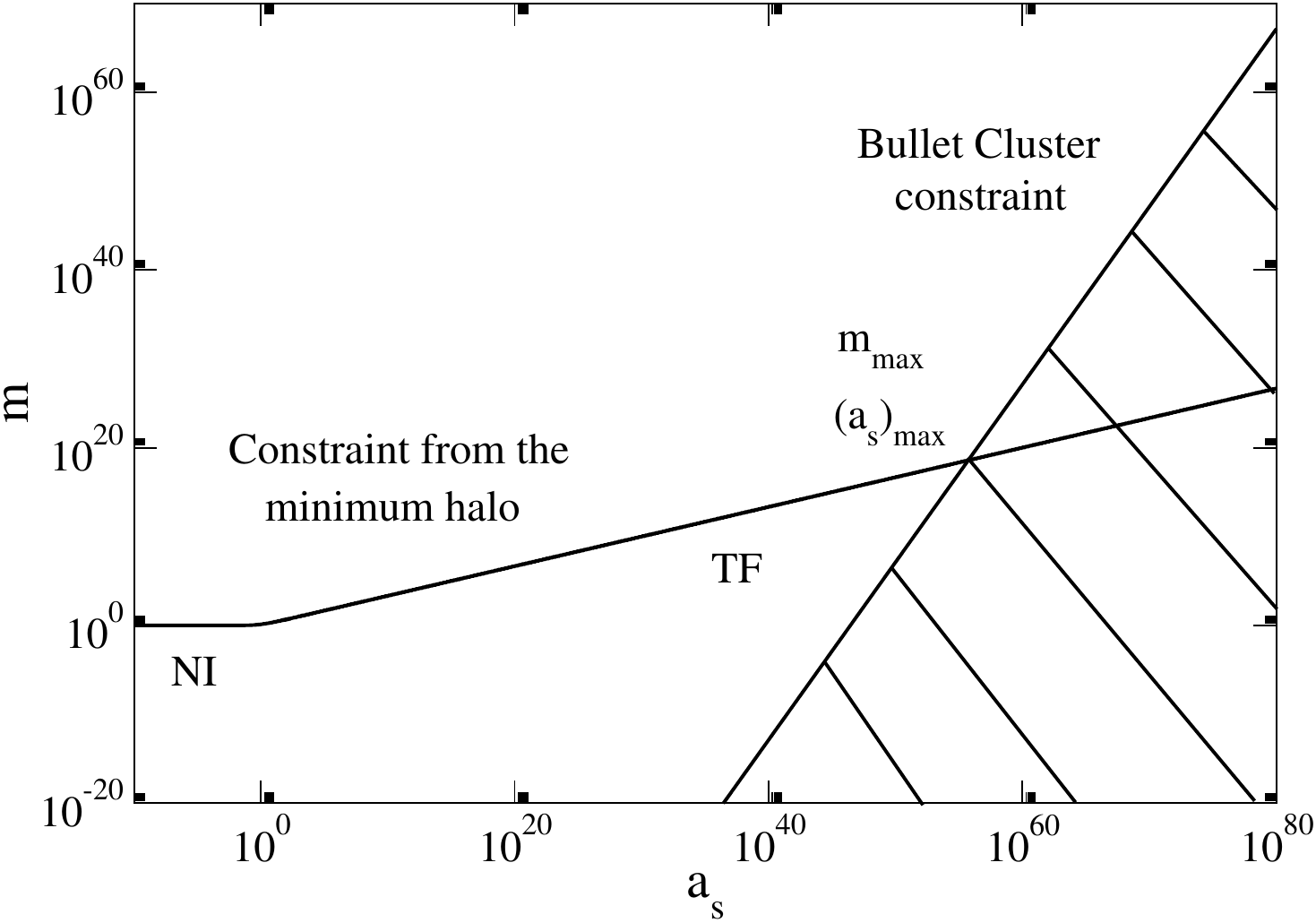}
\caption{The initially upper curve gives the DM particle mass versus
scattering length relation 
in order to match the characteristics of the minimum halo [see Eq.
(\ref{mas2})]. The mass is normalized by $m_0$ and the scattering length by
$a'_*$. The initially lower curve gives the Bullet Cluster constraint from Eq.
(\ref{mas8}). Only the region above this curve is allowed by the observations.
The intersection between these two curves determines the maximum DM particle
mass $m_{\rm max}/m_0=3.79\times 10^{18}$ and its maximum 
scattering length $(a_s)_{\rm max}/a'_*=5.45\times 10^{55}$, 
leading to the results of Eq. (\ref{mas9}). We note that the intersection occurs
in the TF
regime where Eq. (\ref{mas2}) can be approximated by Eq.
(\ref{mas6}).}
\label{ambc}
\end{center}
\end{figure}

For bosons with a repulsive self-interaction ($a_s>0$),  $a'_*$ determines the transition between the NI regime ($a_s\ll a'_*$) where $m\sim m_0$ and the TF regime ($a_s\gg a'_*$) where
\begin{equation}
\label{mas6}
\frac{m}{m_0}\sim \left (\frac{a_s}{a'_*}\right )^{1/3}.
\end{equation}
When the self-interaction is repulsive, we have seen that all the equilibrium
states are stable. 
Therefore, in principle, all the scattering lengths $a_s>0$ and the
corresponding masses $m>m_0$ are possible. In the TF regime, the mass-scattering
length relation (\ref{mas6}) can be written as
\begin{equation}
\label{mas7}
\frac{a_s}{m^3}\sim \frac{GR^2}{b^2\hbar^2},
\end{equation}
which is equivalent to Eq. (\ref{pa5}).
The minimum halo [Eq. (\ref{mas1})] just determines the ratio 
\begin{equation}
\frac{a_s}{m^3}=3.28\times 10^3\, {\rm
fm}/({\rm eV/c^2})^3.
\end{equation}
Note that only the radius $R$ of the minimum
halo matters in this
determination. In order to determine $m$ and $a_s$ individually, we need another
equation. Observations of the Bullet Cluster give the  constraint $\sigma/m\le
1.25\, {\rm cm^2/g}$ where $\sigma=4\pi a_s^2$ is the self-interaction cross
section \cite{bullet}. This can be written as
\begin{equation}
\label{mas8}
\frac{4\pi a_s^2}{m}\le 1.25\, {\rm cm^2/g}\qquad  \Leftrightarrow 
\qquad \frac{(a_s/a'_*)^2}{m/m_0}\le 7.83\times 10^{92}.
\end{equation}
If we replace the inequality by an equality, and combine Eq. (\ref{mas8}) with
Eq. (\ref{mas6}), we find that the mass and scattering 
length of the DM particle are given by\footnote{Craciun and
Harko \cite{craciun} obtained a similar
estimate.
However, they
took a larger BECDM radius $R=10\, {\rm kpc}$ instead of $R=1\, {\rm kpc}$
because they modeled large DM halos by a pure BEC at
$T=0$ in its ground state  while we argue that the ground state
solution leading to Eq. (\ref{dm15}) only applies to the  minimum halo with
$M=10^8\, M_{\odot}$ and $R=1\, {\rm kpc}$ and to the
quantum  core of size $R_c=1\, {\rm kpc}$ of larger DM halos
(recall that
the quantum core is
surrounded by an approximately isothermal atmosphere due to the quantum
interferences of excited states) \cite{modeldm,mcmh}. Since they
applied Eq. (\ref{dm15}) to the whole DM halo instead of just its core as we do,
they found a smaller maximum mass
$m_{\rm max}=0.1791\, {\rm meV}/c^2$ instead of $m_{\rm max}=1.10\, {\rm
meV}/c^2$.} 
\begin{eqnarray}
\label{mas9}
m_{\rm max}=1.10\times
10^{-3}\, {\rm eV}/c^2,\quad (a_s)_{\rm
max}=4.41\times 10^{-6}\, {\rm fm} \nonumber\\
({\rm BECTF})\qquad
\end{eqnarray}
More generally, because of the Bullet Cluster constraint, the scattering length
of the DM boson must lie in the range $0\le a_s\le (a_s)_{\rm max}$ and its mass
must lie in the range $m_0\le m\le m_{\rm max}$ (see Fig. \ref{ambc}).
Therefore, when we account for
a repulsive self-interaction, the mass $m$ of the boson required to match the
observations of the minimum halo 
can increase by $\sim 18$ orders of magnitude as compared to
its value $m_0$ in the NI case [see Eqs. (\ref{mas5}) and (\ref{mas9})].

The BECTF model discussed
previously corresponds 
to the case where the bound fixed by the  Bullet Cluster is reached. For
comparison, we can consider a BECt model which corresponds to the transition
between the NI limit and the TF limit. It is
obtained by substituting Eq. (\ref{mas5}) into Eq. (\ref{mas6}), or
Eq. (\ref{mas7qw}) into Eq. (\ref{mas7}),
giving 
\begin{eqnarray}
\label{mas10}
m_{\rm t}=2.92\times 10^{-22}\, {\rm eV}/c^2,\quad (a_s)_{\rm t}=8.13\times
10^{-62}\, {\rm fm}\nonumber\\
 ({\rm BECt})\qquad
\end{eqnarray}
This corresponds to the scales $m_0$ and $a'_*$
defined
by Eqs. (\ref{mas3}) and (\ref{mas4}).

\subsection{Attractive self-interaction}

For bosons with an attractive self-interaction ($a_s<0$), the relation
(\ref{mas2}) reveals 
the existence of a minimum scattering length
\begin{equation}
\label{mas11}
\frac{(a_s)_{\rm min}}{a'_*}=-\frac{2}{3\sqrt{3}}\qquad {\rm at \quad
which}\qquad \frac{m_*}{m_0}=\frac{1}{\sqrt{3}}.
\end{equation}
We find $(a_s)_{\rm min}=-1.20\times 10^{-62}\, {\rm fm}$ and
$m_*=1.78\times 10^{-22}\, {\rm eV}/c^2$. The NI regime corresponds to $|a_s|\ll
a'_*$ and $m\sim m_0$. The NG regime corresponds to $|a_s|\ll a'_*$ and $m\ll
m_0$ such that
\begin{equation}
\label{mas12}
\frac{m}{m_0}\sim \frac{|a_s|}{a'_*}.
\end{equation}
In the NG regime, the mass-scattering length relation
(\ref{mas12}) can
be written as
\begin{equation}
\label{mas12qw}
\frac{|a_s|}{m}=\frac{aR}{b^2M}=1.01\times 10^{-40}\, {\rm
fm}/({\rm eV/c^2}),
\end{equation}
which is equivalent to Eq. (\ref{pa9}). It is
important to note that the minimum 
scattering length $(a_s)_{\rm min}$ does {\it not} correspond to the critical
point (associated with the maximum mass $M_{\rm max}$) separating stable from
unstable equilibrium states. This latter is located at
\begin{equation}
\label{mas13b}
\frac{(a_s)_{c}}{a'_*}=-\frac{1}{2\sqrt{2}}, \qquad      \frac{m_c}{m_0}=\frac{1}{\sqrt{2}}.
\end{equation}
The equilibrium states with $m<m_c$ are unstable (they correspond to 
configurations with $R<R_*$) so that only the equilibrium states with $m>m_c$
are stable (they correspond to configurations with $R>R_*$). Therefore, in the
attractive case, the scattering length of the DM boson must lie in the range
$(a_s)_c<a_s<0$ and its mass must lie in the range $m_c<m<m_0$,
with
\begin{eqnarray}
m_{\rm c}=2.19\times 10^{-22}\, {\rm eV}/c^2,\quad (a_s)_{c}=-1.11\times
10^{-62}\, {\rm fm}\nonumber\\
 ({\rm BECcrit})\qquad
\label{beccrit}
\end{eqnarray}
There
is no equilibrium state with $a_s<(a_s)_{\rm min}$ and
the equilibrium states with $(a_s)_{\rm min}<a_s<(a_s)_{c}$ are unstable. We
note that, in the attractive case, the mass $m$ does not change substantially
from its value $m_0$ in the NI limit. The BECcrit model
from Eq. (\ref{beccrit}) corresponds to the case where the minimum halo is
critical (i.e. its mass $M=10^8\, M_{\odot}$ is equal to $M_{\rm max}$).

\subsection{Constraints from particle physics and
cosmology}
\label{sec_cpp}

For bosons with an attractive self-interaction, like the axion
\cite{marshrevue}, it is more
convenient to express the results in terms of the decay constant $f$ 
instead of the scattering length $a_s$. They are related by (see, e.g.,
\cite{phi6})
\begin{equation}
\label{mrr19}
f=\left (\frac{\hbar c^3 m}{32\pi |a_s|}\right )^{1/2}.
\end{equation}
 Particle physics and
cosmology lead to the following relation between $f$ and $m$ \cite{hui}:
\begin{equation}
\label{hui1}
\Omega_{\rm axion}\sim 0.1 \left (\frac{f}{10^{17}\, {\rm GeV}}\right )^2\left
(\frac{m}{10^{-22}\, {\rm eV}/c^2}\right )^{1/2},
\end{equation}
where $\Omega_{\rm axion}$ is the present fraction of axions in the
universe. Taking $\Omega_{\rm axion}\sim \Omega_{\rm m,0}=0.3089$ (assuming
that DM is exclusively made of axions), this relation can be
rewritten as
\begin{equation}
\label{hui2}
\frac{m^{3/2}}{|a_s|}=1.57\times 10^{35}\, ({\rm eV/c^2})^{3/2}/{\rm fm}
\end{equation}
or, in dimensionless form, as
\begin{equation}
\label{hui3}
\frac{(m/m_0)^{3/2}}{|a_s|/a'_*}=9.06\times 10^{5}.
\end{equation}
Considering the intersection between the curves defined by Eqs.
(\ref{mas2}) and (\ref{hui3}), we find that $m\simeq m_0$. Then, taking 
$m=m_0=2.92\times
10^{-22}\, {\rm eV}/c^2$ [see Eq. (\ref{mas5})] and using Eq. (\ref{hui2}) we
get $a_s=-3.18\times 10^{-68}\, {\rm fm}$. Therefore, we can determine $a_s$ and
$m$ {\it individually}. We find
\begin{eqnarray}
m_{\rm th}=2.92\times 10^{-22}\, {\rm eV}/c^2,\quad (a_s)_{\rm th}=-3.18\times
10^{-68}\, {\rm fm}\nonumber\\
 ({\rm BECth})\qquad
\end{eqnarray}
We note that $m$ has approximately the same value as in the
noninteracting model while $a_s$ has a nonzero value.
It corresponds to a decay constant $f_{\rm th}=1.34\times 10^{17}\, {\rm GeV}$.
Interestingly, $f$ lies in the range
$10^{16}\, {\rm GeV}\le f\le 10^{18}\, {\rm GeV}$ expected in particle physics
 ($f$ is bounded above by the reduced Planck mass and below by the
grand unified scale of particle physics) \cite{hui}. We note that $|a_s|_{\rm
th}\ll a'_*$ so we are essentially in the NI regime. This is confirmed by the
following discussion.

The maximum mass \cite{prd1} of a self-gravitating BEC made of bosons with mass
$m_{\rm
th}=2.92\times 10^{-22}\, {\rm eV}/c^2$ and scattering length $(a_s)_{\rm
th}=-3.18\times 10^{-68}\, {\rm fm}$  is $M_{\rm max}=5.10\times 10^{10}\,
M_{\odot}$ and the corresponding radius is $R_{99}^*=1.09\, {\rm pc}$. The
minimum halo ($M=10^8\, M_{\odot}$, $R=1\, {\rm kpc}$) has a mass much smaller
than the maximum mass, so it is stable ($M<M_{\rm
max}$). Larger halos have a core-halo structure
with a quantum core and an approximately isothermal halo. The mass $M_c$ of the
core increases
with the halo mass $M_h$. Therefore, above a critical halo mass value
$(M_h)_{\rm max}$, the core mass passes above the maximum mass ($M_c>M_{\rm
max}$) and collapses. The outcome of the collapse (dense
axion star, black hole, bosenova...) is discussed in
\cite{braaten,cotner,bectcoll,ebycollapse,tkachevprl,helfer,phi6,visinelli,
moss}. The
core mass -- halo mass relation of self-interacting bosons has been determined
in \cite{mcmh,mcmhbh}
(without or with the presence of a central black hole). It is found that the
maximum halo mass (at which $M_c=M_{\rm max}$) is given by \cite{mcmh}
\begin{equation}
\label{mrr20}
(M_h)_{\rm max}=2255
\frac{f^4}{\hbar^2 c^6 \Sigma_0},
\end{equation}
where $\Sigma_0=141\, M_{\odot}/{\rm pc}^2$ is the universal surface density of
DM halos \cite{kormendy,spano,donato}. We note that the maximum halo
mass $(M_h)_{\rm max}$ depends only on $f$ while the  maximum core mass $M_{\rm
max}$ depends on $f$ and $m$ [see Eq. (\ref{alf7})].  For $f_{\rm
th}=1.34\times
10^{17}\,
{\rm GeV}$, we find $(M_h)_{\rm
max}=1.01\times 10^{20}\, M_{\odot}$. Since the largest DM halos observed in
the Universe have a much smaller mass,
$M_h\sim 10^{14}\, M_{\odot}\ll (M_h)_{\rm
max}$, these results suggest that the effect of an
attractive self-interaction is negligible for what concerns the
structure of DM halos in the nonlinear regime: Everything
happens {\it as if} the bosons were not self-interacting. This conclusion
assumes
that Eq. (\ref{hui1}) is fulfilled \cite{hui} (see also the Remark below).

In conclusion, bosons with an attractive
self-interaction are essentially equivalent to noninteracting bosons in
situations of astrophysical interest while
bosons with a repulsive self-interaction can be very
different from noninteracting bosons (their mass $m$ may be $18$ orders
of magnitude larger).

{\it Remark:} As shown above (in line with \cite{mcmh}), the quantum cores of
BECDM halos with an attractive self-interaction are always stable  ($M_c<M_{\rm
max}$ of \cite{prd1}). Because of the constraints from particle physics and
cosmology
\cite{hui}, an attractive self-interaction is usually negligible. An attractive
self-interaction would be important in sufficiently
large DM halos, and lead to the collapse of the quantum core ($M_c>M_{\rm
max}$ of \cite{prd1}), if $f<4.22\times 10^{15}\, {\rm GeV}$ [the bound
corresponds
to
$(M_h)_{\rm max}=10^{14}\, M_{\odot}$ in Eq. (\ref{mrr20})]. This is outside of
the range 
$10^{16}\, {\rm GeV}\le f\le 10^{18}\, {\rm GeV}$ expected in particle physics
\cite{hui} [NB: if $f\sim 10^{15}\, {\rm GeV}$ is allowed then the quantum core
of sufficiently large DM halos, with  mass $M_h\gtrsim (M_h)_{\rm max}=10^{14}\,
M_{\odot}$, can collapse since $M_c>M_{\rm max}$]. On the other hand, it is
shown in Appendix C of \cite{mcmhbh} that
the quantum cores of BECDM halos with a vanishing or a repulsive
self-interaction are always Newtonian, i.e., their mass is always much smaller
than the
general relativistic maximum mass ($M_c\ll M_{\rm max}^{\rm
GR}$ of \cite{kaup,rb,colpi,chavharko}) so they cannot collapse towards a black
hole. In conclusion, the cores of BECDM halos are expected to be stable in all
cases of astrophysical interest. They represent large quantum bulges. They may,
however, evolve collisionally on a secular timescale and ultimately collapse
towards a supermassive black hole via the process of gravothermal catastrophe
followed by a dynamical
instability of general relativistic origin \cite{balberg} if the halo mass
$M_h$ is sufficiently large (above the microcanonical critical point), as
advocated in \cite{modeldm}.

\subsection{QCD axions}
\label{sec_qcda}

In the previous sections, we have determined some constraints
that the mass $m$ and the scattering length $a_s$ of the bosons possibly
composing  DM must satisfy so that they are able to form giant BECs of mass
$M\sim 10^8\, M_{\odot}$ and radius $R\sim 1\, {\rm kpc}$ comparable to dSphs
like Fornax. This leads to ULAs with a very small mass that are allowed by
particle physics (in connection to string theory) but that have not been fully
characterized yet \cite{marshrevue}. On the other hand, the characteristics of
the QCD axion are precisely known from cosmology and particle physics and we can
see how they enter into the problem.

QCD axions have a mass $m=10^{-4}\,
{\rm eV}/c^2$ and a negative scattering length
$a_s=-5.8\times 10^{-53}\, {\rm m}$ \cite{kc}, corresponding to a dimensionless
self-interaction constant 
$\lambda=-7.39\times
10^{-49}$ and a decay constant $f=5.82\times 10^{19}\, {\rm eV}$ (see Appendix
\ref{sec_c}). The maximum mass of QCD axion
stars  is $M_{\rm max}^{\rm exact}=6.46\times 10^{-14}\,
M_{\odot}$ and their minimum stable radius is
$(R_{99}^*)^{\rm exact}=227\, {\rm km}$ (their
average maximum density is $\overline{\rho}=2.62\times 10^3\, {\rm g/m^3}$
and the maximum number of axions in an axion star is $N_{\rm max}=M_{\rm
max}/m=7.21\times 10^{56}$).

These values of $M_{\rm max}$ and $R_{99}^*$ correspond to the typical size of
asteroids.
Obviously, QCD axions cannot form giant BECs with the dimension of DM halos like
Fornax. However, they can form mini boson stars (mini axion stars or dark matter
stars) of very low
mass -- axteroids -- that could be the constituents of DM halos under the form
of mini massive compact halo objects (mini MACHOs) \cite{bectcoll,phi6}. These
mini axion
stars are Newtonian self-gravitating BECs of QCD axions with an attractive
self-interaction stabilized by the quantum pressure (Heisenberg uncertainty
principle).  They may cluster into structures similar to standard CDM
halos.  They
might play a role as
DM components (i.e. DM halos could be made of mini axion
stars interpreted as MACHOs instead of WIMPs) if they exist in the Universe in
abundance. However, mini axion
stars (MACHOs)  behave essentially as
CDM and do not solve the small-scale crisis of CDM.

{\it Remark:} The collapse of axion stars above the limiting mass  $M_{\rm
max}$ \cite{prd1} has been discussed by several authors
\cite{braaten,cotner,bectcoll,ebycollapse,tkachevprl,helfer,phi6,visinelli,
moss}. The collapse may lead to the formation of a dense
axion star or a black hole. It may also be accompanied by an explosion with an
ejection of relativistic axions (bosenova).

\section{Jeans mass-radius relation}
\label{sec_jmr}

In this section, we study how the Jeans length $\lambda_J$ 
and the Jeans mass $M_J$ of self-gravitating BECs depend on the density $\rho$.
We apply these results in a cosmological context, during the matter era, where
the density of BECDM evolves with time as (see, e.g.,
\cite{suarezchavanisprd3} for more
details)
\begin{equation}
\label{jmr1}
\frac{\rho}{\rm g/m^3}=2.25\times 10^{-24}\, a^{-3},
\end{equation}
where $a$ is the scale factor. The
beginning of the matter era, which can be identified with the epoch of
radiation-matter equality (i.e. the transition between the radiation era and the
matter era) occurs at $a_{\rm
eq}=2.95\times 10^{-4}$. At that
moment, the DM density is $\rho_{\rm
eq}=8.77\times 10^{-14}\, {\rm
g}/{\rm m}^3$. In comparison, the present
density of DM
is $\rho_0=2.25\times
10^{-24}\, {\rm g/m^3}$ (corresponding to $a_0=1$). In the following, we
compute the Jeans scales $\lambda_J$ and $M_J$ for any value of the density
between the epoch of radiation-matter equality $\rho_{\rm eq}$  and the present
epoch $\rho_0$.

The Jeans instability analysis is valid during the linear regime
of structure formation (it describes their initiation) which is expected to be
close to the epoch of radiation-matter equality which marks the beginning of
the matter era. By contrast, at the present epoch, nonlinear effects have become
important (the DM halos are already formed) and the Jeans instability analysis
is not valid anymore except at very large scales.  We stress that the Jeans
scales can
only give an order of magnitude of the size and mass of the DM halos since these
objects result from a very nonlinear process of free fall and  violent
relaxation which extends far beyond the linear regime. It is therefore not
straightforward to relate quantitatively the characteristic sizes, masses and
densities of DM halos to the Jeans scales. Nevertheless, the Jeans approach
provides a simple first step towards the problem of structure formation.

Let us consider a standard BEC at $T=0$ with an equation of state given by Eq.
(\ref{mad10}). Using the corresponding expression of the speed of
sound, the Jeans wavenumber (\ref{j7}) can be written as \cite{prd1}
\begin{equation}
\label{jmr2}
k_J^2=\frac{8\pi |a_s|\rho}{m}\left\lbrack \sqrt{1+\frac{Gm^4}{4\pi
a_s^2\hbar^2\rho}}-{\rm sgn}(a_s)\right\rbrack.
\end{equation}
The Jeans radius and the Jeans mass defined by Eq. (\ref{j8}) are then given by
\begin{equation}
\label{jmr3}
R_J=\frac{\left (\frac{\pi m}{8|a_s|\rho}\right
)^{1/2}}{\left\lbrack \sqrt{1+\frac{Gm^4}{4\pi a_s^2\hbar^2\rho}}-{\rm
sgn}(a_s)\right\rbrack^{1/2}},
\end{equation}
\begin{equation}
\label{jmr4}
M_J=\frac{\frac{4}{3}\pi \left (\frac{\pi m}{8|a_s|}\right
)^{3/2}}{\rho^{1/2}\left\lbrack \sqrt{1+\frac{Gm^4}{4\pi a_s^2\hbar^2\rho}}-{\rm
sgn}(a_s)\right\rbrack^{3/2}}.
\end{equation}
Eliminating the density between Eqs. (\ref{jmr3}) and (\ref{jmr4}), we obtain the Jeans mass-radius relation
\begin{eqnarray}
\label{jmr5}
M_J=\frac{\frac{\pi^4}{12}\frac{\hbar^2}{Gm^2
R_J}}{1-\pi^2\frac{a_s\hbar^2}{Gm^3R_J^2}}.
\end{eqnarray}
As noted in \cite{prd1}, this expression is similar 
to the approximate mass-radius relation of BECDM halos given by Eq.
(\ref{pa1}).\footnote{The similarity between the mass-radius relation obtained
from the $f$-ansatz and from the Jeans instability study is discussed at a
general level in Appendix \ref{sec_sim}.} Comparing Eqs. (\ref{pa1}) and
(\ref{jmr5}), we get $a_J=\pi^4/12\simeq 8.12$ and $b_J=\pi$ which are close to
the values of $a$ and $b$ obtained in Sec. \ref{sec_pa}. This agreement is
striking because the Jeans mass-radius relation [Eq.
(\ref{jmr5})] is valid in the linear regime of structure formation
close to spatial homogeneity while the
mass-radius relation of BECDM halos [Eq. (\ref{pa1})] is valid in the strongly
nonlinear regime of structure formation (after free fall and violent
relaxation) for very inhomogeneous objects. Before studying the relations
(\ref{jmr3})-(\ref{jmr5}) in the
general case, we consider particular limits of these relations.

\subsection{NI limit}
\label{sec_nib}

In the NI limit ($a_s=0$), the Jeans length and the Jeans mass are
given by \cite{khlopov,bianchi,hu,sikivie,prd1}
\begin{equation}
\lambda_J=2\pi\left (\frac{\hbar^2}{16\pi G\rho m^2}\right )^{1/4},\quad M_J=\frac{1}{6}\pi\left
(\frac{\pi^3\hbar^2\rho^{1/3}}{Gm^2}\right
)^{3/4}.
\label{jj7}
\end{equation}
They can be written as
\begin{eqnarray}
\frac{\lambda_J}{\rm pc}=1.16\times 10^{-12}\, \left (\frac{{\rm
eV}/c^2}{m}\right )^{1/2}\left (\frac{\rm g/m^3}{\rho}\right )^{1/4},
\label{jj8}
\end{eqnarray}
\begin{eqnarray}
\frac{M_J}{M_{\odot}}= 1.20\times 10^{-20}\, \left (\frac{{\rm
eV}/c^2}{m}\right )^{3/2}\left (\frac{\rho}{\rm g/m^3}\right )^{1/4}.
\label{jj9}
\end{eqnarray}
Using Eq. (\ref{jmr1}), we find that the Jeans length
increases as
$a^{3/4}$ while the
Jeans mass
decreases as $a^{-3/4}$.  Eliminating the
density between the relations of Eq.
(\ref{jj7}), we obtain  
\begin{eqnarray}
M_{J}\lambda_J=\frac{\pi^4}{6} \frac{\hbar^2}{Gm^2}
.
\label{jj10}
\end{eqnarray}
This relation  is similar to the mass-radius
relation (\ref{dm12}) of Newtonian BECDM halos
made of noninteracting bosons \cite{membrado,prd1,prd2}.

\subsection{TF limit}
\label{sec_tfb}

Let us consider bosons with a repulsive self-interaction ($a_s>0$). In the TF limit ($\hbar=0$), the Jeans
length and the Jeans mass are given by \cite{prd1} 
\begin{equation}
\lambda_J=2\pi\left (\frac{a_s\hbar^2}{G m^3}\right )^{1/2},\quad M_J=\frac{1}{6}\pi\rho \left (\frac{4\pi^2 a_s\hbar^2}{G
m^3}\right )^{3/2}.
\label{jj11}
\end{equation}
They can be written as
\begin{eqnarray}
\frac{\lambda_J}{\rm pc}=34.9\, \left (\frac{a_s}{\rm
 fm}\right )^{1/2}\left (\frac{{\rm
eV}/c^2}{m}\right )^{3/2},
\label{jj12}
\end{eqnarray}
\begin{eqnarray}
\frac{M_J}{M_{\odot}}=3.30\times 10^{20}\, \left
(\frac{a_s}{\rm
 fm}\right )^{3/2}\left (\frac{{\rm
eV}/c^2}{m}\right )^{9/2} \frac{\rho}{\rm g/m^3}.
\label{jj13}
\end{eqnarray}
We note that the Jeans length is independent of the density \cite{prd1}. Using
Eq. (\ref{jmr1}), we find that the Jeans mass decreases as $a^{-3}$. The
relation from
Eq. (\ref{jj11}) is
similar to the relation (\ref{dm15})
determining the radius of a self-interacting BECDM halo in the TF
approximation \cite{tkachev,maps,leekoh,goodman,arbey,bohmer,prd1}.

\subsection{NG limit}
\label{sec_ngb}

Let us consider bosons with an attractive self-interaction ($a_s<0$). In the
nongravitational limit ($G=0$), the Jeans length and the Jeans mass\footnote{We
call them  ``Jeans length'' and ``Jeans mass'' by an abuse of language since
there is no gravity in the present situation. The
instability is a purely ``hydrodynamic instability'' (also called ``tachyonic
instability'') due to the attractive
self-interaction ($a_s<0$) which yields a negative squared speed of sound
($c_s^2<0$). This
terminology will make
sense, however, in the general case (see Sec. \ref{sec_att}) where the
instability is due
to the combined effect of self-gravity and self-interaction. } are given
by \cite{prd1}
\begin{equation}
\lambda_J=2\pi\left (\frac{m}{16\pi|a_s|\rho}\right )^{1/2},\quad M_J=\frac{\pi}{6}\frac{1}{\rho^{1/2}} \left (\frac{\pi m}{4|a_s|}\right )^{3/2}.
\label{jj14}
\end{equation}
They can be written as
\begin{equation}
\frac{\lambda_J}{\rm pc}=3.83\times 10^{-26}\, \left (\frac{\rm
 fm}{|a_s|}\right )^{1/2}\left (\frac{m}{{\rm
eV}/c^2}\right )^{1/2}\left (\frac{\rm g/m^3}{\rho}\right )^{1/2},
\label{jj15}
\end{equation}
\begin{equation}
\frac{M_J}{M_{\odot}}=4.36\times 10^{-61}\, \left
(\frac{\rm
 fm}{|a_s|}\right )^{3/2}\left (\frac{m}{{\rm
eV}/c^2}\right )^{3/2} \left (\frac{\rm g/m^3}{\rho}\right )^{1/2}.
\label{jj16}
\end{equation}
Using Eq. (\ref{jmr1}), we find that the
Jeans length and the Jeans mass both
increase
as $a^{3/2}$.
Eliminating the density between the relations of Eq. (\ref{jj14}), we
obtain
\begin{eqnarray}
M_J=\frac{\pi^2}{24}\frac{m}{|a_s|} \lambda_J.
\label{jj17}
\end{eqnarray}
This relation is similar to the
mass-radius relation of nongravitational BECDM halos with an attractive
self-interaction given by Eq. (\ref{dm19}) \cite{prd2}. We
recall, however, that these equilibrium states (valid in the nonlinear regime
of structure formation)
are unstable so they should not be observed in practice (see \cite{prd1} for
detail). Therefore, only the relations (\ref{jj14})-(\ref{jj17})
obtained from the Jeans analysis in the linear regime of structure formation are
physically meaningful. They determine the initiation of structures (clumps) in a
homogeneous BEC due to the attractive self-interaction of the bosons. Their
evolution in the nonlinear regime requires a specific
analysis. Since these clumps cannot evolve
towards stable DM halos with mass  $M\sim M_J$ and radius $R\sim R_J$, they
are expected to collapse towards smaller structures until repulsive terms in
the self-interaction potential (not considered here) come into play
\cite{phi6}.

\subsection{Repulsive self-interaction}
\label{sec_rew}

\begin{figure}
\begin{center}
\includegraphics[clip,scale=0.3]{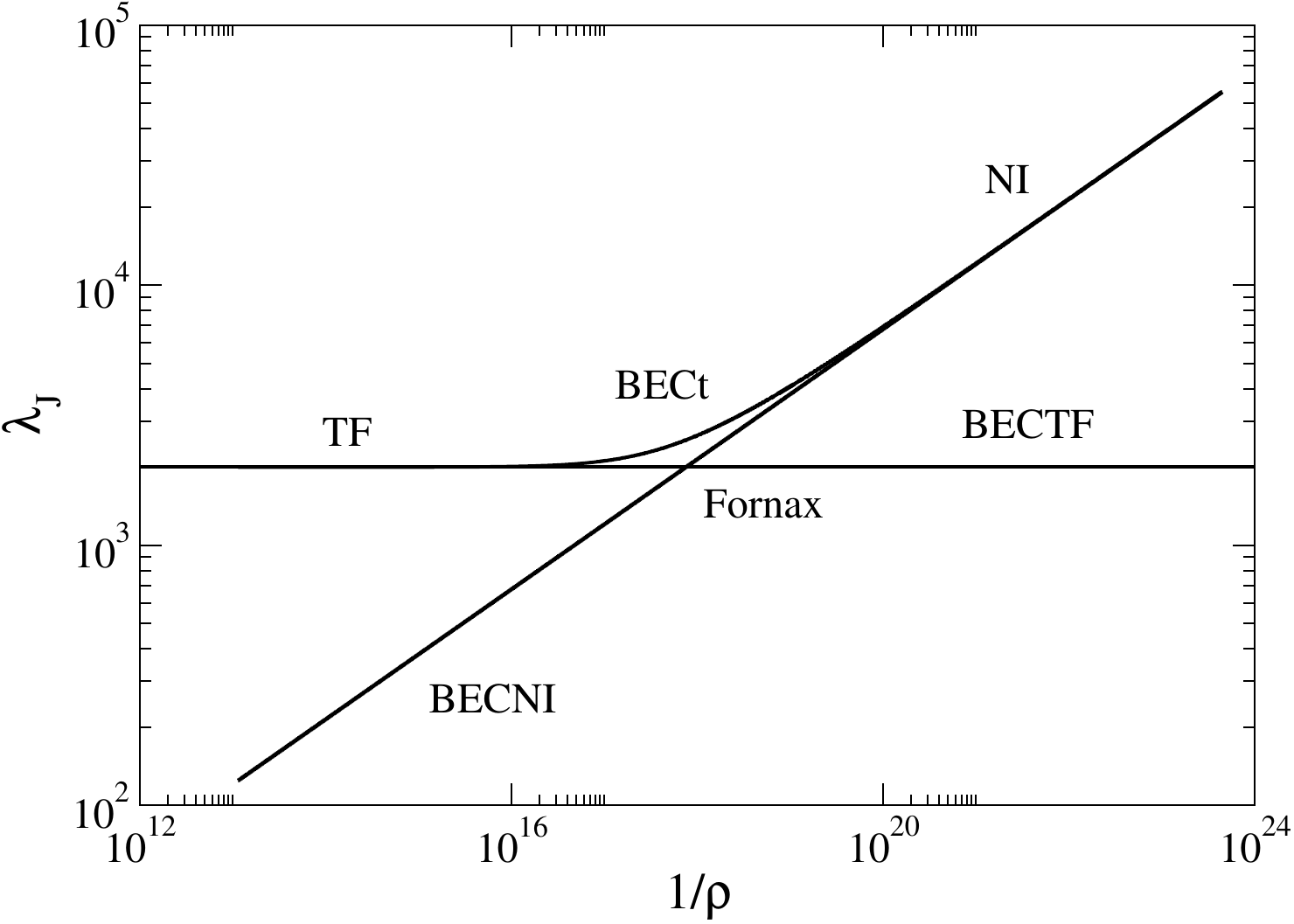}
\includegraphics[clip,scale=0.3]{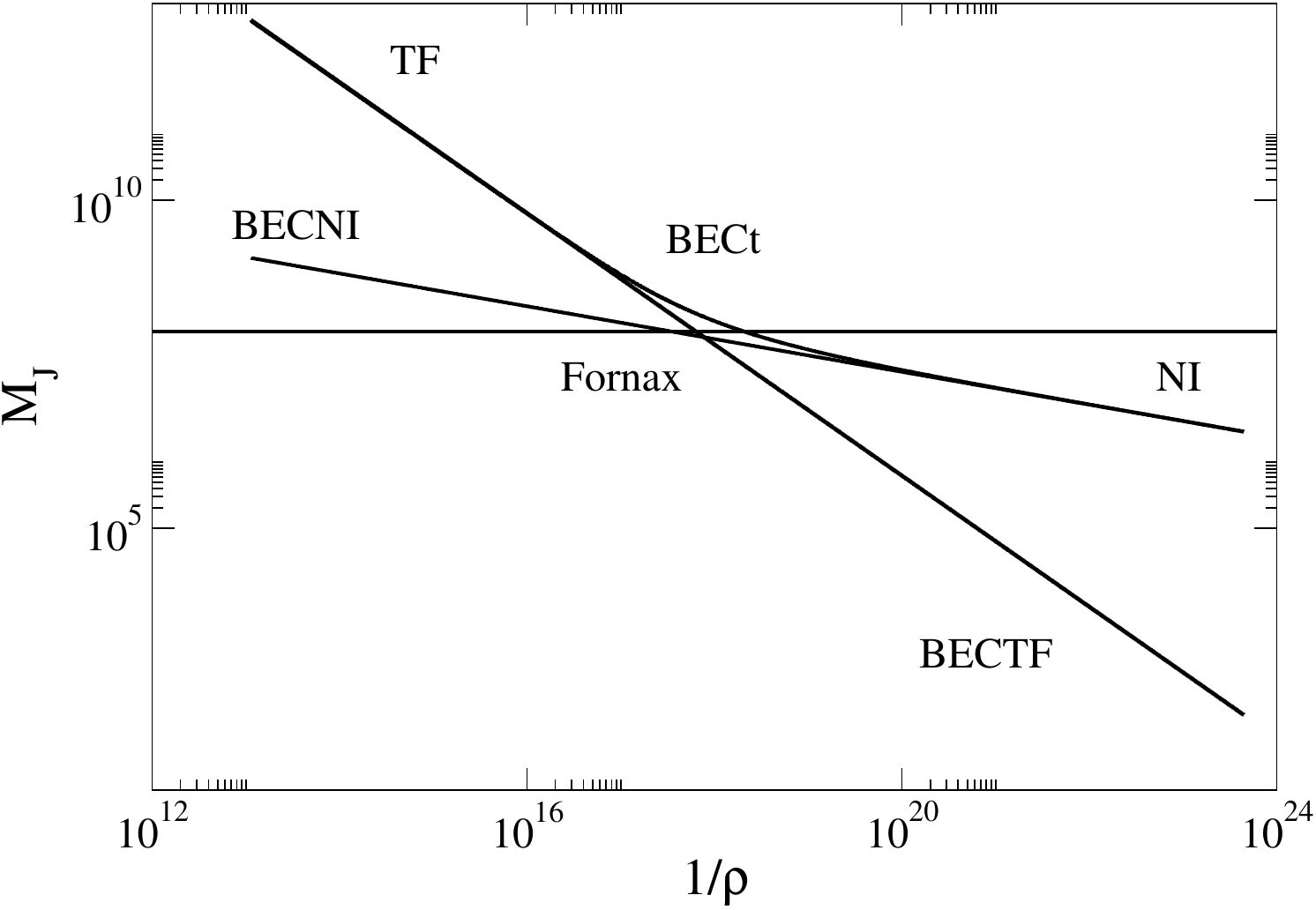}
\caption{Evolution of the Jeans length and Jeans mass with the inverse density
of the universe for bosons with a repulsive self-interaction ($\lambda_J$ is in
${\rm pc}$, $M_J$ is in solar masses $M_\odot$, and $\rho$ is
in ${\rm g/m^3}$) for the three models BECNI, BECTF and BECt considered in the
text. Here and in the following figures we have indicated the values
$\lambda_J/2=1\, {\rm kpc}$, $M_J=10^8\, M_{\odot}$ and $\rho=1.62\times
10^{-18}\, {\rm g/m^3}$ corresponding to the minimum halo (Fornax) for
reference (see Sec. \ref{sec_optimal}).}
\label{jrep}
\end{center}
\end{figure}

\begin{figure}
\begin{center}
\includegraphics[clip,scale=0.3]{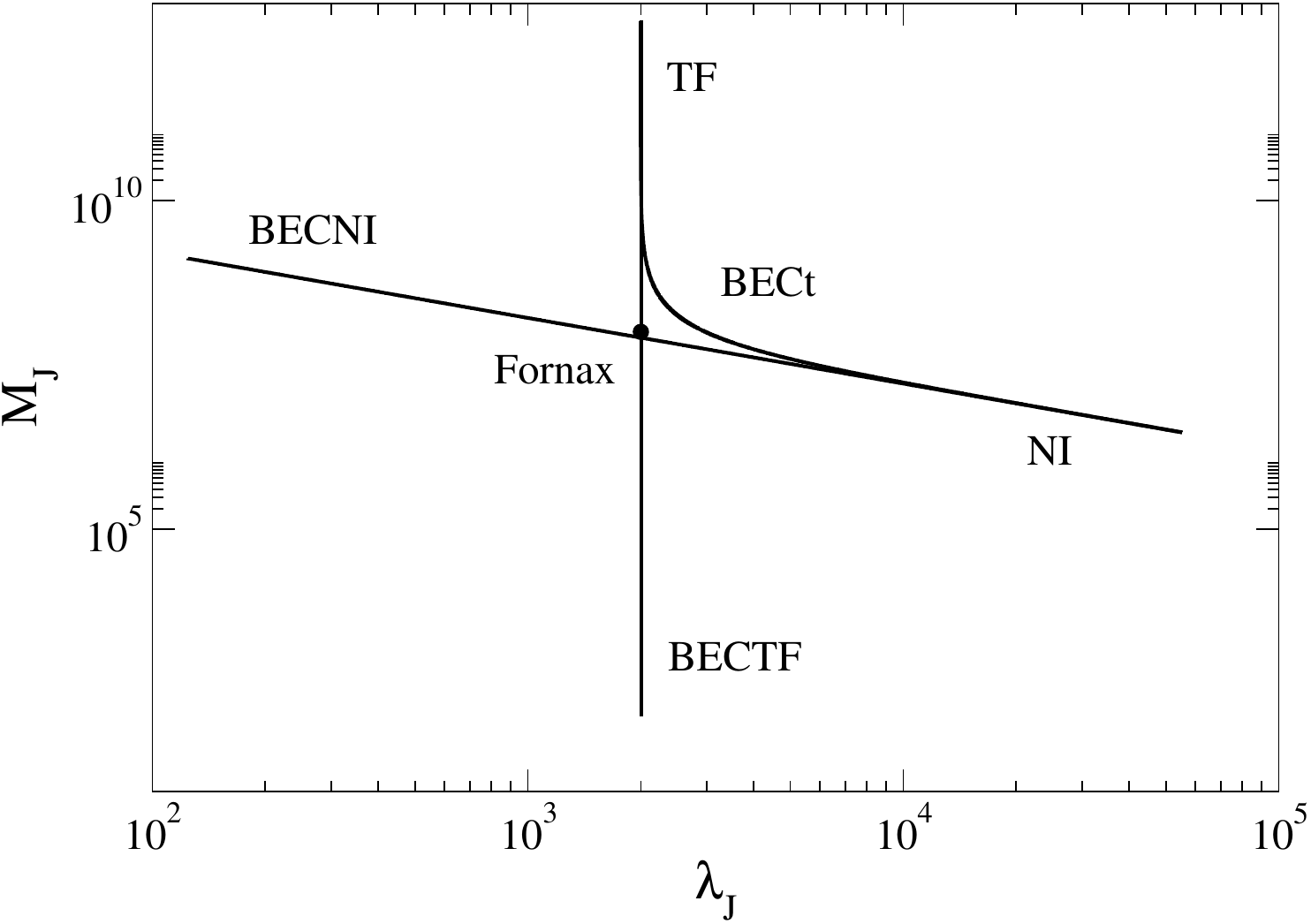}
\caption{Jeans-mass radius relation for bosons with a repulsive 
self-interaction for the three models  BECNI, BECTF and BECt  described in the
text. We see that the
Jeans mass-radius relation (linear regime) is similar to the mass-radius
relation of DM halos (nonlinear regime) represented in Fig.
\ref{M-R-chi-pos-part1}.}
\label{mrrep}
\end{center}
\end{figure}

In order to determine the evolution of the Jeans scales with the 
density, we need to specify the parameters of the DM particle. For illustration,
we use the parameters obtained in Sec. \ref{sec_mas} (see also Appendix D
of \cite{abrilphas} and Sec. II of \cite{mcmh}). They have been obtained in
order to match the characteristics of a ``minimum halo'' of radius $R\sim 1\,
{\rm kpc}$ and mass $M\sim 10^8\, M_{\odot}$, similar to Fornax, interpreted as
the ground state of the self-gravitating BEC. We  use this procedure to
determine the parameters of the DM particle, then compute the Jeans scales at
the  epoch of radiation-matter equality and at the present epoch, instead of
trying to determine the parameters of the DM particle directly from the Jeans
scales.\footnote{We believe that this alternative procedure, often used in the
literature, is less
accurate.} In the present section, we consider the case of bosons with a
repulsive self-interaction (or no self-interaction). We consider
different types of DM particles denoted BECNI, BECTF and BECt in Sec.
\ref{sec_mas}. For each of these particles, the evolution of the Jeans length
$\lambda_J$ and Jeans mass $M_J$ as a function of the inverse density
$1/\rho$ (which increases with time
as the Universe expands) is plotted in Fig. \ref{jrep}. The Jeans mass-radius
relation (parameterized by the density) is plotted in Fig. \ref{mrrep}.
The curves start from the epoch of matter-radiation equality and end at the present epoch.

Generically, as the density of the universe decreases, the BEC is 
first in the TF regime then in the NI regime. In the TF regime the Jeans length
is constant while the Jeans mass decreases like  $\rho$ (see Sec.
\ref{sec_tfb}). In the NI regime the Jeans length increases like $\rho^{-1/4}$
while the Jeans mass decreases like $\rho^{1/4}$ (see Sec. \ref{sec_nib}).  The
transition between the TF regime and the NI regime occurs at a typical density
\begin{eqnarray}
\rho_s=\frac{Gm^4}{16\pi\hbar^2 a_s^2}
\label{jj17q}
\end{eqnarray}
obtained by equating Eqs. (\ref{jj7}) and (\ref{jj11}). At that
point
\begin{equation}
\label{jpa8h}
(M_J)_{s}=\frac{\pi^3}{12} \frac{\hbar}{\sqrt{Gm a_s}}
\end{equation}
and 
\begin{equation}
(\lambda_J)_s=2\pi\left (\frac{a_s\hbar^2}{Gm^3}\right )^{1/2}.
\label{jj17bh}
\end{equation}
The BEC is always in the NI regime (during the period going from 
the epoch of matter-radiation equality to the present epoch) if
$1/\rho_s<1/\rho_{\rm eq}$, i.e., if
\begin{equation}
\frac{a_s}{m^2}<\left (\frac{G}{16\pi\hbar^2\rho_{\rm eq}}\right
)^{1/2}=3.71\times 10^{-21}\, \frac{\rm fm}{({\rm eV}/c^2)^2}.
\label{jj17j}
\end{equation}
Combining this inequality with the $m(a_s)$ relation of Sec. \ref{sec_mas}, we
find that the BEC is always in the NI regime if $0\le a_s\le 3.16\times
10^{-64}\, {\rm fm}$ (and $m\sim 2.92\times
10^{-22}\, {\rm eV}/c^2$). This corresponds to $0\le
\lambda\le 1.17\times 10^{-92}$. On the other hand, the BEC is
always in the TF regime (during the same period) if $1/\rho_s>1/\rho_{0}$, i.e.,
if
\begin{eqnarray}
\frac{a_s}{m^2}>\left (\frac{G}{16\pi\hbar^2\rho_{0}}\right
)^{1/2}=7.32\times 10^{-16}\, \frac{\rm fm}{({\rm eV}/c^2)^2}. 
\label{jj17k}
\end{eqnarray}
Combining this inequality  with the $m(a_s)$ relation of Sec. \ref{sec_mas}, we
find that the BEC is always in the TF regime if $3.65\times 10^{-53}\, {\rm
fm}\le a_s\le (a_s)_{\rm max}=4.41\times 10^{-6}\, {\rm fm}$ (and $2.23\times
10^{-19}\, {\rm eV}/c^2\le m\le m_{\rm max}=1.10\times
10^{-3}\, {\rm eV}/c^2$). This corresponds to $1.04\times
10^{-78}\le
\lambda\le \lambda_{\rm max}=6.18\times
10^{-16}$.

{\it BECNI:} Let us consider noninteracting ULAs with a mass $m=2.92\times
10^{-22}\, {\rm eV}/c^2$ determined by the characteristics of the minimum halo
(see Sec. \ref{sec_mas}). At
the epoch of
radiation-matter equality, we find $\lambda_J=124\, {\rm pc}$ and
$M_J=1.31\times 10^9\, M_{\odot}$ (the comoving Jeans length
is $\lambda_J^c=\lambda_J/a=0.420\, {\rm Mpc}$). At the present epoch, we find
$\lambda_J=55.3\, {\rm kpc}$ and
$M_J=2.94\times 10^6\, M_{\odot}$.

{\it BECTF:} Let us consider self-interacting bosons with a mass $m=1.10\times
10^{-3}\, {\rm eV}/c^2$ and a
scattering length $a_s=4.41\times 10^{-6}\, {\rm
fm}$ (this yields $\lambda=6.18\times 10^{-16}$).
This corresponds to a ratio $a_s/m^3=3.28\times
10^3\, {\rm fm}/({\rm eV/c^2})^3$ determined by the
radius of the minimum halo and to a ratio $4\pi a_s^2/m=
1.25\, {\rm cm^2/g}$ determined by the constraint set by the Bullet Cluster
assuming that the bound is reached (see Sec. \ref{sec_mas}). For the period
considered, the BEC is always in the TF regime. At the epoch of radiation-matter
equality, we find $\lambda_J=2.01\, {\rm
kpc}$ and $M_J=5.51\times 10^{12}\, M_{\odot}$ (the comoving
Jeans length
is $\lambda_J^c=\lambda_J/a=6.81\, {\rm Mpc}$). At the present epoch we find
$\lambda_J=2.01\, {\rm kpc}$ and $M_J=141\, M_{\odot}$. 

{\it BECt:}  Let us consider self-interacting bosons with a
mass $m=2.92\times 10^{-22}\, {\rm eV}/c^2$ and a scattering length
$a_s=8.13\times 10^{-62}\, {\rm fm}$ (this yields
$\lambda=3.02\times 10^{-90}$).
This corresponds to a ratio
$a_s/m^3=3.28\times
10^3\, {\rm fm}/({\rm eV/c^2})^3$ determined by the
radius of the minimum halo and to a scattering length
 chosen such that the minimum halo is just at the
transition
between the TF regime and the NI regime (see Sec. \ref{sec_mas}).
For the period considered, the BEC is first in
the TF regime then in the NI regime (the transition occurs at a
typical density $\rho_s=1.33\times 10^{-18}\, {\rm g/m^3}$). At the epoch
of
radiation-matter equality, we find $\lambda_J=2.00\, {\rm
kpc}$ and $M_J=5.39\times 10^{12}\, M_{\odot}$ (the comoving
Jeans length
is $\lambda_J^c=\lambda_J/a=6.78\, {\rm Mpc}$). At
the present
epoch we find $\lambda_J=55.3\, {\rm kpc}$ and
$M_J=2.94\times 10^6\, M_{\odot}$.
In the TF regime, the BECt model behaves as the BECTF model (because they have
the same ratio $a_s/m^3$) and in the NI regime, 
the BECt model behaves as the BECNI model corresponding to noninteracting ULAs
(because they have the same mass $m$).

{\it (iv)  BECf:} Let us
consider
self-interacting bosons with a mass $m=3\times 10^{-21}\, {\rm eV}/c^2$ and
a scattering length $a_s=1.11\times 10^{-58}\, {\rm fm}$ 
(this yields $\lambda=4.24\times 10^{-86}$). This fiducial model
is
motivated by cosmological considerations \cite{shapiro}. It is similar to the
BECt
model.  For the period considered, the BEC is first in
the TF regime then in the NI regime (the
transition occurs at a
typical density $\rho_s=7.93\times 10^{-21}\, {\rm g/m^3}$). At the epoch of
radiation-matter equality, we find $\lambda_J=2.24\, {\rm
kpc}$ and $M_J=7.61\times 10^{12}\, M_{\odot}$ (the comoving
Jeans length
is $\lambda_J^c=\lambda_J/a=7.59\, {\rm Mpc}$). At
the present
epoch we find $\lambda_J=17.3\, {\rm kpc}$ and
$M_J=9.05\times 10^4\, M_{\odot}$.

{\it Remark:} ULA clumps formed in the linear regime by Jeans instability may
evolve, in the nonlinear regime, into stable DM halos  with mass  $M\sim M_J$
and radius $R\sim R_J$ (since self-gravitating BECs
with a repulsive self-interaction are stable).  They can then
increase their mass by mergings and accretion (or possibly loose mass) leading
to the DM halos observed in the universe. Large DM halos have a
core-halo structure resulting from violent relaxation and gravitational cooling.
The core mass -- halo mass of self-interacting BECs has been determined in
\cite{mcmh,mcmhbh}.

\subsection{Attractive self-interaction}
\label{sec_att}

\begin{figure}
\begin{center}
\includegraphics[clip,scale=0.3]{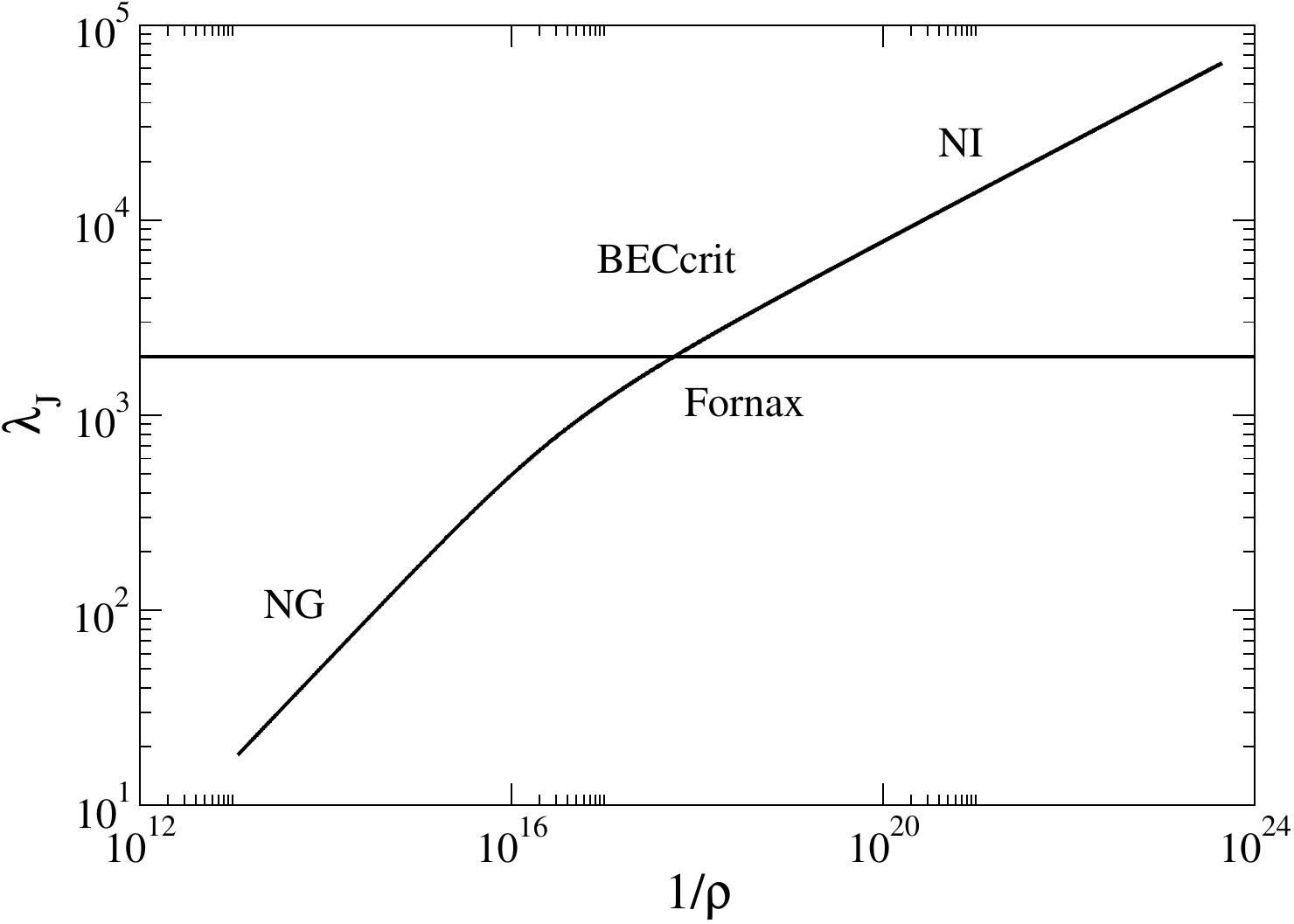}
\includegraphics[clip,scale=0.3]{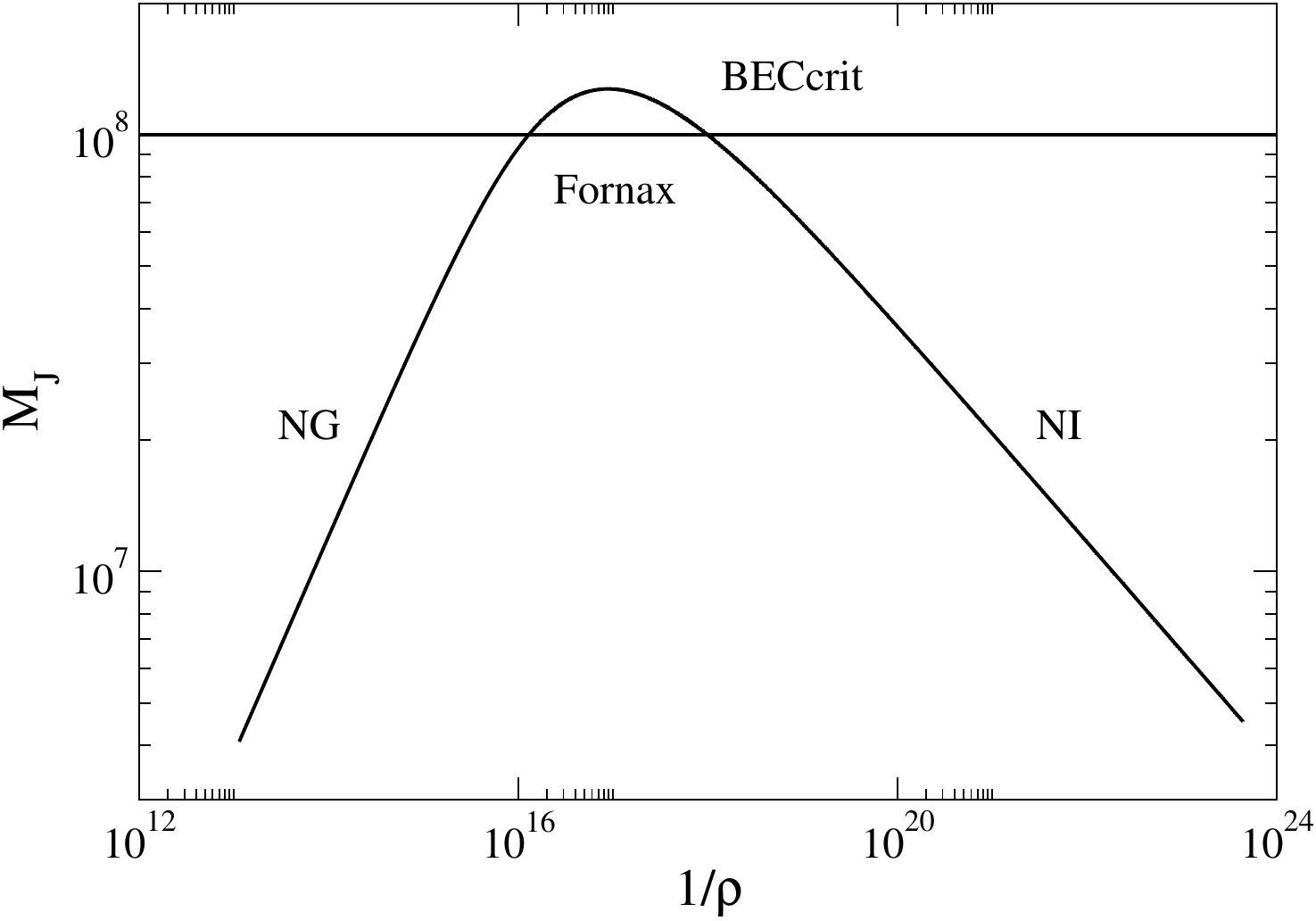}
\caption{Evolution of the Jeans length and Jeans mass with the inverse density
of the universe  for bosons with an attractive self-interaction ($\lambda_J$ is
in ${\rm pc}$, $M_J$ is in solar masses $M_\odot$, and $\rho$ is in ${\rm
g/m^3}$) for the BECcrit model described in the
text.}
\label{wrgG0N}
\end{center}
\end{figure}

\begin{figure}
\begin{center}
\includegraphics[clip,scale=0.3]{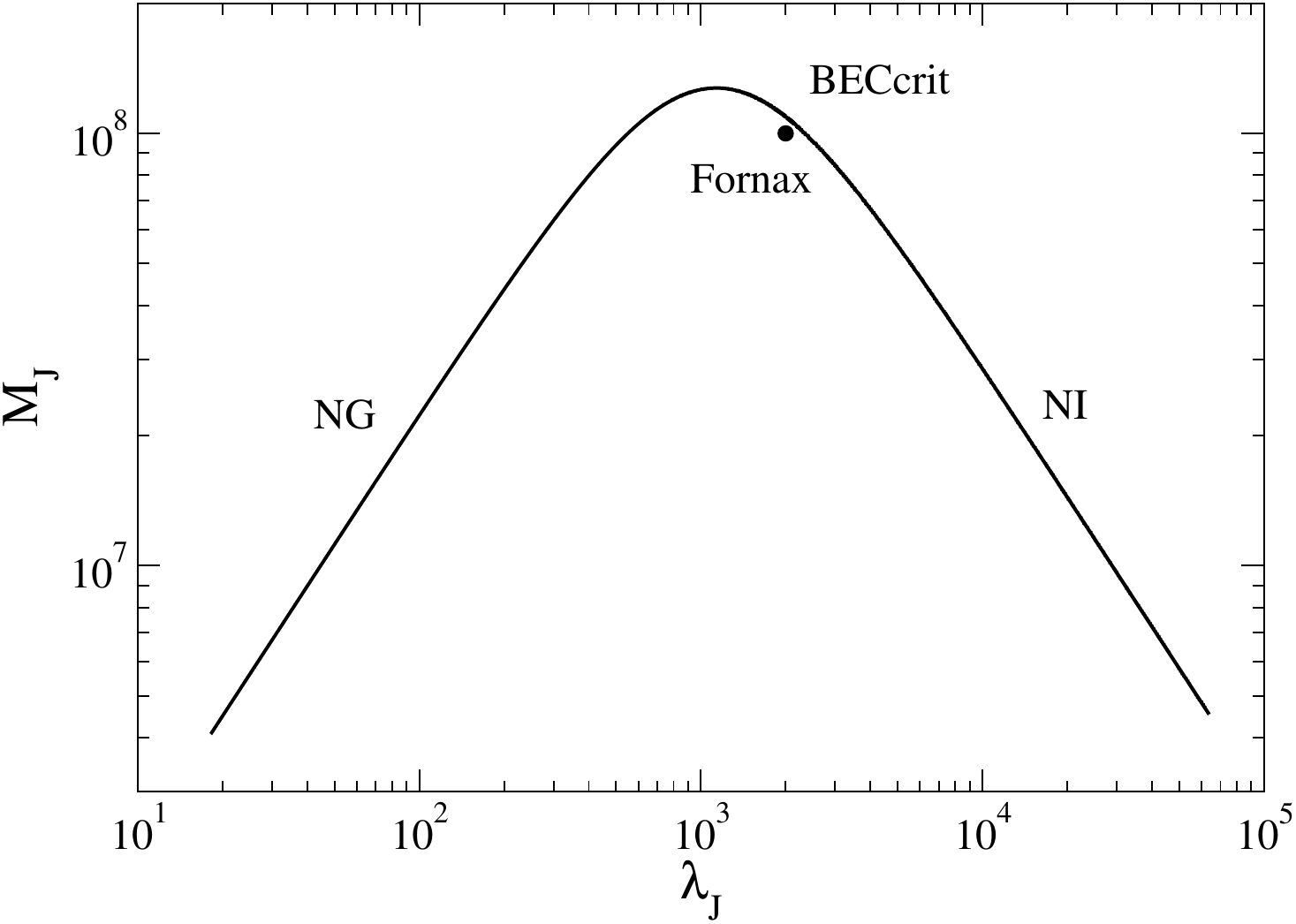}
\caption{Jeans-mass radius relation for bosons with an attractive
self-interaction for the  BECcrit model described in the
text. We note
that the
Jeans mass-radius relation (linear regime) is similar to the mass-radius
relation of DM halos (nonlinear regime) represented in Fig.
\ref{M-R-chi-neg-part1}.}
\label{LambdaMATTRACTIFexact}
\end{center}
\end{figure}

In this section, we consider the 
case of bosons with an attractive self-interaction. We consider different types
of DM particles denoted BECcrit, BECth and QCD axions in
Sec. \ref{sec_mas}. For each of these particles,
the evolution of the Jeans length $\lambda_J$ and Jeans mass $M_J$ as a function
of the inverse
density $1/\rho$ (which increases with time as the universe expands) is plotted
in Fig.
\ref{wrgG0N}. The Jeans mass-radius relation (parameterized by the density) is
plotted in Fig. \ref{LambdaMATTRACTIFexact}.
The curves start from the epoch of matter-radiation equality and end at the present epoch.

Generically, as the density of the universe decreases, 
the BEC is first in the NG regime then in the NI regime. In the NG regime the
Jeans length and the Jeans mass both increase like $\rho^{-1/2}$ (see Sec.
\ref{sec_ngb}).  In the NI regime the Jeans length increases like $\rho^{-1/4}$
while the Jeans mass decreases like $\rho^{1/4}$ (see Sec. \ref{sec_nib}). There
is a maximum Jeans mass
\begin{equation}
\label{aw1}
(M_J)_{\rm max}=\frac{\pi^3}{24} \frac{\hbar}{\sqrt{Gm|a_s|}},
\end{equation}
corresponding to a Jeans length 
\begin{equation}
(\lambda_J)_*=2\pi\left (\frac{|a_s|\hbar^2}{Gm^3}\right )^{1/2},
\label{aw2}
\end{equation}
at the density
\begin{equation}
\rho_*=\frac{Gm^4}{32\pi\hbar^2 a_s^2}.
\label{aw3}
\end{equation}
The
transition between the NG regime and the NI regime occurs at a typical density
\begin{equation}
\rho_s=\frac{Gm^4}{16\pi\hbar^2 a_s^2},
\label{aw4}
\end{equation}
obtained by equating Eqs. (\ref{jj7}) and (\ref{jj14}). At that
point
\begin{equation}
\label{aw5}
(M_J)_{s}=\frac{\pi^3}{12} \frac{\hbar}{\sqrt{Gm|a_s|}}
\end{equation}
and 
\begin{equation}
(\lambda_J)_s=2\pi\left (\frac{|a_s|\hbar^2}{Gm^3}\right )^{1/2}.
\label{aw6}
\end{equation}
These scales are similar to those corresponding to the maximum mass (we have
$\rho_s=2\rho_*$, $(M_J)_{s}=2(M_J)_{\rm max}$ and
$(\lambda_J)_s=(\lambda_J)_*$). The BEC is always in the
NI regime (during the period going from the epoch of matter-radiation equality
to
the present epoch) if $1/\rho_s<1/\rho_{\rm eq}$, i.e., if
\begin{equation}
\frac{|a_s|}{m^2}<\left (\frac{G}{16\pi\hbar^2\rho_{\rm eq}}\right
)^{1/2}=3.71\times 10^{-21}\, \frac{\rm fm}{({\rm
eV}/c^2)^2}.
\label{aw7}
\end{equation}
Combining this inequality with the $m(a_s)$ relation of Sec. \ref{sec_mas}, we
find that the BEC is always in the NI regime if $-3.16\times
10^{-64}\, {\rm fm}\le a_s\le 0$ and $m\sim 2.92\times
10^{-22}\, {\rm eV}/c^2$. This corresponds to
$-1.18\times 10^{-92}\le\lambda\le 0$ and $f\ge 1.35\times 10^{15}\, {\rm GeV}$.
On the other hand, the BEC is always in
the NG regime (during the same period) if $1/\rho_s>1/\rho_{0}$, i.e., if
\begin{equation}
\frac{|a_s|}{m^2}>\left (\frac{G}{16\pi\hbar^2\rho_{0}}\right
)^{1/2}=7.32\times 10^{-16}\, \frac{\rm fm}{({\rm eV}/c^2)^2}. 
\label{aw8}
\end{equation}
Combining this inequality with the $m(a_s)$ relation 
of Sec. \ref{sec_mas}, we find that the BEC is always in the NG
regime if $-1.39\times 10^{-65}\, {\rm fm}\le a_s\le 0$ and $m\le
1.38\times 10^{-25}\, {\rm eV/c^2}$. This corresponds to
$-2.44\times 10^{-97}\le\lambda\le 0$ and $f\sim f_{\rm min}=1.39\times
10^{14}\, {\rm GeV}$.

{\it BECcrit:} Let us consider self-interacting  bosons with a mass
$m=2.19\times 10^{-22}\, {\rm eV}/c^2$ and a
scattering length $a_s=-1.11\times
10^{-62}\, {\rm fm}$ (this yields $\lambda=-3.10\times 10^{-91}$
and $f=1.97\times 10^{14}\, {\rm GeV}$).
These values are obtained by requiring that 
the minimum halo is critical (see Sec. \ref{sec_mas}).  For the period
considered, the BEC is first in the NG regime
then in the NI regime (the transition occurs at a
typical density $\rho_s=2.25\times 10^{-17}\, {\rm
g/m^3}$). The Jeans mass is maximal at the density
$\rho_*=1.13\times 10^{-17}\, {\rm g/m^3}$. At that density $(M_J)_{\rm
max}=1.27\times
10^{8}\, M_{\odot}$ and $(\lambda_J)_*=1.13\, {\rm kpc}$. At the epoch of
radiation-matter equality, we find $\lambda_J=18.2\, {\rm pc}$ and
$M_J=4.08\times 10^6\, M_{\odot}$ (the comoving
Jeans length
is $\lambda_J^c=\lambda_J/a=0.0617\, {\rm Mpc}$). At the present
epoch, we find $\lambda_J=63.8\, {\rm kpc}$ and $M_J=4.53\times
10^{6}\, M_{\odot}$.

{\it BECth:} Let us consider self-interacting  bosons with a mass $m=2.92\times
10^{-22}\, {\rm eV}/c^2$ and a
scattering length $a_s=-3.18\times
10^{-68}\, {\rm fm}$  (this yields $\lambda=-1.18\times 10^{-96}$
and $f=1.34\times 10^{17}\, {\rm GeV}$).
These values are obtained by using constraints from
particle physics and cosmology (see Sec. \ref{sec_cpp}).  For the period
considered, the BEC is always in the NI
regime (see the BECNI case studied above). At the epoch of
radiation-matter equality, we find $\lambda_J=124\, {\rm pc}$ and
$M_J=1.31\times 10^9\, M_{\odot}$ (the comoving
Jeans length
is $\lambda_J^c=\lambda_J/a=0.420\, {\rm Mpc}$). At the
present epoch, we find
$\lambda_J=55.3\, {\rm kpc}$ and
$M_J=2.94\times 10^6\, M_{\odot}$.   The Jeans mass is
always much below the maximum Jeans mass $(M_J)_{\rm max}=6.52\times
10^{10}\, M_{\odot}$  reached at a density
$\rho_*=4.33\times 10^{-6}\, {\rm g/m^3}$. These results
show that the effect
of an attractive self-interaction is negligible for what concerns the
formation of structures in the linear regime: Everything
happens {\it as if} the bosons were not self-interacting.

{\it QCD axions:} Let us consider QCD axions with a mass $m=10^{-4}\, {\rm
eV}/c^2$ and a scattering length $a_s=-5.8\times 10^{-53}\, {\rm m}$ (this
yields $\lambda=-7.39\times 10^{-49}$
and $f=5.82\times 10^{10}\, {\rm GeV}$). For the
period considered (matter era), the axions are always in
the NI regime. At the epoch of
radiation-matter equality, we find $\lambda_J=2.13\times 10^{-7}\, {\rm pc}$ and
$M_J=6.52\times 10^{-18}\, M_{\odot}$ (the comoving
Jeans length
is $\lambda_J^c=\lambda_J/a=7.22\times 10^{-10}\, {\rm Mpc}$). At
the present
epoch we find $\lambda_J=9.45\times 10^{-5}\, {\rm pc}$ and $M_J=1.47\times
10^{-20}\, M_{\odot}$.   The Jeans mass is
always much below the maximum Jeans mass $(M_J)_{\rm
max}=8.25\times 10^{-14}\, M_{\odot}$ reached at a density
$\rho_*=1.79\times 10^{4}\, {\rm g/m^3}$. These results show that the
effect
of an attractive self-interaction is negligible for what concerns the
formation of structures in the linear regime: Everything
happens {\it as if} the QCD axions were not self-interacting. Furthermore, the
Jeans scales computed above are much 
smaller than the galactic scales indicating the QCD axions essentially behave as
CDM.

{\it Remark:} QCD axions clumps  formed in the linear
regime by Jeans instability  may evolve, in the
nonlinear regime, into stable
dilute axion stars  (since noninteracting self-gravitating BECs are stable).
They can then increase their mass by mergings and accretion (or possibly loose
mass). If their mass passes above the maximum mass $M_{\rm max}=6.46\times
10^{-14}\, M_{\odot}$ they undergo gravitational collapse, leading to a
bosenova or a dense axion star (see Sec. \ref{sec_qcda}). Clumps of DM particles
corresponding to the BECcrit parameters formed in the linear regime by Jeans
instability  cannot evolve, in the nonlinear regime,
into stable configurations 
since nongravitational BECs are unstable (see Sec. \ref{sec_ngb}). Therefore,
they are expected to directly collapse, leading to bosenovae or dense axion
stars. Clumps of DM particles corresponding to the  BECth parameters formed in
the linear
regime by Jeans instability  may evolve, in
the nonlinear regime, into stable DM halos with a core-halo
profile  (since noninteracting self-gravitating BECs are stable). 
They can then increase their mass by mergings and accretion. We have seen in
Sec. \ref{sec_cpp} that, for realistic DM halos, the core mass is always smaller
than the critical mass so the quantum core is always stable.

\subsection{An optimal cosmological density}
\label{sec_optimal}

Except for QCD axions, all the models that we have considered above
are based on values of $m$ and $a_s$ that are consistent with the properties
of the minimum halo (see Sec. \ref{sec_mas}). Therefore, by construction, we
have $M_J\sim 10^8\, M_{\odot}$
and $R_J\sim 1\, {\rm kpc}$ at a particular density $\rho=3M/4\pi R^3=1.62\times
10^{-18}\, {\rm g/m^3}$ in the evolution of the universe. This ``optimal''
density corresponds to a scale factor $a=0.0111$ and 
a redshift $z=1/a-1=88.6$. If the structures formed at this epoch they
would have a Jeans  mass and a Jeans radius comparable to the mass and size of
the minimum halo ($M\sim 10^8\, M_{\odot}$
and $R\sim 1\, {\rm kpc}$). Actually, structures may form at a different epoch
and
evolve by accreting or loosing mass during the nonlinear regime. The relation
between the Jeans scales (in
the linear regime) and the actual scales of DM halos (in the nonlinear regime)
is not straightforward and usually requires to study the nonlinear process of
structure
formation numerically.

\section{Conclusion}

In this paper, following previous works on the subject, we have 
considered the possibility that DM is made of bosons in the form of
self-gravitating BECs. This model is interesting because it may solve the
small-scale problems of the standard CDM model such as the core-cusp problem and
the missing satellite problem. Indeed, in the linear regime of structure
formation due to the Jeans instability, quantum mechanics (Heisenberg
uncertainty principle) or a
repulsive self-interaction ($a_s>0$) leads to a finite Jeans length $\lambda_J$
even at $T=0$. Therefore, gravitational collapse can take place only above a
sufficiently large size and a sufficiently large mass (i.e. above $\lambda_J$
and $M_J$). The existence of a minimum size and a minimum mass is in agreement
with the observations. By contrast, in
the classical pressureless CDM model ($\hbar=P=0$), the Jeans length and the
Jeans mass vanish ($\lambda_J=M_J=0$), or are very small,
implying the possibility of formation of structures at all scales in
contradiction with the observations. On the other hand, in the nonlinear regime
of structure formation (after the system has experienced free fall, violent
relaxation, gravitational cooling, and virialization), the BECDM model leads to
DM
halos with a core, i.e., the central density is finite instead of diverging as
$r^{-1}$ like in the CDM model. The prediction of DM halos with a core rather
than a
cusp is again in agreement with the observations.

According to the above results, the BECDM model
predicts the existence of a ``minimum DM halo'' which corresponds to the ground
state of the self-gravitating BEC at $T=0$. We have identified this minimum
(ultracompact) halo
with dSphs like Fornax with a typical radius $R_{\rm min}=1\, {\rm kpc}$ and a
typical mass $M_{\rm min}=10^8\, M_{\odot}$.\footnote{We have taken these values
for convenience. The numerical applications of our model could be refined by
considering more accurate values of $M_{\rm min}$ and $R_{\rm min}$ but the
order of magnitude of our results should be correct.} The ground state of the
self-gravitating BEC also describes the quantum core of larger halos
with $M>M_{\rm min}$. This quantum core is surrounded by an approximately
isothermal atmosphere (mimicking the NFW profile) yielding flat rotation curves
at large distances as discussed in, e.g., \cite{modeldm}.

We have first determined an accurate expression of the mass-radius relation
$M(R)$ of self-gravitating BECs by combining approximate analytical results
obtained from the Gaussian ansatz \cite{prd1} with exact asymptotic results
obtained by solving the GPP equations numerically \cite{prd2}.
Assuming that this mass-radius relation describes the minimum DM halo 
with  $R_{\rm min}=1\, {\rm kpc}$ and $M_{\rm min}=10^8\, M_{\odot}$ (as well as
the cores of larger DM halos) we have obtained an accurate expression of the DM
mass-scattering length relation $m(a_s)$. This relation determines the mass $m$
that the DM particle with a scattering length $a_s$ should have in order to
yield results that are consistent with the mass and the size of the minimum
halo typically representing a dSph.

For noninteracting bosons, we found
\begin{eqnarray}
m=2.92\times 10^{-22}\, {\rm eV}/c^2,
\label{con1}
\end{eqnarray}
which is the typical mass of the DM particle considered in FDM scenarios.

For bosons with an attractive self-interaction, we found that the mass of the DM
particle  is restricted by the inequality
\begin{eqnarray}
2.19\times 10^{-22}\, {\rm eV}/c^2 < m < 2.92\times 10^{-22}\, {\rm eV}/c^2,
\label{con2}
\end{eqnarray}
otherwise dSphs like Fornax would be unstable (their mass would be above the 
maximum mass $M_{\rm max}$ found in \cite{prd1}). Therefore, an attractive
self-interaction almost
does not change the typical mass of the DM particle required to match the
characteristics of the minimum halo (the boson mass is just a little smaller
than the value from Eq. (\ref{con1}) in the noninteracting
model).  In addition, in line with our previous works
\cite{suarezchavanisprd3,mcmh,mcmhbh},
we have shown that, in situations of
astrophysical interest, the effect of an attractive self-interaction is
negligible both in the linear (see Sec. \ref{sec_att}) and nonlinear (see
Sec. \ref{sec_cpp}) regimes of structure formation. Therefore, in practice,
bosons with an attractive self-interaction can be considered as
noninteracting.\footnote{These conclusions are valid for ULAs with 
$m_{\rm th}=2.92\times 10^{-22}\, {\rm eV}/c^2$ and $(a_s)_{\rm th}=-3.18\times
10^{-68}\, {\rm fm}$ (BECth) that
may form DM halos while fulfilling the constraints from particle
physics and cosmology (see
Sec. \ref{sec_cpp}). By contrast, the attractive
self-interaction of QCD axions is crucial in the context of QCD axion stars (see
Sec. \ref{sec_qcda}) while being negligible in the linear
regime of structure formation  (see Sec. \ref{sec_att}).
This suggests that the attractive
self-interaction of QCD axions  becomes important in the nonlinear
regime of structure formation.} 

For bosons with a repulsive self-interaction, we found that the mass of the DM
particle is restricted by the inequality
\begin{eqnarray}
2.92\times 10^{-22}\, {\rm eV}/c^2<m<1.10\times 10^{-3}\, {\rm eV}/c^2,
\label{con3}
\end{eqnarray}
where the maximum bound arises from the Bullet Cluster constraint.  Therefore, a
repulsive self-interaction can increase 
the typical mass of the DM particle by $18$ orders of magnitude
with respect to its value in the noninteracting case. As
noted in Appendix D.4 of \cite{abrilphas}, a mass larger than
$2.92\times 10^{-22}\, {\rm eV/c^2}$ could alleviate some
tensions with the observations of the Lyman-$\alpha$ forest
encountered in the noninteracting model. We have considered two typical
models corresponding to $m=1.10\times
10^{-3}\, {\rm eV}/c^2$ and $a_s=4.41\times 10^{-6}\, {\rm fm}$ (BECTF) and
$m=2.92\times 10^{-22}\, {\rm eV}/c^2$ and $a_s=8.13\times
10^{-62}\, {\rm fm}$ (BECt).

We have then shown that the Jeans mass-radius relation $M_J(R_J)$, which 
is valid in the linear regime of structure formation, is similar to
the mass-radius relation $M(R)$ of the minimum BECDM halo (or the core of large
halos), corresponding to the ground state of the GPP equations, which is valid
in the nonlinear regime of structure formation. This
analogy allows us to directly apply some results obtained in the context of
(nonlinear) self-gravitating BECs to the Jeans (linear) instability problem and
{\it vice versa}.

The two curves $M_J(R_J)$ and  $M(R)$ are parameterized by a typical density 
(the density of the universe for the $M_J(R_J)$ relation and the average -- or
central -- density
of the BEC for the $M(R)$ relation) going from high values to low
values.\footnote{In cosmology, it is natural to follow the
series of equilibria $M_J(R_J)$ from high to low values of the density because
this corresponds to the temporal evolution of the universe (from early to late
epochs). In the context of DM halos, as in the case of compact stars
\cite{htww}, it may be more relevant to follow the
series of equilibria $M(R)$ from low to high values of the density because
this corresponds to their natural evolution.} For
noninteracting
bosons, the mass decreases as the radius increases. For bosons with a repulsive
self-interaction, there is a minimum radius at which $M\rightarrow +\infty$
corresponding to the TF limit. The mass decreases as the radius increases, going
from the TF limit (high densities) to the NI limit (low densities). For  bosons
with an attractive self-interaction, the mass first increases as the radius
increases, reaches a maximum value $M_{\rm max}$, and then decreases, going from
the NG limit
(high densities) to the NI limit (low densities).

Despite these analogies, the curves $M_J(R_J)$ and  $M(R)$ have a very different
physical interpretation. 
The curve $M_J(R_J)$ determines the Jeans mass and the Jeans radius at different
epochs in the history of the universe characterized by its density $\rho$ (in
that case it is more relevant to plot $M_J(\rho)$ and $R_J(\rho)$ individually).
The Jeans
scales determine the minimum mass and the minimum size of a condensation that
can become unstable and form a clump. We must be careful, however, that the
Jeans instability study is
valid only in the linear regime of structure formation.
As a result, the
interpretation of the curve $M_J(R_J)$  and its domain of validity is not
straightforward. In principle, the results of the linear Jeans instability study
are valid only in a sufficiently young universe (typically the beginning of the
matter era) where $\rho_{\rm eq}=8.77\times 10^{-14}\, {\rm g}/{\rm m}^3$. It
is not clear if we can apply the results of the Jeans instability study at
later epochs. On
the other hand, the curve $M(R)$ determines the mass-radius relation of DM
halos that are formed in the nonlinear regime of structure formation after
having experienced free fall, violent relaxation, gravitational cooling, and
virialization. It applies to the minimum halo or to the quantum core of larger
halos.\footnote{In principle, even if we know the parameters of the DM particle
(mass $m$ and scattering length $a_s$) we
cannot determine the mass $M$ and the radius  $R$ of the minimum halo
individually. We just know its mass-radius relation $M(R)$. However, if we
assume a universal value $\Sigma_0=141\, M_{\odot}/{\rm pc}^2$  of the surface
density of DM halos compatible with the
observations, then we can determine the mass $(M_h)_{\rm min}$ and the radius 
$(R_h)_{\rm min}$ of the
minimum halo individually.  This is done in \cite{modeldm} and in Sec.
II of \cite{mcmh} where the mass $(M_h)_{\rm min}$  and the radius $(R_h)_{\rm
min}$ of the
minimum halo are expressed in terms of $m$, $a_s$ and $\Sigma_0$. We can then
see if they coincide with the Jeans scales (see Appendix I of \cite{mcmh}).}
We expect that the
mass and the size of the minimum halo is of the order of the Jeans mass and
Jeans radius ($M\sim M_J$ and $R\sim R_J$) calculated at the ``relevant'' epoch
of structure formation. There is, however, an uncertainty about what this epoch
is (see Sec. \ref{sec_optimal}). Furthermore, the relation between the Jeans
scales and the characteristics of DM halos is not straightforward. In practice,
the linear Jeans instability occurs at a certain epoch, leading to weakly
inhomogeneous clumps of mass $M_J$ and radius $R_J$. Then, these clumps
evolve in the nonlinear regime ultimately leading to DM halos of minimum mass 
$M\sim M_J$ and minimum radius $R\sim R_J$. The DM halos may also merge (or
inversely loose mass) so that their actual mass $M$ and radius $R$ may be
different
from $M_J$ and $R_J$. 
On the other hand, the $M(R)$ relation may
present regions of instability such as the NG branch of Fig.
 \ref{M-R-chi-neg-part1}. The solutions on these branches
cannot correspond to observable DM halos since they are unstable. These
branches are therefore forbidden in the nonlinear problem of structure
formation. However, the corresponding branches in the Jeans
mass-radius relation $M_J(R_J)$ have their usual interpretation. They
determine the mass and size triggering the gravitational instability in the
linear regime.
The existence of stable or unstable branches in the mass-radius relation
$M(R)$ of DM halos leads to the different
possibilities described at the end of Secs. \ref{sec_rew} and \ref{sec_att}.
 Typically, we have two possibilities:

(i) Consider first a branch of the $M_J(R_J)$ relation such that the
corresponding branch of the $M(R)$ relation is stable (for example the
NI branch or the TF branch). In that case, clumps of mass $M_J$ and $R_J$ formed
in the linear regime by Jeans instability evolve, in the nonlinear regime, into
stable DM halos of mass $M\sim M_J$ and radius $R\sim R_J$. They can then
increase their mass by mergings and accretion (or possibly loose mass) leading
to the DM halos observed in the universe.

(ii) Consider now a branch of the $M_J(R_J)$ relation such that the
corresponding branch of the $M(R)$ relation is unstable (for example the
NG branch). In that case, clumps of mass $M_J$ and $R_J$ formed in the linear
regime by Jeans instability cannot evolve, in the nonlinear regime, into
DM halos of mass  $M\sim M_J$ and radius $R\sim R_J$ (such halos are unstable).
They
rather undergo an explosion \cite{tkachevprl} or a  violent (nonlinear)
gravitational collapse leading presumably to smaller objects
determined by higher order repulsive terms in the self-interaction potential
\cite{braaten,ebycollapse,phi6}.

In the present paper, we have illustrated our results for bosons 
interacting via a $\phi^4$ potential. We have considered the case of a
vanishing ($a_s=0$), repulsive ($a_s>0$) or attractive ($a_s<0$)
self-interaction. The model of noninteracting bosons (FDM) leads to a boson mass
$m\sim 10^{-22}\, {\rm eV/c^2}$ that creates some
tensions with the observations of the Lyman-$\alpha$ forest \cite{hui}. These
observations
require a larger mass of at least one order of magnitude. We have shown 
that the model of bosons with an attractive self-interaction necessitates a
mass even smaller than $m\sim 10^{-22}\, {\rm eV/c^2}$ (according to Fig.
\ref{abism} the mass $m$ decreases as $|a_s|$ increases when $a_s<0$). This
model is therefore also in tension with the observations. By contrast, the model
of bosons with a repulsive self-interaction allows a boson mass which can be up
to $18$ orders of magnitude larger than $m\sim 10^{-22}\, {\rm eV/c^2}$
(according to Fig. \ref{abism} the
mass $m$ increases as $a_s$ increases when $a_s>0$). As noted in
\cite{abrilphas} this model could alleviate some tensions with the
observations of the Lyman-$\alpha$ forest
encountered in the noninteracting model. As a result, a
repulsive self-interaction ($a_s>0$) is priviledged over an attractive
self-interaction ($a_s<0$) \cite{modeldm}. A repulsive self-interaction is
also favored by
cosmological constraints \cite{shapiro,abrilphas} which yield a
fiducial model with a mass $m=3\times 10^{-21}\, {\rm eV}/c^2$ and
a scattering length $a_s=1.11\times 10^{-58}\, {\rm fm}$ (BECf). We recall that
theoretical models of particle physics  usually lead to particles with
an attractive self-interaction (e.g., the QCD axion). However, some authors
\cite{fan,reig} have pointed out the possible existence of particles with a
repulsive self-interaction (e.g., the light majoron).

\appendix

\section{Derivation of Schr\"odinger's equation}

In this Appendix, we briefly recall the derivation of the 
Schr\"odinger equation from the
formalism of scale relativity \cite{nottale}. We follow
the presentation given in Ref. \cite{epjpnottale}.

Nottale  \cite{nottale} has shown
that the Schr\"odinger equation
is equivalent to the fundamental equation of dynamics
\begin{equation}
\label{w1}
\frac{D{\bf U}}{Dt}=-\nabla\Phi,
\end{equation}
where ${\bf F}=-\nabla\Phi$ is the force by unit of mass exerted on a
particle, provided that ${\bf U}({\bf r},t)$ is interpreted as a complex
velocity field
and $D/Dt$ as a complex time derivative operator (or covariant derivative)
defined by
\begin{equation}
\label{w2}
\frac{D}{Dt}=\frac{\partial}{\partial t}+{\bf U}\cdot \nabla-i {\cal D}\Delta,
\end{equation}
where
\begin{equation}
\label{w3}
{\cal D}=\frac{\hbar}{2m}
\end{equation}
is the Nelson \cite{nelson} diffusion coefficient of quantum
mechanics or the fractal fluctuation parameter in the theory of scale
relativity \cite{nottale}. 
Using the
expression (\ref{w2}) of the covariant derivative, Eq. (\ref{w1}) can be
rewritten as a complex viscous Burgers equation
\begin{equation}
\label{w5}
\frac{\partial {\bf U}}{\partial t}+({\bf U}\cdot \nabla){\bf U}=i {\cal
D}\Delta{\bf U}-\nabla\Phi
\end{equation}
with an imaginary viscosity $\nu=i{\cal D}$.
It can be shown \cite{nottale} that the complex velocity field can be
written as the gradient of a complex action:
\begin{equation}
\label{w6}
{\bf U}=\frac{\nabla {\cal S}}{m}.
\end{equation}
This defines a potential flow. As a consequence, the
flow
is irrotational: $\nabla\times {\bf U}={\bf 0}$. Using the well-known identities
of fluid mechanics $({\bf U}\cdot \nabla){\bf U}=\nabla ({{\bf U}^2}/{2})-{\bf
U}\times (\nabla\times {\bf U})$ and $\Delta{\bf U}=\nabla(\nabla\cdot {\bf
U})-\nabla\times(\nabla\times{\bf U})$  which reduce to $({\bf U}\cdot
\nabla){\bf U}=\nabla ({{\bf U}^2}/{2})$ and $\Delta{\bf U}=\nabla (\nabla\cdot
{\bf U})$ for an irrotational flow, and using the identity $\nabla\cdot {\bf
U}=\Delta{\cal S}/m$ resulting from Eq. (\ref{w6}), we find that Eq. (\ref{w5})
is equivalent to the complex quantum Hamilton-Jacobi (or Bernoulli) equation
\begin{equation}
\label{w7}
\frac{\partial {\cal S}}{\partial t}+\frac{1}{2m}(\nabla {\cal S})^2-i{\cal
D}\Delta{\cal S}+m\Phi=0.
\end{equation}
We
now define the wave function $\psi({\bf r},t)$ through the complex Cole-Hopf
transformation
\begin{equation}
\label{w8}
{\cal S}=-2im{\cal D}\ln\psi,
\end{equation}
which is equivalently to the WKB formula
\begin{equation}
\label{w9}
\psi=e^{i{\cal S}/\hbar}.
\end{equation}
Substituting Eq. (\ref{w8}) into Eq. (\ref{w7}), and using the
identity
\begin{equation}
\label{w10}
\Delta(\ln\psi)=\frac{\Delta\psi}{\psi}-\frac{1}{\psi^2}(\nabla\psi)^2,
\end{equation}
 we obtain the 
Schr\"odinger equation 
\begin{equation}
\label{w11}
i\hbar\frac{\partial\psi}{\partial
t}=-\frac{\hbar^2}{2m}\Delta\psi+m\Phi\psi.
\end{equation}

\section{Equivalence between the stability criteria based on the equation of
pulsation and on the energy principle}
\label{sec_alt}

\subsection{Energy principle}
\label{sec_en}

The GPP equations (\ref{gpp1}) and (\ref{gpp2}), or equivalently the quantum
Euler equations (\ref{mad3})-(\ref{mad5}), conserve the mass $M$ and the energy
$E_{\rm tot}$ defined by Eqs. (\ref{mad11}) and (\ref{mad12}) (see, e.g.,
Appendix E of \cite{prd1}). Using very general arguments \cite{holm}, this
implies that:

(i) An equilibrium state of the GPP equations is an extremum of
energy at fixed mass. This result can be established easily. Let us
write the variational problem for the first variations as 
\begin{eqnarray}
\label{en1}
\delta E_{\rm tot}-\frac{\mu}{m}\delta M=0,
\end{eqnarray}
where $\mu$ (global chemical potential) is a Lagrange multiplier taking into
account the mass constraint. Using Eqs. (\ref{ui0})-(\ref{ui3}), and treating
the perturbations $\delta{\bf u}$ and $\delta\rho$ independently, we
obtain ${\bf u}={\bf 0}$ (the equilibrium state is static) and the quantum Gibbs
condition (see footnote 6)
\begin{eqnarray}
\label{en2}
Q+m h+m \Phi=\mu,
\end{eqnarray}
which is equivalent to the condition of quantum hydrostatic equilibrium (see
Sec. \ref{sec_dm} and Appendix \ref{sec_ep}).

(ii) An equilibrium state of the GPP equations is stable
if, and only if, it is a minimum of energy at fixed mass. We will establish
this result in Appendix \ref{sec_eq} directly from the equation of pulsation.
Since
$\delta^2\Theta_c$ depends only on $\delta {\bf u}$ and is positive [see Eq.
(\ref{ui4a}) with ${\bf u}={\bf 0}$], and since the perturbations $\delta{\bf
u}$ and $\delta\rho$ are treated independently, we can
equivalently claim that an equilibrium state of the GPP equations is stable
if, and only if, it is a minimum of the reduced energy $E^*_{\rm
tot}=\Theta_Q+U+W$
(excluding the classical kinetic energy) at fixed mass. The
condition of
dynamical stability based on the energy principle is therefore
\begin{eqnarray}
\label{en3}
\delta^2E^*_{\rm tot}>0
\end{eqnarray}
for all perturbations $\delta\rho$ that conserve mass ($\delta M=0$).
Using the identities of Appendix \ref{sec_ui}, we have
\begin{eqnarray}
\label{en4}
\delta^2E_{\rm tot}^*=\frac{1}{2}\int  \left (\frac{\delta
Q}{m}+\delta h+\delta\Phi\right )\delta\rho\, d{\bf r}
\end{eqnarray}
or, equivalently,
\begin{eqnarray}
\label{en5}
\delta^2E_{\rm tot}^*=\frac{1}{2}\int h'(\rho)(\delta\rho)^2\, d{\bf
r}+\frac{1}{2}\int \delta\Phi\delta\rho\, d{\bf
r}\nonumber\\
+\frac{\hbar^2}{8m^2}\int \frac{1}{\rho}\left \lbrack
(\nabla\delta\rho)^2+\left
(\frac{\Delta\rho}{\rho}-\frac{(\nabla\rho)^2}{\rho^2}\right
)(\delta\rho)^2\right\rbrack \, d{\bf r},
\end{eqnarray}
where we recall that $h'(\rho)=P'(\rho)/\rho$ (see Appendix \ref{sec_ti}). 

{\it Remark:} The minimization problem (\ref{dm8}) expressing the energy
principle 
is a criterion of nonlinear dynamical
stability resulting from the fact that $E_{\rm tot}$ and $M$ are conserved by 
the GPP equations \cite{holm}. It provides a necessary and
sufficient condition of dynamical stability since it takes into
account all the invariants of the GPP equations.

\subsection{Equation of pulsations}
\label{sec_ep}

The quantum Euler-Poisson equations (\ref{mad3})-(\ref{mad5}) may be written as
\begin{equation}
{\partial\rho\over\partial t}+\nabla \cdot (\rho {\bf u})=0,
\label{ep1}
\end{equation}
\begin{equation}
{\partial {\bf u}\over\partial t}+({\bf u}\cdot \nabla) {\bf
u}=-\frac{1}{m}\nabla Q-\nabla h-\nabla\Phi,
\label{ep2}
\end{equation}
\begin{equation}
\Delta\Phi=4\pi G\rho,
\label{ep3}
\end{equation}
where we have introduced the enthalpy $h(\rho)=V'(\rho)$ through the
relation (see Appendix \ref{sec_ti})
\begin{equation}
\nabla h=\frac{\nabla P}{\rho}.
\label{ep4}
\end{equation}
A steady state of the quantum Euler equation (\ref{ep2})
satisfies
the condition of quantum  hydrostatic equilibrium (see Sec. \ref{sec_dm})
\begin{equation}
\frac{\nabla Q}{m}+\nabla h+\nabla\Phi={\bf 0},
\label{ep5}
\end{equation}
which is equivalent to Eq. (\ref{en2}).
Combined with the Poisson equation (\ref{ep3}) we obtain the fundamental
differential equation determining the equilibrium structure of a
self-gravitating BEC
\begin{equation}
\label{ep6}
\frac{\Delta Q}{m}+\Delta h=-4\pi G\rho,
\end{equation}
where $\rho=\rho(h)$ according to Eq. (\ref{ep4}). This equation is equivalent
to Eq.
(\ref{dm6}).

Let us consider a stationary solution of the quantum Euler-Poisson equations
(\ref{ep1})-(\ref{ep3}) satisfying ${\bf u}={\bf 0}$ and the
condition of quantum hydrostatic equilibrium (\ref{ep5}). The linearized
quantum
Euler-Poisson equations around this equilibrium state are
\begin{eqnarray}
\label{ep7} {\partial\delta \rho\over\partial t} +\nabla\cdot
(\rho\delta {\bf u})=0,
\end{eqnarray}
\begin{eqnarray}
\label{ep8} {\partial \delta {\bf u}\over\partial t}=-\frac{1}{m}\nabla \delta
Q
-\nabla\delta h-\nabla\delta\Phi,
\end{eqnarray}
\begin{eqnarray}
\label{ep9} \Delta\delta\Phi=4\pi G\delta\rho.
\end{eqnarray}
It is convenient to introduce the Lagrangian displacement
$\vec{\zeta}=\delta {\bf
r}$ such that
\begin{eqnarray}
\label{ep10} \delta {\bf u}=\frac{\partial \vec{\zeta}}{\partial t}.
\end{eqnarray}
The linearized continuity equation (\ref{ep7}) leads to the relation
\begin{eqnarray}
\label{ep11} \delta \rho=-\nabla\cdot
(\rho \vec{\zeta}).
\end{eqnarray}
Writing the evolution of
the perturbation as $e^{-i\omega t}$, Eq. (\ref{ep10}) implies that $\delta{\bf
u}=-i\omega
\vec{\zeta}$. 
On the other hand, the
linearized quantum  Euler equation (\ref{ep8}) becomes
\begin{eqnarray}
\label{ep12}
\omega^2 \vec{\zeta}=\frac{1}{m}\nabla \delta
Q+\nabla\delta h+\nabla\delta\Phi.
\end{eqnarray}
Using  Eq. (\ref{ep11}) and Eqs. (\ref{ui7})-(\ref{ui10}), the first order
variations $\delta Q$, $\delta h$ and $\delta\Phi$ can be  expressed in terms of
 $\vec{\zeta}$. In this manner, Eq. (\ref{ep12}) represents the quantum
generalization of the equation of
pulsations in the form given
by Chandrasekhar \cite{chandraVP}. This is
an eigenvalue equation determining the possible pulsations of the system. The
equilibrium state is stable if $\omega^2>0$ for all modes (in that case the
perturbation oscillates) and unstable if $\omega^2<0$ for some modes  (in that
case the perturbation grows exponentially rapidly).
Using Eqs. (\ref{ep11}), (\ref{ui7}) and (\ref{ep13}), we can rewrite Eq.
(\ref{ep12}) more explicitly as
\begin{eqnarray}
\label{ep15}
\omega^2\vec{\zeta}=\frac{1}{m}\nabla \delta
Q-\nabla \left \lbrack \frac{P'(\rho)}{\rho}\nabla\cdot (\rho\vec{\zeta})\right
\rbrack-4\pi G\rho \vec{\zeta}.
\end{eqnarray}
Alternatively, combining
Eqs.  (\ref{ep11}) and (\ref{ep12}), we can write the quantum
equation of pulsations under the form
\begin{eqnarray}
\label{ep16}
-\omega^2\delta\rho=\nabla\cdot\left\lbrack \rho \left
(\frac{\nabla\delta Q}{m}+\nabla\delta h+\nabla\delta\Phi\right
)\right\rbrack,
\end{eqnarray}
where $\delta Q$, $\delta h$ and
$\delta\Phi$ are expressed in terms of
$\delta\rho$ through Eqs. (\ref{ui7})-(\ref{ui10}).

If we consider a spherically symmetric distribution of matter, and consider
radial perturbations, it is convenient to
introduce the quantity $q$ from the relation \cite{aaantonov}
\begin{eqnarray}
\label{ep17}
\delta\rho=\frac{1}{4\pi r^2}\frac{dq}{dr}.
\end{eqnarray}
Physically, $q(r,t)=\delta M(r,t)=\int_0^r \delta\rho(r',t)4\pi {r'}^2\, dr'$
represents the perturbed mass within the sphere of radius $r$. The
perturbed Newton equation takes the form
\begin{eqnarray}
\label{ep18}
\frac{d\delta\Phi}{dr}=\frac{Gq}{r^2}.
\end{eqnarray}
Since
\begin{eqnarray}
\label{ep18b}
\delta\rho=-\frac{1}{r^2}\frac{d}{dr}(r^2\rho\zeta),
\end{eqnarray}
we obtain the relation
\begin{eqnarray}
\label{ep19}
\zeta=-\frac{q}{4\pi\rho r^2}.
\end{eqnarray}
Starting from Eq. (\ref{ep15}) or from Eq. (\ref{ep16}), and using Eq.
(\ref{ep19}) or Eq. (\ref{ep17}), we obtain 
\begin{eqnarray}
\label{ep20}
\frac{d}{dr}\left (\frac{P'(\rho)}{4\pi\rho r^2}\frac{dq}{dr}\right
)+\frac{Gq}{r^2}+\frac{1}{m}\frac{d\delta Q}{dr}=-\frac{\omega^2}{4\pi\rho
r^2}q.
\end{eqnarray}
This is the quantum generalization of the equation of pulsation in the form
given in Appendix A of \cite{aaantonov}. Starting from Eq.
(\ref{ep15}) or from Eq. (\ref{ep20}), we obtain after some calculations
\begin{eqnarray}
\label{ep22}
\frac{d}{dr}\left\lbrack \gamma P
\frac{1}{r^2}\frac{d}{dr}(r^2\zeta)\right\rbrack
-\frac{4}{r}\frac{dP}
{ dr } \zeta\nonumber\\
-\frac{\rho}{m}\Delta Q\zeta-\frac{\rho}{m}\frac{d\delta
Q}{dr}=-\omega^2\rho\zeta,
\end{eqnarray}
where we have defined
\begin{eqnarray}
\label{ep23}
\gamma(r)=\frac{d\ln P}{d\ln\rho}=\frac{\rho}{P}P'(\rho).
\end{eqnarray}
Introducing the variable $\xi=\zeta/r$, we can transform Eq. (\ref{ep22})
into
\begin{eqnarray}
\label{ep24}
\frac{d}{dr}\left ( \gamma
P r^4\frac{d\xi}{dr}\right )
+r^3 \frac{d}{dr}\left \lbrack (3\gamma-4)P\right 
\rbrack \xi\nonumber\\
-\frac{\rho}{m}\Delta Q r^4 \xi-\frac{\rho}{m}r^3\frac{d\delta
Q}{dr}=-\omega^2\rho r^4 \xi.
\end{eqnarray}
This is the quantum generalization of the equation of pulsations in the form
given by Eddington \cite{eddington18}.

\subsection{Equivalence between $\omega^2>0$ and $\delta^2E_{\rm tot}^*>0$}
\label{sec_eq}

Taking the scalar product of Eq. (\ref{ep12}) with $\rho\vec{\zeta}$ and
integrating over the whole domain we obtain
\begin{eqnarray}
\label{eq1}
\omega^2 \int \rho \zeta^2 \, d{\bf r}=\int \rho \vec{\zeta}\cdot
\nabla\left (\frac{\delta
Q}{m}+\delta
h+\delta\Phi\right )\, d{\bf r}.
\end{eqnarray}
Integrating the second integral by parts and using Eq. (\ref{ep11}), we can
rewrite Eq. (\ref{eq1}) as
\begin{eqnarray}
\label{eq2}
\omega^2 \int \rho \zeta^2 \, d{\bf r}=\int \delta\rho \left (\frac{\delta
Q}{m}+\delta h+\delta\Phi\right )\, d{\bf r}.
\end{eqnarray}
Comparing the right hand side of this expression with the second variations of
the energy functional
from Eq. (\ref{en4}), we
obtain the identity
\begin{eqnarray}
\label{eq3}
\frac{1}{2}\omega^2 \int \rho \zeta^2 \, d{\bf r}=\delta^2 E_{\rm tot}^*.
\end{eqnarray}
Since the integral is positive, this identity shows that an equilibrium state of
the GPP equations is
dynamically stable ($\omega^2>0$) if, and only if, it is a minimum of energy at
fixed mass ($\delta^2 E_{\rm tot}^*>0$). Therefore, the
stability criteria based on the equation of
pulsation and on the energy principle are equivalent. This identity also
provides  the basis for a quantum generalization of the Chandrasekhar
variational principle \cite{chandraVP}.

{\it Remark:} Since the total energy is conserved, we have $\delta^2E_{\rm
tot}=0$ or, equivalently, $\delta^2\Theta_c+\delta^2E^*_{\rm
tot}=0$. Using Eq. (\ref{ui4a}) with ${\bf u}={\bf 0}$ and $\delta{\bf
u}=-i\omega
\vec{\zeta}$, we see that this identity is equivalent to Eq. (\ref{eq3}).
On the other hand, since  $\delta^2\Theta_c\ge 0$, the identity
$\delta^2\Theta_c+\delta^2E^*_{\rm
tot}=0$ implies the following results: (i) If
$\delta^2E^*_{\rm
tot}>0$, we cannot have growing modes so the system is stable. (ii) If
$\delta^2E^*_{\rm
tot}<0$, we can have a growing mode so the system is unstable. This directly
establishes the stability result based on the energy principle (see Appendix
\ref{sec_en}).

\subsection{Useful identities}
\label{sec_ui}

The first order variations of the functionals defined by Eqs.
(\ref{mad12b})-(\ref{mad12d}) are 
\begin{eqnarray}
\label{ui0}
\delta\Theta_c=\int \delta\rho \frac{{\bf u}^2}{2}\, d{\bf r}+\int \rho {\bf
u}\cdot \delta {\bf u}\, d{\bf r},
\end{eqnarray}
\begin{eqnarray}
\label{ui1}
\delta\Theta_Q=\int \frac{Q}{m}\delta\rho\, d{\bf r},
\end{eqnarray}
\begin{eqnarray}
\label{ui2}
\delta U=\int V'(\rho)\delta\rho\, d{\bf r}=\int h(\rho)\delta\rho\, d{\bf
r},
\end{eqnarray}
\begin{eqnarray}
\label{ui3}
\delta W=\int \Phi\delta\rho\, d{\bf r},
\end{eqnarray}
where we have used $h(\rho)=V'(\rho)$ (see Appendix
\ref{sec_ti}). The second order variations of
these functionals are
\begin{eqnarray}
\label{ui4a}
\delta^2\Theta_c=\int \rho \frac{(\delta{\bf u})^2}{2}\, d{\bf r}+\int
\delta\rho {\bf
u}\cdot \delta {\bf u}\, d{\bf r},
\end{eqnarray}
\begin{eqnarray}
\label{ui4}
\delta^2\Theta_Q=\frac{1}{2}\int \delta\rho \frac{\delta Q}{m}\, d{\bf r},
\end{eqnarray}
\begin{eqnarray}
\label{ui5}
\delta^2 U=\frac{1}{2}\int V''(\rho)(\delta\rho)^2\, d{\bf r}&=&\frac{1}{2}\int
h'(\rho)(\delta\rho)^2\, d{\bf r}\nonumber\\
&=&\frac{1}{2}\int \delta h
\delta\rho\, d{\bf r},
\end{eqnarray}
\begin{eqnarray}
\label{ui6}
\delta^2 W=\frac{1}{2}\int \delta\Phi\delta\rho\, d{\bf r}.
\end{eqnarray}
We also note that $\delta h$, $\delta\Phi$ and $\delta Q$ are related to
$\delta\rho$ by
\begin{eqnarray}
\label{ui7}
\delta h=h'(\rho)\delta\rho=\frac{P'(\rho)}{\rho}\delta\rho,
\end{eqnarray}
\begin{eqnarray}
\label{ui8}
\delta\Phi=-G\int \frac{\delta\rho({\bf r}')}{|{\bf r}-{\bf r}'|}\, d{\bf r}',
\end{eqnarray}
\begin{equation}
\label{ui9}
\delta Q=\frac{\hbar^2}{4m}\frac{1}{\sqrt{\rho}}\left \lbrack
\frac{\delta\rho}{\rho}\Delta\sqrt{\rho}-\Delta \left
(\frac{\delta\rho}{\sqrt{\rho}}\right )\right\rbrack,
\end{equation}
or
\begin{equation}
\label{ui10}
\delta Q=\frac{\hbar^2}{4m}\left \lbrack
\frac{\Delta\rho}{\rho^2}\delta\rho-\frac{\Delta\delta\rho}{\rho}
-\frac{1}{\rho^3}
(\nabla\rho)^2\delta\rho+(\nabla\rho\cdot\nabla\delta\rho)
\frac{1}{\rho^2}\right\rbrack,
\end{equation}
where we have used
$h'(\rho)=P'(\rho)/\rho$ (see Appendix
\ref{sec_ti}). Eq. (\ref{ui9}) has been obtained by starting
from the first equality of Eq. (\ref{mad6}) and Eq. (\ref{ui10}) has been
obtained by starting
from the second equality of Eq. (\ref{mad6}). Other
expressions of $\delta Q$ are provided in Appendix C of \cite{ggpp}.
The identities (\ref{ui1})-(\ref{ui10}) are straightforward except, maybe, Eqs.
(\ref{ui1}) and (\ref{ui4}). Therefore, we give a short derivation of
these identities below.

Starting from the first equality of Eq. (\ref{mad12b}), we get at first order
\begin{equation}
\label{ui11}
\delta\Theta_Q=\frac{\hbar^2}{8m^2}\int \left\lbrack
-\left (\frac{\nabla\rho}{\rho}\right )^2
\delta\rho+2\frac{\nabla\rho}{\rho}\cdot\nabla\delta\rho\right\rbrack\, d{\bf
r}.
\end{equation}
Integrating the second term by parts, the foregoing equation can be rewritten as
\begin{equation}
\label{ui12}
\delta\Theta_Q=-\frac{\hbar^2}{4m^2}\int \left\lbrack
\frac{\Delta\rho}{\rho}-\frac{1}{2} \left (\frac{\nabla\rho}{\rho}\right
)^2\right\rbrack \delta\rho\, d{\bf r}.
\end{equation}
Together with Eq. (\ref{mad6}), it yields Eq. (\ref{ui1}). 
At second order,
we have
\begin{eqnarray}
\label{ui13}
\delta^2\Theta_Q=\frac{\hbar^2}{8m^2}\int \frac{1}{\rho}\biggl \lbrack
(\nabla\delta\rho)^2-2(\nabla\rho\cdot\nabla\delta\rho)\frac{\delta\rho}{\rho}
\nonumber\\
 +(\nabla\rho)^2\left (\frac{\delta\rho}{\rho}\right
)^2\biggr\rbrack\, d{\bf r}.
\end{eqnarray}
Integrating the middle term by parts, we can rewrite Eq. (\ref{ui13}) as
\begin{equation}
\label{ui14}
\delta^2\Theta_Q=\frac{\hbar^2}{8m^2}\int \frac{1}{\rho}\left \lbrack
(\nabla\delta\rho)^2+\left
(\frac{\Delta\rho}{\rho}-\frac{(\nabla\rho)^2}{\rho^2}\right
)(\delta\rho)^2\right\rbrack \, d{\bf r},
\end{equation}
which is the result quoted in Appendix C of Ref. \cite{ggpp}. On the other
hand, multiplying  Eq.
(\ref{ui10}) by $\delta\rho$ and integrating over the whole domain, we obtain
\begin{eqnarray}
\label{ui16}
\frac{1}{2}\int\delta\rho\frac{\delta Q}{m}\, d{\bf
r}=\frac{\hbar^2}{8m^2}\int \biggl \lbrack
\frac{\Delta\rho}{\rho^2}(\delta\rho)^2-\frac{\Delta\delta\rho}{\rho}
\delta\rho\nonumber\\
-\frac{1}{\rho^3}
(\nabla\rho)^2(\delta\rho)^2+(\nabla\rho\cdot\nabla\delta\rho)
\frac{1}{\rho^2}\delta\rho\biggr\rbrack\, d{\bf r}.
\end{eqnarray}
Integrating the second term by parts, and comparing the resulting expression
with Eq. (\ref{ui14}), we obtain Eq. (\ref{ui4}).

Finally, we establish the identity
\begin{eqnarray}
\label{ep13}
\nabla\delta\Phi=-4\pi G\rho \vec{\zeta},
\end{eqnarray}
used in Eq. (\ref{ep15}). This identity results from the following steps:
\begin{eqnarray}
\label{ui8b}
\nabla\delta\Phi&=&-G\int \nabla\left (\frac{1}{|{\bf r}-{\bf
r}'|}\right )\delta\rho({\bf r}')\, d{\bf r}'\nonumber\\
&=&G\int \nabla'\left (\frac{1}{|{\bf r}-{\bf
r}'|}\right )\delta\rho({\bf r}')\, d{\bf r}'\nonumber\\
&=&-G\int \nabla'\left (\frac{1}{|{\bf r}-{\bf
r}'|}\right )\nabla'\cdot (\rho \vec{\zeta})\, d{\bf r}'\nonumber\\
&=&G\int \Delta'\left (\frac{1}{|{\bf r}-{\bf
r}'|}\right ) (\rho \vec{\zeta})'\, d{\bf r}'\nonumber\\
&=&-4\pi G\int \delta({\bf r}-{\bf
r}') (\rho \vec{\zeta})'\, d{\bf r}'\nonumber\\
&=&-4\pi G \rho \vec{\zeta}.
\end{eqnarray}

\section{Dimensionless self-interaction constant $\lambda$ and decay
constant $f$}
\label{sec_c}

In this Appendix, we introduce the dimensionless self-interaction
constant $\lambda$ and decay constant $f$ and regroup in a compact manner the
main formulae of the paper for a better visualisation. A detailed explanation
of these formulae is given in the main text and in Appendices \ref{sec_alf} and
\ref{sec_lf}.

The dimensionless self-interaction
constant is defined by (see, e.g.,
Appendix A of \cite{bectcoll})
\begin{equation}
\frac{\lambda}{8\pi}=\frac{a_s m c}{\hbar}.
\label{alf1}
\end{equation}
On the other hand, for bosons with an attractive self-interaction ($a_s<0$), the
decay constant is defined by (see, e.g.,
\cite{phi6})
\begin{equation}
\label{alf2}
f=\left (\frac{\hbar c^3 m}{32\pi |a_s|}\right )^{1/2}.
\end{equation}
We have the relation
\begin{eqnarray}
\label{alf6}
f=\frac{mc^2}{2|\lambda|^{1/2}}.
\end{eqnarray}

\subsection{Noninteracting bosons}

For noninteracting bosons
\begin{equation}
M=a\frac{\hbar^2}{Gm^2R}\quad \Rightarrow \quad m=\left
(\frac{a\hbar^2}{GMR}\right )^{1/2}.
\label{c1}
\end{equation}

\subsection{Repulsive self-interaction}

In the NI regime:
\begin{equation}
M=a\frac{\hbar^2}{Gm^2R}\quad\Rightarrow\quad m=\left
(\frac{a\hbar^2}{GMR}\right )^{1/2}.
\label{c2}
\end{equation}
In the TF regime:
\begin{equation}
R=b\left (\frac{a_s\hbar^2}{Gm^3}\right )^{1/2} \quad\Rightarrow\quad
\frac{a_s}{m^3}=\frac{GR^2}{b^2\hbar^2};
\label{c3}
\end{equation}
\begin{eqnarray}
\label{c3b}
R=b\left (\frac{\lambda\hbar^3}{8\pi Gm^4 c}\right
)^{1/2}\quad\Rightarrow\quad\frac{\lambda}{8\pi m^4}=
\frac{GR^2c}{b^2\hbar^3}.
\end{eqnarray}

For given $(m,a_s)$ the transition between the NI regime and the TF regime
corresponds to
\begin{equation}
\label{c4}
M_{t}=\frac{a}{b} \frac{\hbar}{\sqrt{Gm a_s}},\quad
R_t=b\left (\frac{a_s\hbar^2}{Gm^3}\right )^{1/2};
\end{equation}
\begin{eqnarray}
\label{c4bh}
M_t=\frac{a}{b}\left (\frac{8\pi\hbar c}{G\lambda}\right
)^{1/2},\quad R_t=b\left
(\frac{\lambda\hbar^3}{8\pi Gm^4 c}\right
)^{1/2}.
\end{eqnarray}

For given $(M,R)$ the transition between the NI regime and the TF regime
corresponds to
\begin{equation}
\label{c5}
m_0=\left (\frac{a\hbar^2}{GMR}\right )^{1/2},\quad
a'_*=\frac{a^{3/2}}{b^2}\left (\frac{\hbar^2 R}{GM^3}\right
)^{1/2},
\end{equation}
\begin{equation}
\frac{\lambda'_*}{8\pi}=\frac{a^2\hbar c}{b^2GM^2}.
\label{c7}
\end{equation}

\subsection{Attractive self-interaction}

In the NI regime:
\begin{equation}
M=a\frac{\hbar^2}{Gm^2R}\quad\Rightarrow \quad m=\left
(\frac{a\hbar^2}{GMR}\right )^{1/2}.
\label{c8}
\end{equation}
In the NG regime:
\begin{equation}
M=\frac{a}{b^2}\frac{mR}{|a_s|} \quad\Rightarrow\quad
\frac{|a_s|}{m}=\frac{aR}{b^2M};
\label{c9}
\end{equation}
\begin{equation}
M=\frac{a}{b^2}\frac{8\pi m^2Rc}{|\lambda|\hbar} \quad\Rightarrow\quad
\frac{|\lambda|}{8\pi m^2}=\frac{a}{b^2}\frac{Rc}{M\hbar};
\label{c9b}
\end{equation}
\begin{equation}
M=\frac{a}{b^2}\frac{32\pi Rf^2}{\hbar c^3} \quad\Rightarrow\quad
f^2=\frac{b^2}{a}\frac{M\hbar c^3}{32\pi R}.
\label{c9c}
\end{equation}

For given $(m,a_s)$ the transition between the NI regime and the NG regime
corresponds to
\begin{equation}
\label{c10}
M_{t}=\frac{a}{b} \frac{\hbar}{\sqrt{Gm |a_s|}},\qquad 
R_t=b\left (\frac{|a_s|\hbar^2}{Gm^3}\right )^{1/2};
\end{equation}
\begin{eqnarray}
\label{c4b}
M_t=\frac{a}{b}\left (\frac{8\pi\hbar c}{G|\lambda|}\right
)^{1/2},\quad R_t=b\left
(\frac{|\lambda|\hbar^3}{8\pi Gm^4 c}\right
)^{1/2};
\end{eqnarray}
\begin{equation}
\label{c4c}
M_t=\frac{a}{b}\left (\frac{32\pi\hbar f^2}{Gm^2c^3}\right
)^{1/2},\quad R_t=b\left
(\frac{\hbar^3c^3}{32\pi Gm^2 f^2}\right
)^{1/2}.
\end{equation}

For given $(M,R)$ the transition between the NI regime and the NG regime
corresponds to
\begin{equation}
\label{c11}
m_0=\left (\frac{a\hbar^2}{GMR}\right )^{1/2},\quad
a'_*=\frac{a^{3/2}}{b^2}\left (\frac{\hbar^2 R}{GM^3}\right
)^{1/2},
\end{equation}
\begin{equation}
\frac{\lambda'_*}{8\pi}=\frac{a^2\hbar c}{b^2GM^2},\quad
f'_*=\left (\frac{b^2}{a}\frac{\hbar c^3M}{32\pi R}\right )^{1/2}.
\label{c13}
\end{equation}

{\it Remark:} We note that the transition scales between the NI regime and the
NG regime in the attractive case coincide with the transition scales between the
NI regime and the TF regime in the repulsive case provided
that $a_s$ is replaced by $|a_s|$. We also note that the formulae expressed in
terms of $\lambda$ and $f$ involve
the speed of light $c$. This is purely artificial since our
results apply to nonrelativistic systems. The occurrence of $c$ is
due to the definitions of $\lambda$ and $f$ in Eqs. (\ref{alf1}) and
(\ref{alf2}).

\section{Reformulation of the results of Sec. \ref{sec_dm} in terms of
$\lambda$ and $f$}
\label{sec_alf}

In this Appendix, we reformulate the results of Sec. \ref{sec_dm} in terms of
the dimensionless self-interaction constant $\lambda$ and decay constant
$f$ (see Appendix \ref{sec_c}) instead of the scattering length $a_s$.

Using Eqs. (\ref{dm15}) and (\ref{alf1}), the radius of self-gravitating BECs
with a repulsive self-interaction in the TF regime can be written as
\begin{eqnarray}
\label{alf3}
R_{\rm TF}=\pi\left (\frac{\lambda\hbar^3}{8\pi Gm^4 c}\right
)^{1/2}=0.627\sqrt{\lambda}\frac{M_P}{m}
\lambda_C,
\end{eqnarray}
where $M_{\rm P}=(\hbar c/G)^{1/2}=2.18\times 10^{-5}\, {\rm
g}$ is the Planck mass and $\lambda_C=\hbar/mc$ is the Compton
wavelength of the particle. 

Using Eqs. (\ref{dm16}), (\ref{dm17}) and (\ref{alf1}), the maximum mass and the
corresponding radius  of self-gravitating BECs
with an attractive self-interaction can be written as
\begin{eqnarray}
\label{alf4}
M_{\rm max}=1.012\, \left (\frac{8\pi\hbar c}{G|\lambda|}\right
)^{1/2}=5.073\frac{M_P}{\sqrt{|\lambda|}},
\end{eqnarray}
\begin{eqnarray}
\label{alf5}
R_{99}^*=5.5\, \left (\frac{|\lambda|\hbar^3}{8\pi Gm^4 c}\right
)^{1/2}=1.1\sqrt{|\lambda|}\frac{M_P}{m}
\lambda_C.
\end{eqnarray}
We note that $M_{\rm max}$ depends only on
$\lambda$. Using Eq. (\ref{alf6}),
we also have
\begin{equation}
\label{alf7}
M_{\rm max}=1.012\, \left (\frac{8\pi\hbar c}{G}\right
)^{1/2}\frac{2f}{mc^2}=10.15 \frac{f}{M_P c^2}
\frac{M_{\rm P}^2}{m},
\end{equation}
\begin{equation}
\label{alf8}
R_{99}^*=5.5\, \left (\frac{\hbar^3}{8\pi Gm^4 c}\right
)^{1/2}\frac{mc^2}{2f}=0.55
\frac{M_Pc^2}{f}\lambda_C.
\end{equation}

{\it Remark:} A self-gravitating BEC of mass $M$ can be in equilibrium only if
$\lambda>\lambda_{\rm min}$ or $f>f_{\rm min}$ with 
\begin{eqnarray}
\lambda_{\rm min}=-25.7\,\frac{\hbar c}{GM^2}=-25.74\left
(\frac{M_P}{M}\right )^2,
\label{maxsl2b}
\end{eqnarray}
\begin{eqnarray}
f_{\rm min}&=&9.86\times 10^{-2} mMc^2\left (\frac{G}{\hbar c}\right
)^{1/2}\nonumber\\
&=&9.86\times 10^{-2} \frac{mM}{M_P^2}M_P c^2.
\label{maxsl2}
\end{eqnarray}

\section{Reformulation of the results of Sec. \ref{sec_mas} in terms of
$\lambda$ and $f$}
\label{sec_lf}

In this Appendix, we reformulate the results of Sec. \ref{sec_mas} in terms of
the dimensionless self-interaction constant $\lambda$ and decay constant
$f$ instead of the scattering length $a_s$.

The dimensionless self-interaction constant is defined by Eq. (\ref{alf1}). This
relation may be rewritten as
\begin{equation}
\frac{\lambda}{\lambda'_*}=\frac{a_s}{a'_*}\frac{m}{m_0},
\label{lf2}
\end{equation}
where we have introduced the scales from Eqs. (\ref{mas3}) and (\ref{mas4}), and
the new scale
\begin{equation}
\frac{\lambda'_*}{8\pi}=\frac{a^2\hbar c}{b^2GM^2}.
\label{lf3}
\end{equation}
We note that this scales depends only on the mass $M$ of the minimum halo (not
on its radius $R$).

\begin{figure}[!h]
\begin{center}
\includegraphics[clip,scale=0.3]{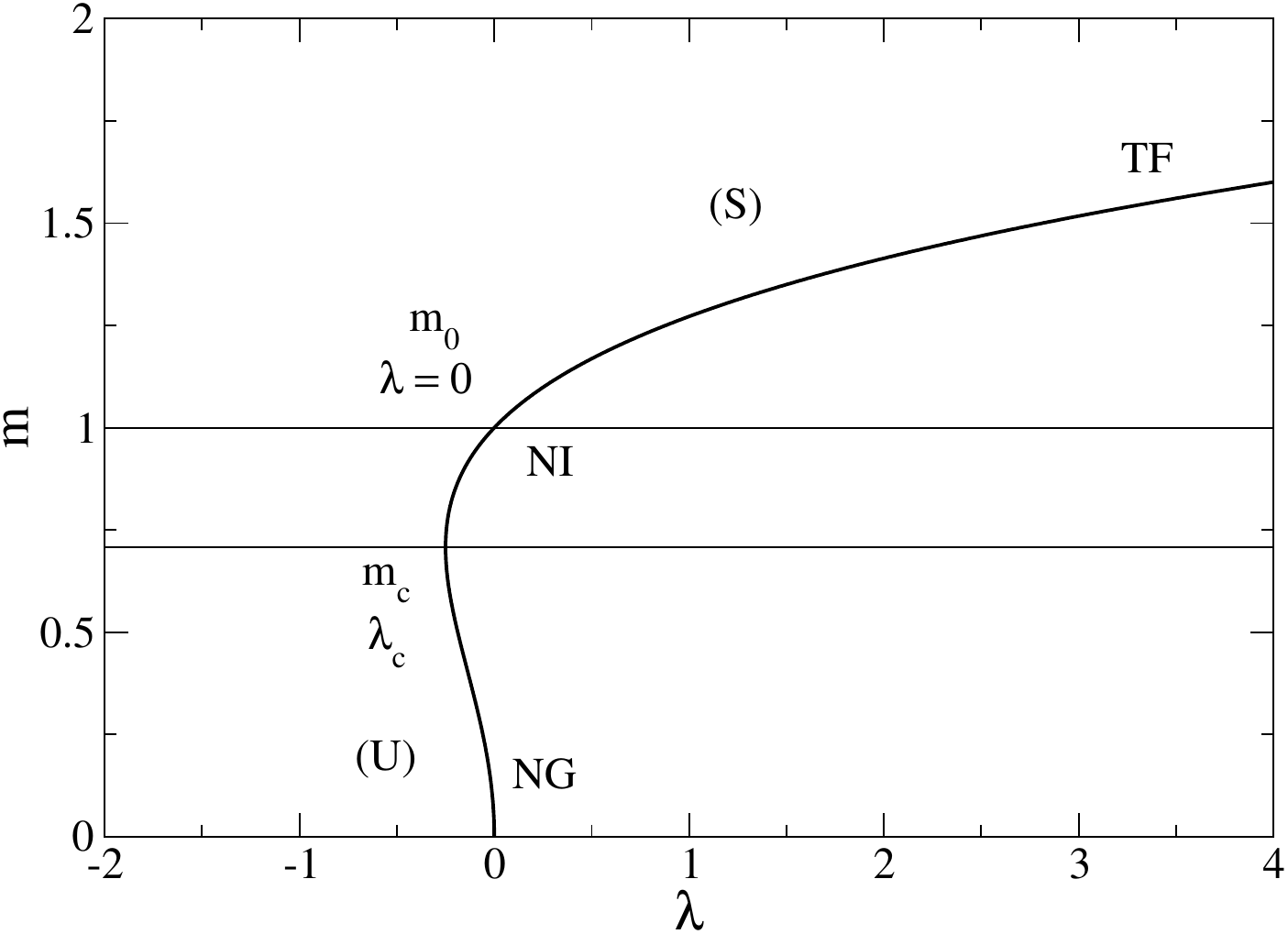}
\caption{Mass $m$ of the DM particle as a function of the dimensionless
self-interaction constant $\lambda$ in order to match the characteristics of the
minimum halo. The
mass is normalized by
$m_0$ and the dimensionless
self-interaction constant  by $\lambda'_*$. The stable part of
the curve starts at the critical minimum halo point ($\lambda_c,m_{c}$) which
is also the minimum of the curve $\lambda(m)$.}
\label{lambdam}
\end{center}
\end{figure}

Using Eqs. (\ref{asm1}) and (\ref{alf1}), the relation between the particle
mass
$m$ and the dimensionless self-interaction constant $\lambda$ required to
match the characteristics of the minimum halo [see Eq. (\ref{mas1})] is given by
\begin{equation}
\lambda=\frac{8\pi a R m^2 c}{b^2\hbar M}\left
(\frac{GMm^2R}{a\hbar^2}-1\right ).
\end{equation}
Alternatively, using Eqs. (\ref{mas2}) and (\ref{lf2}), we obtain in
dimensionless form
\begin{equation}
\label{lf4}
\frac{\lambda}{\lambda'_*}=\left (\frac{m}{m_0}\right
)^2\left\lbrack \left (\frac{m}{m_0}\right
)^2-1\right\rbrack.
\end{equation}
This is a second degree equation whose solutions are
\begin{equation}
\label{lf5}
\frac{m}{m_0}=\left (\frac{1\pm\sqrt{1+4\lambda/\lambda'_*}}{2}\right )^{1/2}
\end{equation}
with the sign $+$ if $\lambda>0$ and the sign $\pm$ if $\lambda<0$.
The curve $m(\lambda)$ is plotted in Fig. \ref{lambdam}.
Taking $a=9.946$
and $b=\pi$ (see Secs. \ref{sec_panb} and \ref{sec_parb}) adapted to bosons
with a repulsive self-interaction (or no interaction), we get $m_0=2.92\times
10^{-22}\, {\rm eV}/c^2$ and $\lambda'_*=3.02\times 10^{-90}$. Taking
$a=11.1$ and $b=5.5$ (see Sec. \ref{sec_paab}) adapted to bosons with an
attractive
self-interaction, we get $m_0=3.08\times 10^{-22}\,
{\rm eV}/c^2$ and $\lambda'_*=1.23\times 10^{-90}$. 

{\it Remark:} we note that the characteristic scale $\lambda'_*\sim  10^{-90}$
is extremely small. We will see below that the NI limit is valid
for $|\lambda|\ll \lambda'_*$. Therefore, the dimensionless self-interaction
constant $|\lambda|$ must be small with respect to $10^{-90}$, not with respect
to $1$. For example, an apparently small value of $|\lambda|$ such as
$|\lambda|= 10^{-80}$ actually corresponds to a strong self-interaction. In
other words, $|\lambda|=10^{-80}$  is very different from $\lambda=0$. The
extraordinarily small value of  $\lambda'_*$ was first noted in
\cite{prd2} (see also \cite{abrilphas,phi6,bectcoll,suarezchavanisprd3}).

\subsection{Noninteracting bosons}

For noninteracting bosons ($\lambda=0$), we get
\begin{equation}
\label{lf6}
m_{0}=2.92\times
10^{-22}\, {\rm eV}/c^2 \qquad ({\rm BECNI})
\end{equation}

\subsection{Repulsive self-interaction}

For bosons with a repulsive self-interaction ($\lambda>0$), 
$\lambda'_*$ determines the transition between the NI regime ($\lambda\ll
\lambda'_*$) where $m\sim m_0$ and the TF regime ($\lambda\gg \lambda'_*$) where
\begin{equation}
\label{lf7}
\frac{m}{m_0}\sim \left (\frac{\lambda}{\lambda'_*}\right )^{1/4}.
\end{equation}
When the self-interaction is repulsive, all
the equilibrium states  are stable. Therefore, in principle, all
the values of $\lambda\ge 0$ and the corresponding masses $m\ge m_0$ are
possible. In the TF regime, the $m(\lambda)$ relation (\ref{lf7}) can be written
as
\begin{equation}
\label{lf7b}
\frac{\lambda}{8\pi m^4}\sim \frac{GR^2c}{b^2\hbar^3},
\end{equation}
which is equivalent to Eq. (\ref{mas7}).
The minimum halo just determines the ratio
\begin{equation}
\label{lf7c}
\frac{\lambda}{8\pi m^4}=1.66\times 10^{-5}\, ({\rm eV}/c^2)^{-4}.
\end{equation}
Only the radius $R$ of the minimum
halo matters in this
determination. In order to determine $m$ and $\lambda$ individually, we need
another
equation. Repeating the argument from Sec. \ref{sec_rsi}, the Bullet Cluster
constraint implies that $\lambda$ must lie
in the range $0\le \lambda\le \lambda_{\rm max}$ and that the particle mass
must lie in the range $m_0\le m\le m_{\rm max}$
where\footnote{More
generally, if we define $\beta=a_s^2/m=\sigma/(4\pi m)$, we obtain $m_{\rm
max}=(\beta\pi^4\hbar^4/G^2R^4)^{1/5}$,
$(a_{s})_{\rm max}=(\beta^3\pi^2\hbar^2/GR^2)^{1/5}$ and $\lambda_{\rm
max}/8\pi=(\beta^4\pi^6\hbar c^5/G^3R^6)^{1/5}$.} 
\begin{eqnarray}
\label{lf8}
m_{\rm max}=1.10\times
10^{-3}\, {\rm eV}/c^2,\quad \lambda_{\rm
max}=6.18\times 10^{-16}\nonumber\\
({\rm BECTF})\qquad
\end{eqnarray}
Although the value of $\lambda_{\rm
max}=6.18\times 10^{-16}$ corresponding to the BECTF model may seem small, it is
much larger than $\lambda'_*=3.02\times 10^{-90}$, implying that we are deep
into
the TF regime (see the Remark above).

On the other hand, the BECt model corresponding to the transition
between the NI limit and the TF limit,  is
obtained by substituting Eq. (\ref{lf6}) into Eq. (\ref{lf7}), or 
Eq. (\ref{mas7qw}) into Eq. (\ref{lf7b}). This gives
\begin{eqnarray}
\label{lf9}
m_{\rm t}=2.92\times 10^{-22}\, {\rm eV}/c^2,\quad \lambda'_*=3.02\times
10^{-90}\nonumber\\
 ({\rm BECt})\qquad
\end{eqnarray}
This corresponds to the scales $m_0$ and $\lambda'_*$ defined
by Eqs. (\ref{mas3}) and (\ref{lf3}).

\subsection{Attractive self-interaction in terms of $\lambda$}

For bosons with an attractive self-interaction ($\lambda<0$), the relation
(\ref{lf4}) reveals the existence of a minimum value of the dimensionless
self-interaction constant
\begin{equation}
\label{lf10}
\frac{\lambda_{c}}{\lambda'_*}=-\frac{1}{4}, \qquad   {\rm at \quad
which}\qquad    
\frac{m_c}{m_0}=\frac{1}{\sqrt{2}}.
\end{equation}
It turns out that this minimum value also corresponds to the critical point
(associated with the maximum
mass $M_{\rm max}$) separating stable from unstable equilibrium states. The NI
regime corresponds to $|\lambda|\ll \lambda'_*$ and $m\sim m_0$.  The NG regime
corresponds to $|\lambda|\ll \lambda'_*$ and $m\ll m_0$ such that
\begin{equation}
\label{lf11}
\frac{m}{m_0}\sim \left (\frac{|\lambda|}{\lambda'_*}\right )^{1/2}.
\end{equation}
In the NG regime, the relation
(\ref{lf11}) between $m$ and $\lambda$ can
be written as
\begin{equation}
\label{lf11b}
\frac{|\lambda|}{8\pi m^2}=\frac{a}{b^2}\frac{Rc}{M\hbar}=5.12\times
10^{-49}\, ({\rm
eV/c^2})^{-2},
\end{equation}
which is equivalent to Eq. (\ref{pa9}). The equilibrium states with $m<m_c$ are
unstable (they
correspond to 
configurations with $R<R_*$) so that only the equilibrium states with $m>m_c$
are stable (they correspond to configurations with $R>R_*$). Therefore, in the
attractive case, the scattering length of the DM boson must lie in the range
$\lambda_c<\lambda<0$ and its mass must lie in the range $m_c<m<m_0$,
with
\begin{eqnarray}
m_{\rm c}=2.19\times 10^{-22}\, {\rm eV}/c^2,\quad
\lambda_{c}=-3.07\times 10^{-91}\nonumber\\
 ({\rm BECcrit})\qquad
\end{eqnarray}
There is no equilibrium state with $\lambda<\lambda_{c}$.
Finally, using the constraints from particle physics and cosmology (see
Sec. \ref{sec_cpp}) we find
\begin{eqnarray}
m_{\rm th}=2.92\times 10^{-22}\, {\rm eV}/c^2,\quad \lambda_{\rm
th}=-1.18\times 10^{-96}\nonumber\\
 ({\rm BECth})\qquad
\end{eqnarray}

\subsection{Attractive self-interaction in terms of $f$}

The decay constant is defined by Eq. (\ref{alf2}). This relation may be
rewritten as
\begin{equation}
\frac{f}{f'_*}=\left (\frac{m}{m_0}\right )^{1/2}\left
(\frac{a'_*}{|a_s|}\right )^{1/2},
\label{lf13}
\end{equation}
where we have introduced the scales from Eqs. (\ref{mas3}) and (\ref{mas4}), and
the new scale
\begin{equation}
f'_*=\left (\frac{b^2}{a}\frac{\hbar c^3M}{32\pi R}\right )^{1/2}.
\label{lf14}
\end{equation}

\begin{figure}[!h]
\begin{center}
\includegraphics[clip,scale=0.3]{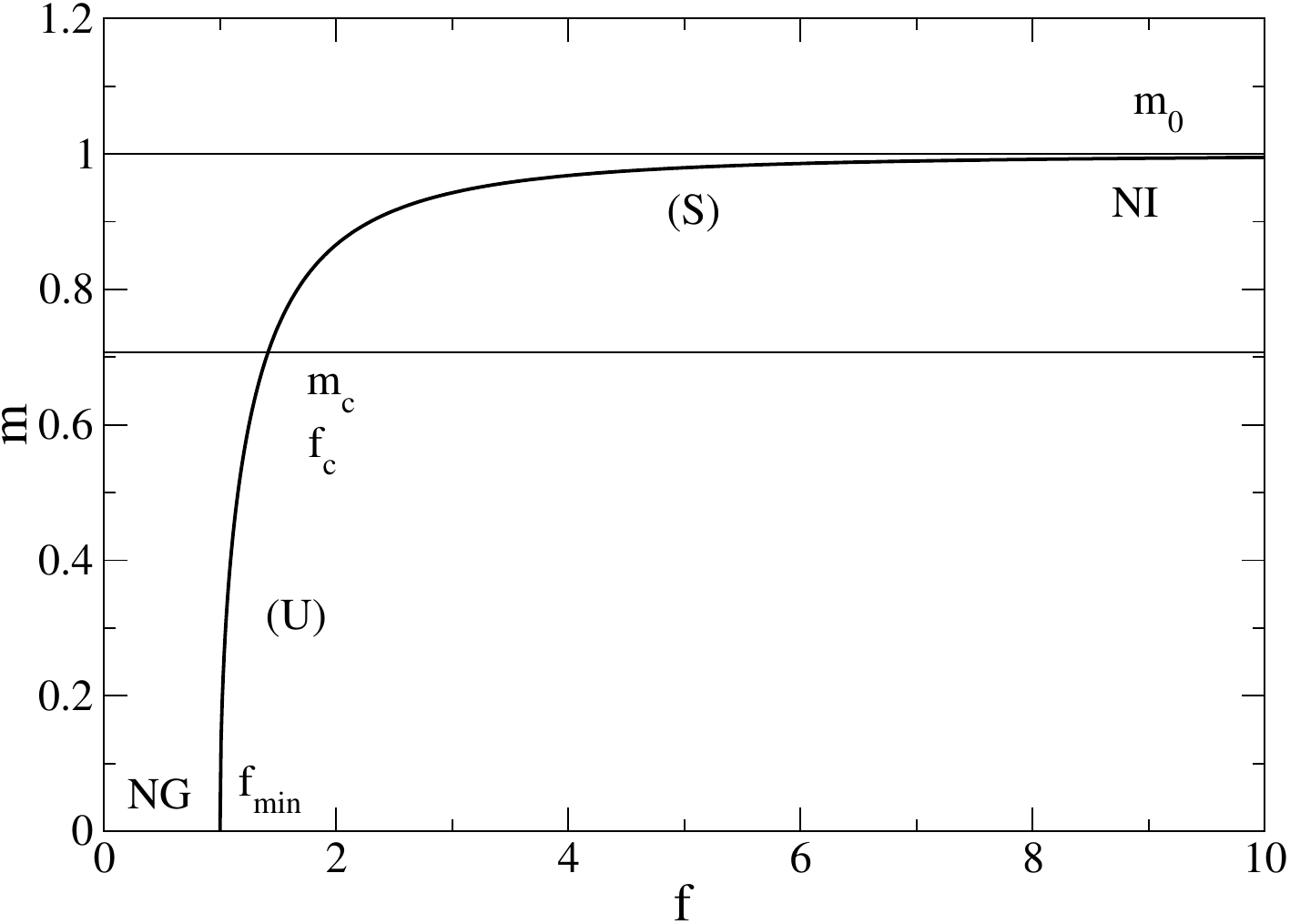}
\caption{Mass $m$ of the DM particle as a function of its decay
constant $f$ in order to match the characteristics of the minimum halo. The
mass is normalized by
$m_0$ and the decay constant by $f'_*$. The stable part of
the curve starts at the critical minimum halo point ($f_c,m_{c}$). It differs
from the absolute minimum value $f_{\rm min}$ of the decay
constant.}
\label{fm}
\end{center}
\end{figure}

Using Eqs. (\ref{asm1}) and (\ref{alf2}), the relation between the particle mass
and the decay constant is given by
\begin{equation}
\frac{1}{f^2}=\frac{32\pi a R}{b^2\hbar c^3 M}\left
(1-\frac{GMm^2R}{a\hbar^2}\right ).
\end{equation}
Alternatively, using Eqs. (\ref{mas2}) and (\ref{lf13}), we obtain in
dimensionless form
\begin{equation}
\label{lf15}
\frac{m}{m_0}=\sqrt{1-\left (\frac{f'_*}{f}\right )^2}.
\end{equation}
The curve $m(f)$ is plotted in Fig. \ref{fm}.
Taking
$a=11.1$ and $b=5.5$ (see Sec. \ref{sec_paab}) adapted to bosons with an
attractive
self-interaction, we get $m_0=3.08\times 10^{-22}\,
{\rm eV}/c^2$ and $f'_*=1.39\times 10^{14}\, {\rm GeV}$.

The relation from Eq. (\ref{lf15}) reveals 
the existence of a minimum decay constant $f_{\rm min}=f'_*=1.39\times
10^{14}\, {\rm GeV}$ at which $m=0$.\footnote{This value arises from the fact
that both the mass-radius relation  in the NG regime from Eq. (\ref{pa9}) and
the decay constant from Eq. (\ref{alf2}) present a scaling in $m/|a_s|$.}
However, this minimum scattering length $f_{\rm
min}$ does {\it not} correspond to the critical point (associated with the
maximum mass $M_{\rm max}$) separating stable from unstable equilibrium states.
This latter is located at
\begin{equation}
\label{mas13}
\frac{f_{c}}{f'_*}=\sqrt{2}, \qquad     
\frac{m_c}{m_0}=\frac{1}{\sqrt{2}}.
\end{equation}
The NI regime corresponds to
$f\gg f_{\rm min}$ and $m\sim m_0$. The NG regime
corresponds to $f\sim f_{\rm min}$ and $m\ll m_0$. 

The equilibrium states with $m<m_c$ are unstable (they
correspond to 
configurations with $R<R_*$) so that only the equilibrium states with $m>m_c$
are stable (they correspond to configurations with $R>R_*$). Therefore, the
decay constant of the DM boson must lie in the range
$f>f_c$ and its mass must lie in the range $m_c<m<m_0$,
with\footnote{We note that when the self-interaction
is attractive $m$ almost does not change (it is of the order of the mass of
noninteracting bosons) while $f$ can change by several orders of
magnitude.}
\begin{eqnarray}
m_{\rm c}=2.19\times 10^{-22}\, {\rm eV}/c^2,\quad
f_{c}=1.97\times 10^{14}\, {\rm GeV}\nonumber\\
 ({\rm BECcrit})\qquad
\end{eqnarray}
There is no equilibrium state with $f<f_{\rm min}$. On the other hand, the
equilibrium states
with $f_{\rm min}\le f<f_c$ are unstable. Using the constraints from particle
physics and cosmology (see
Sec. \ref{sec_cpp}) we find
\begin{eqnarray}
m_{\rm th}=2.92\times 10^{-22}\, {\rm eV}/c^2,\quad f_{\rm
th}=1.34\times 10^{17}\, {\rm GeV}\nonumber\\
 ({\rm BECth})\qquad
\end{eqnarray}
We note that $f_{\rm th}<M_P c^2=1.22\times 10^{19}\, {\rm GeV}$. 

\section{Pulsation of self-gravitating BECs}
\label{sec_pul}

\subsection{General case}

The square pulsation of the standard self-gravitating BEC is 
approximately 
given by (see Appendix
\ref{sec_mra})
\begin{equation}
\omega^2=\frac{6\Theta_Q+12U+2W}{I}.
\label{pul1}
\end{equation}
This equation can be written in different forms by using the virial theorem
$2\Theta_Q+3U+W=0$ (see Sec. \ref{sec_mad}).

With the $f$-ansatz (see Appendix
\ref{sec_mra}), the square pulsation is given by
\begin{equation}
\omega^2=\frac{6\sigma}{\alpha}\frac{\hbar^2}{m^2R^4}-\frac{2\nu}{\alpha}
\frac{GM}{R^3}+\frac{24\pi\zeta}{\alpha}\frac{a_s\hbar^2 M}{m^3R^5}.
\label{pul2}
\end{equation}
On the other hand, the mass-radius relation writes
\begin{eqnarray}
-2\sigma\frac{\hbar^2M}{m^2R^3}+\nu \frac{GM^2}{R^2}-6\pi\zeta
\frac{a_s\hbar^2M^2}{m^3R^{4}}=0
\label{pul3}
\end{eqnarray}
or, equivalently,
\begin{equation}
M=\frac{\frac{2\sigma}{\nu}\frac{\hbar^2}{Gm^2R}}{1-\frac{6\pi\zeta}{\nu}\frac{
a_s\hbar^2 } { Gm^3R^2 } }.
\label{mrga}
\end{equation}
If we use a Gaussian ansatz,
the values 
of the coefficients are $\alpha_{\rm G}=3/2$,  $\sigma_{\rm G}=3/4$, $\zeta_{\rm
G}=1/(2\pi)^{3/2}$, and  $\nu_{\rm G}=1/\sqrt{2\pi}$ \cite{prd1}.
Furthermore, the relation between the radius $R$ and the radius $R_{99}$
containing $99\%$ of the mass is $R_{99}=2.38167\, R$ \cite{prd1}. The square
pulsation is
plotted as a function of the BEC radius $R$ in
\cite{prd1}.

\subsection{Noninteracting case}

For noninteracting bosons ($a_s=0$), the square pulsation from Eq. (\ref{pul1}) reduces to
\begin{equation}
\omega^2=\frac{6\Theta_Q+2W}{I}=\frac{2\Theta_Q}{I}=-\frac{W}{I},
\label{pul4}
\end{equation}
where we have used the virial theorem $2\Theta_Q+W=0$ to get the last two
equalities.

With the $f$-ansatz, the
square pulsation is given by
\begin{equation}
\omega^2=\frac{6\sigma}{\alpha}\frac{\hbar^2}{m^2R^4}-\frac{2\nu}{\alpha}
\frac{GM}{R^3}.
\label{pul5}
\end{equation}
On the other hand,  the mass-radius relation writes
\begin{eqnarray}
R=\frac{2\sigma}{\nu}\frac{\hbar^2}{GMm^2}.
\label{pul6}
\end{eqnarray}
Combining these two relations we obtain
\begin{eqnarray}
\label{pul7}
\omega^2=\frac{\nu}{\alpha}\frac{GM}{R^3}=\frac{2\sigma}{\alpha}\frac{\hbar^2}{m^2R^4}=
\frac{\nu^4}{8\alpha\sigma^3}\frac{G^4M^4m^6}{\hbar^6}.
\end{eqnarray}
If we use a Gaussian ansatz, the prefactors are $0.266$, $1$ and $5.00\times
10^{-3}$. The
first relation of Eq. (\ref{pul7}) shows that the pulsation period
$T=2\pi/\omega$ is about $12.2$ dynamical times $t_d=(R^3/GM)^{1/2}$. For the
minimum halo with
$M=10^8\, M_{\odot}$ and $R=(1/2.38167)\, {\rm kpc}$, we get $t_d=12.8\, {\rm
Myrs}$ and $T=156\, {\rm Myrs}$. 

\subsection{TF limit}

For bosons with a repulsive self-interaction ($a_s>0$) in the TF limit  ($\hbar=0$), the square pulsation from Eq. (\ref{pul1}) reduces to
\begin{equation}
\omega^2=\frac{12U+2W}{I}=\frac{6U}{I}=-\frac{2W}{I},
\label{pul8}
\end{equation}
where we have used the virial theorem $3U+W=0$ to get the last two equalities.

With the $f$-ansatz, the square pulsation is given by
\begin{equation}
\omega^2=-\frac{2\nu}{\alpha}
\frac{GM}{R^3}+\frac{24\pi\zeta}{\alpha}\frac{a_s\hbar^2 M}{m^3R^5}.
\label{pul9}
\end{equation}
On the other hand, the radius of the BEC is
\begin{eqnarray}
R=\left (\frac{6\pi\zeta}{\nu}\right )^{1/2}\left (\frac{a_s\hbar^2}{Gm^3}\right )^{1/2}.
\label{pul10}
\end{eqnarray}
Combining these two relations we obtain
\begin{eqnarray}
\label{pul11}
\omega^2=\frac{2\nu}{\alpha}\frac{GM}{R^3}
=\frac{2\nu^{5/2}}{\alpha(6\pi\zeta)^{3/2}}\frac{G^{5/2}Mm^{9/2}}{a_s^{3/2}\hbar^3}.
\end{eqnarray}
If we use a Gaussian ansatz, the prefactors are $0.532$ and  $0.102$.
The pulsation period $T=2\pi/\omega$
is about $8.62$ dynamical times $t_d=(R^3/GM)^{1/2}$. For the
minimum halo with
$M=10^8\, M_{\odot}$ and $R=(1/2.38167)\, {\rm kpc}$, we get $t_d=12.8\, {\rm
Myrs}$ and $T=111\, {\rm Myrs}$.

{\it Remark:} In the TF approximation, the density profile of the
BECDM halo is known analytically. In that case, one can obtain the exact
expression of the pulsation (see \cite{prd1} and Appendices H and I of
\cite{epjpbh}).

\subsection{Maximum mass and maximum pulsation}

For bosons with an attractive self-interaction ($a_s<0$), the pulsation
vanishes at the maximum mass \cite{prd1}:
\begin{eqnarray}
\label{pul12}
\omega=0\quad {\rm at}\quad M=M_{\rm max}.
\end{eqnarray}
With the $f$-ansatz:
\begin{eqnarray}
\label{pul12a}
M_{\rm max}=\left
(\frac{\sigma^2}{6\pi\zeta\nu}\right )^{1/2}\frac{\hbar}{\sqrt{Gm|a_s|}},
\end{eqnarray}
\begin{eqnarray}
\label{pul12b}
R_*=\left (\frac{6\pi\zeta}{\nu}\right )^{1/2}\left
(\frac{|a_s|\hbar^2}{Gm^3}\right )^{1/2}.
\end{eqnarray}
On the other hand, there is a maximum pulsation at some ${\tilde M}<M_{\rm
max}$ \cite{prd1}. With
the $f$-ansatz:
\begin{eqnarray}
\label{pul12c}
\omega_{\rm max}=0.4246\, \frac{\nu}{6\pi\zeta}\sqrt{\frac{\sigma}{\alpha}}
\frac{Gm^2}{|a_s|\hbar}
\end{eqnarray}
at
\begin{eqnarray}
\label{pul12d}
{\tilde M}=0.9717\, M_{\rm max},\qquad {\tilde R}=1.272\, R_*.
\end{eqnarray}

\subsection{Nongravitational case}

For bosons with an attractive self-interaction ($a_s<0$) in the NG limit  ($G=0$), the square pulsation from Eq. (\ref{pul1}) reduces to
\begin{equation}
\omega^2=\frac{6\Theta_Q+12U}{I}=-\frac{2\Theta_Q}{I}=\frac{3U}{I},
\label{pul13}
\end{equation}
where we have used the virial theorem $2\Theta_Q+3U=0$ to get the last two
equalities.

With the $f$-ansatz, the square pulsation is given by
\begin{equation}
\omega^2=\frac{6\sigma}{\alpha}\frac{\hbar^2}{m^2R^4}+\frac{24\pi\zeta}{\alpha}
\frac{a_s\hbar^2 M}{m^3R^5}.
\label{pul14}
\end{equation}
On the other hand, the mass-radius relation writes
\begin{eqnarray}
M=\frac{\sigma}{3\pi\zeta}\frac{m R}{|a_s|}.
\label{pul15}
\end{eqnarray}
Combining these two relations we obtain
\begin{eqnarray}
\label{pul16}
\omega^2=-\frac{2\sigma}{\alpha}\frac{\hbar^2}{m^2R^4}=
-\frac{2\sigma^5}{\alpha(3\pi \zeta)^4}\frac{m^2\hbar^2}{M^4a_s^4}.
\end{eqnarray}
If we use a Gaussian ansatz, the prefactors are $1$ and $2.47$. Note that these
configurations are
unstable ($\omega^2<0$) so they should not be observed in practice.

\section{Similarity between the mass-radius relation obtained 
from the $f$-ansatz and from the Jeans instability study}
\label{sec_sim}

In this Appendix, we show at a general level that the mass-radius
relation $M_J(R_J)$ obtained from the Jeans instability study is similar to the
mass-radius relation $M(R)$ of DM halos in their ground state obtained from the
minimization of the energy at fixed mass using an $f$-ansatz. This similarity
was
first observed in Ref. \cite{prd1} in a special case (for a $|\psi|^4$ potential
of interaction and for a Gaussian ansatz) and it is here generalized to an
arbitrary potential of interaction $V(|\psi|^2)$ and an arbitrary ansatz.

\subsection{Mass-radius relation from the Jeans instability study}
\label{sec_mrj}

In this section, we consider the formation of structures 
in the linear regime from the Jeans instability study (see Sec. \ref{sec_j}).
The Jeans wavenumber is determined by the
equation \cite{prd1}
\begin{eqnarray}
\frac{\hbar^2k_J^4}{4m^2}+c_s^2k_J^2-4\pi G\rho=0,
\label{mrj1}
\end{eqnarray}
where $c_s^2$ is the square of the speed of sound. For a barotropic
fluid, this is a function of the density given by Eq. (\ref{mad8b}). Eq.
(\ref{mrj1}) is a second degree equation for $k_J^2$ whose physical solution is
\begin{eqnarray}
k_J^2=\frac{2m^2}{\hbar^2}\left\lbrack
-c_s^2(\rho)+\sqrt{c_s^4(\rho)+\frac{4\pi G\rho\hbar^2}{m^2}}\right\rbrack.
\label{mrj2}
\end{eqnarray}
If we define the Jeans radius and the Jeans mass by
\begin{eqnarray}
R_J=\frac{\lambda_J}{2}=\frac{\pi}{k_J},\qquad M_J=\frac{4}{3}\pi\rho R_J^3,
\label{mrj3}
\end{eqnarray}
we obtain
\begin{eqnarray}
R_J(\rho)=\frac{\frac{\pi\hbar}{\sqrt{2}m}}{\left\lbrack
-c_s^2(\rho)+\sqrt{c_s^4(\rho)+\frac{4\pi
G\rho\hbar^2}{m^2}}\right\rbrack^{1/2}},
\label{mrj4}
\end{eqnarray}
\begin{eqnarray}
M_J(\rho)=\frac{4}{3}\pi\rho R_J(\rho)^3.
\label{mrj5}
\end{eqnarray}
These equations determine the Jeans scales $R_J(\rho)$ and $M_J(\rho)$ as a
function of the density. They also determine the
Jeans mass-radius relation $M_J(R_J)$ in parametric form with parameter
$\rho$.

{\it Remark:} In the nongravitational case, there is a hydrodynamic instability
when $c_s^2<0$ \cite{khlopov,prd1}.\footnote{This
hydrodynamic instability is also
called a tachyonic instability.} In
that case, the ``Jeans''
wavenumber
is determined by the
equation
\begin{eqnarray}
\frac{\hbar^2k_J^2}{4m^2}+c_s^2=0,
\label{mrj6}
\end{eqnarray}
and the parametric equations (\ref{mrj4}) and (\ref{mrj5}) reduce to
\begin{eqnarray}
R_J(\rho)=\frac{\pi\hbar}{2m\sqrt{-c_s^2(\rho)}},
\label{mrj7}
\end{eqnarray}
\begin{eqnarray}
M_J(\rho)=\frac{4}{3}\pi\rho R_J(\rho)^3.
\label{mrj8}
\end{eqnarray}

\subsection{Mass-radius relation from the  $f$-ansatz}
\label{sec_mra}

In this section, we consider BECDM halos that appear in the nonlinear 
regime  of structure formation (see Sec. \ref{sec_dm}). Stable DM halos
correspond to minima of energy $E_{\rm min}$ at fixed mass $M$. We can obtain an
approximate analytical form of the mass-radius relation by making an ansatz for
the wave function. A Gaussian ansatz was considered in \cite{prd1}. To be as
general as possible, we consider here an ansatz of the form (that we call
$f$-ansatz)
\begin{eqnarray}
\rho({\bf r},t)=\frac{M}{R(t)^3}f\left\lbrack \frac{\bf r}{R(t)}\right \rbrack,
\label{mra1}
\end{eqnarray}
where $f({\bf x})$ is an arbitrary function. We impose $\int f({\bf x})\,
d{\bf x}=1$ to satisfy the normalization condition (or the conservation of
mass). We also assume
\begin{eqnarray}
S({\bf r},t)=\frac{1}{2}m H(t)r^2\quad \Rightarrow \quad {\bf u}({\bf
r},t)=H(t){\bf r},
\label{mra2}
\end{eqnarray}
so that the velocity field is proportional to the radial distance. It
can be shown  (see Appendix J of \cite{ggpp}) that Eqs. (\ref{mra1}) and
(\ref{mra2}) yield an exact solution of the continuity equation (\ref{mad3})
provided that
\begin{eqnarray}
H=\frac{\dot R}{R}.
\label{mra3}
\end{eqnarray}
This function is similar to the Hubble parameter in cosmology. On the other
hand, the gravitational potential can be determined from the Poisson equation
(\ref{mad5}). Using Eq. (\ref{mra1}) we obtain
\begin{eqnarray}
\Phi({\bf r},t)=\frac{GM}{R(t)}g\left\lbrack \frac{\bf r}{R(t)}\right \rbrack,
\label{mra4}
\end{eqnarray}
where $g({\bf x})$ is the solution of
\begin{eqnarray}
\Delta g=4\pi f({\bf x}).
\label{mra5}
\end{eqnarray}
We can now use the ansatz (\ref{mra1})-(\ref{mra5}) to determine 
the different functionals that appear in the energy from Eq. (\ref{mad12}). We
find
\begin{equation}
\Theta_c=\frac{1}{2}\alpha M\left (\frac{dR}{dt}\right )^2\quad {\rm with}\quad \alpha=\int f({\bf x}) x^2\, d{\bf x},
\label{mra6}
\end{equation}
\begin{eqnarray}
\Theta_Q=\sigma\frac{\hbar^2M}{m^2R^2}\quad {\rm with}\quad \sigma=\frac{1}{8}\int \frac{(\nabla f)^2}{f}\, d{\bf x},
\label{mra7}
\end{eqnarray}
\begin{eqnarray}
U=\frac{\zeta}{\gamma-1} \frac{KM^{\gamma}}{R^{3(\gamma-1)}}\quad {\rm with}\quad \zeta=\int f^{\gamma}({\bf x})\, d{\bf x},
\label{mra9}
\end{eqnarray}
and
\begin{equation}
W=-\nu \frac{GM^2}{R}\quad {\rm with}\quad \nu=-\frac{1}{2}\int f({\bf x})g({\bf x})\, d{\bf x}.
\label{mra8}
\end{equation}
The expression (\ref{mra9}) of the internal energy $U$ is valid for a 
power-law potential associated with a polytropic equation of state (we will
see later how to generalize the formalism to an arbitrary potential of
interaction or an arbitrary equation of state). The moment of
inertia is
\begin{eqnarray}
I=\alpha MR^2.
\label{mra8b}
\end{eqnarray}
If we use a Gaussian ansatz $f({\bf
x})=\frac{1}{\pi^{3/2}}e^{-x^2}$, the values 
of the coefficients are $\alpha_{\rm G}=3/2$,  $\sigma_{\rm G}=3/4$, $\zeta_{\rm
G}=1/(\gamma\pi^{\gamma-1})^{3/2}$, and  $\nu_{\rm G}=1/\sqrt{2\pi}$
\cite{prd1}.

With the ansatz from Eqs. (\ref{mra1}) and (\ref{mra2}) the total energy can be
written as
\begin{eqnarray}
E_{\rm tot}=\frac{1}{2}\alpha M \left (\frac{d R}{dt}\right )^2+V(R)
\label{mra10}
\end{eqnarray}
with
\begin{eqnarray}
V(R)=\sigma\frac{\hbar^2M}{m^2R^2}-\nu \frac{GM^2}{R}+\frac{\zeta}{\gamma-1} \frac{KM^{\gamma}}{R^{3(\gamma-1)}}.
\label{mra11}
\end{eqnarray}
We have separated the classical kinetic energy $\Theta_c$ from the potential
energy $V=\Theta_Q+U+W$. From the conservation of energy, $\dot E_{\rm
tot}=0$, we obtain
\begin{eqnarray}
\alpha M \frac{d^2R}{dt^2}=-V'(R).
\label{mra12}
\end{eqnarray}
This is similar to the equation of motion of a fictive particle of mass $\alpha
M$ and position $R$ moving in a potential $V(R)$.
At equilibrium, the condition $V'(R)=0$ (extremum of energy) gives the
mass-radius relation
\begin{eqnarray}
-2\sigma\frac{\hbar^2M}{m^2R^3}+\nu \frac{GM^2}{R^2}-3\zeta \frac{KM^{\gamma}}{R^{3(\gamma-1)+1}}=0.
\label{mra13}
\end{eqnarray}
For the standard BEC, we get Eq. (\ref{pul3}). The foregoing  equations
may also be obtained from the virial theorem \cite{prd1,ggpp} or from the
Lagrange equations \cite{bectcoll,ggpp}.

The pulsation of the self-gravitating BEC is given by \cite{prd1,ggpp}
\begin{eqnarray}
\omega^2=\frac{1}{\alpha M}V''(R).
\label{po1}
\end{eqnarray}
The BEC is stable provides that $\omega^2>0$ which is equivalent by Eq.
(\ref{po1}) to the requirement that the equilibrium state is a minimum of
energy. Using Eq.
(\ref{mra11}) we obtain
\begin{equation}
\omega^2=\frac{6\sigma}{\alpha}\frac{\hbar^2}{m^2R^4}-\frac{2\nu}{\alpha} \frac{GM}{R^3}+\lbrack 3(\gamma-1)+1\rbrack\frac{3\zeta}{\alpha} \frac{KM^{\gamma-1}}{R^{3(\gamma-1)+2}}.
\label{po2}
\end{equation}
Using Eqs. (\ref{mra7})-(\ref{mra8b}), the pulsation can also be written 
in terms of the BEC functionals as
\begin{equation}
\omega^2=\frac{6\Theta_Q+\lbrack 3(\gamma-1)+1\rbrack 3(\gamma-1)U+2W}{I}.
\label{po3}
\end{equation}
For the usual BEC, we obtain Eqs. (\ref{pul1}) and (\ref{pul2}).

In order to compute the internal energy $U$ for a general 
self-interaction potential we consider an ansatz based on a  uniform (top-hat)
density
\begin{eqnarray}
\rho({\bf r},t)=\frac{3M}{4\pi R(t)^3}\theta(|{\bf r}|-R(t)),
\label{mra14}
\end{eqnarray}
where $\theta$ is the Heaviside function 
($\theta(x)=1$ if $x<0$ and $\theta(x)=0$ if $x>0$). In that case, the internal
energy is given by
\begin{eqnarray}
U=V\left (\frac{3M}{4\pi R^3}\right )\frac{4}{3}\pi R^3.
\label{mra15}
\end{eqnarray}
We then find that
\begin{eqnarray}
U'(R)=\frac{d}{dR}\left \lbrack \frac{V(\rho)}{\rho} M\right\rbrack=M\frac{d}{d\rho}\left \lbrack \frac{V(\rho)}{\rho}\right\rbrack \frac{d\rho}{dR}\nonumber\\
=-\frac{9M^2}{4\pi R^4}\frac{d}{d\rho}\left \lbrack \frac{V(\rho)}{\rho}\right\rbrack.
\label{mra16}
\end{eqnarray}
Using Eq. (\ref{mad7}) which corresponds 
to the first principle of thermodynamics (see Appendix \ref{sec_ti})
\begin{eqnarray}
\left\lbrack \frac{V(\rho)}{\rho}\right\rbrack'=\frac{P(\rho)}{\rho^2}  \quad  \Leftrightarrow  \quad d\left (\frac{V}{\rho}\right )=-Pd\left(\frac{1}{\rho}\right ),
\label{mra17}
\end{eqnarray}
we obtain
\begin{equation}
U'(R)=-P\left (\frac{3M}{4\pi R^3}\right )4\pi R^2 \quad \Leftrightarrow \quad dU=-P\, d{\cal V},
\label{mra18}
\end{equation}
where ${\cal V}=(4/3)\pi R^3$ denotes 
the volume of the BEC.  For a power-law
self-interaction potential, we recover the expression of $U$ from Eq.
(\ref{mra9}) with a coefficient $\zeta_{\rm C}=(3/4\pi)^{\gamma-1}$. On the
other hand, the coefficients entering in the expressions of $\Theta_c$ and $W$
from Eqs. (\ref{mra6}) and (\ref{mra8}) are $\alpha_{\rm C}=3/5$ and $\nu_{\rm
C}=3/5$. Unfortunately, we cannot use the constant density ansatz to determine
the quantum kinetic energy $\Theta_c$ since it is produced by the gradient of
the density which is infinite for the top-hat profile.

For an arbitrary self-interaction potential, 
we can write the total energy as in Eq. (\ref{mra10}) with an approximate 
potential energy given by
\begin{equation}
V(R)=\sigma\frac{\hbar^2M}{m^2R^2}-\nu\frac{GM^2}{R}+\chi V\left
(\frac{3M}{4\pi R^3}\right )\frac{4}{3}\pi R^3,
\label{mra19}
\end{equation}
where $\chi$ is a tunable coefficient. For a power-law self-interaction
potential, we exactly recover Eq. (\ref{mra11}) with $\chi=\zeta \left
({4\pi}/{3}\right )^{\gamma-1}$. For an  arbitrary self-interaction
potential, using Eqs. (\ref{mra18}) and (\ref{mra19}), we get
\begin{equation}
V'(R)=-2\sigma\frac{\hbar^2M}{m^2R^3}+\nu\frac{GM^2}{R^2}-\chi P\left
(\frac{3M}{4\pi R^3}\right ) 4\pi R^2.
\label{mra20}
\end{equation}
The condition of equilibrium $V'(R)=0$ then yields the mass-radius relation
under the form
\begin{eqnarray}
-2\sigma\frac{\hbar^2M}{m^2R^3}+\nu\frac{GM^2}{R^2}-\chi P\left
(\frac{3M}{4\pi R^3}\right ) 4\pi R^2=0.
\label{mra21}
\end{eqnarray}
If we work with the variables $M$ and $R$ it is usually difficult to solve this
equation explicitly in the general case. However, if we make the change of
variables
\begin{eqnarray}
R=\frac{\pi}{k},\qquad M=\frac{4}{3}\pi\rho R^3,
\label{mra22}
\end{eqnarray}
inspired by Eq. (\ref{mrj3}), we get
\begin{eqnarray}
\frac{2\sigma}{\pi^2}\frac{\hbar^2k^4}{m^2}+3\chi
\frac{P(\rho)}{\rho} k^2- \frac{4}{3}\pi^3\nu G\rho=0.
\label{mra23}
\end{eqnarray}
Remarkably, this equation is similar to the Jeans equation (\ref{mrj1}). 
Therefore, it can be solved easily (this is just a second degree equation for
$k^2$) and the mass-radius relation $M(R)$  can be obtained in parametric form
as in Appendix \ref{sec_mrj}. We get
\begin{equation}
R(\rho)=\frac{\frac{2\sqrt{\sigma}\hbar}{m}}{\left\lbrack
-3\chi\frac{P(\rho)}{\rho}+\sqrt{9\chi^2\frac{P(\rho)^2}{\rho^2}+\frac{32\pi}{3}
\sigma\nu\frac{G\rho\hbar^2}{m^2}} \right\rbrack^{1/2}},
\label{mra24}
\end{equation}
\begin{eqnarray}
M(\rho)=\frac{4}{3}\pi\rho R(\rho)^3.
\label{mra25}
\end{eqnarray}
This shows in full generality that the Jeans mass-radius relation $M_J(R_J)$
valid in the linear regime of structure formation is formally similar to the
mass-radius relation $M(R)$ of DM halos valid in the nonlinear regime of
structure formation. Apart from the precise value of the prefactors, we see
that the difference with the Jeans study is that the pressure derivative 
$P'(\rho)$ (appearing in $c_s^2$) is replaced by the ratio
$P(\rho)/\rho$.  For a polytropic equation of state, the dependence in
the density is the same, i.e., $\rho^{\gamma-1}$ but the prefactor is different.

\section{Thermodynamical identities for cold barotropic gases}
\label{sec_ti}

In this Appendix, we regroup useful thermodynamical identities valid for cold
barotropic gases. 

The first principle of
thermodynamics can be
written under a local form as 
\begin{equation}
\label{ti1}
d\left (\frac{u}{\rho}\right )=-Pd\left
(\frac{1}{\rho}\right )+Td\left (\frac{s}{\rho}\right ),
\end{equation}
where $u$ is the density of internal energy, $\rho$ is the mass density, $P$ is
the pressure, $T$ is the temperature and $s$ is the entropy density. For cold
gases ($T=0$), Eq. (\ref{ti1}) reduces to
\begin{equation}
\label{ti2}
d\left (\frac{u}{\rho}\right )=-Pd\left
(\frac{1}{\rho}\right )=\frac{P}{\rho^2}d\rho.
\end{equation}
If we introduce the enthalpy density
\begin{equation}
h=\frac{P+u}{\rho},
\label{ti3}
\end{equation} 
we obtain the relations
\begin{equation}
\label{ti4}
du=h\, d\rho \qquad {\rm and}\qquad
dh=\frac{dP}{\rho}.
\end{equation}
We also recall that, according to the Gibbs-Duhem relation for a cold gas
($T=0$), the local chemical potential coincides with the enthalpy ($h=\mu_{\rm
loc}/m$).

\subsection{General barotropic equation of state}

For a general barotropic equation of state of the form
$P=P(\rho)$, the foregoing relations lead to the identities
\begin{equation}
\label{ti5}
\left (\frac{u}{\rho}\right )'=\frac{P(\rho)}{\rho^2},
\end{equation}
\begin{equation}
\label{ti6}
h(\rho)=u'(\rho),
\end{equation}
\begin{equation}
\label{ti7}
h'(\rho)=\frac{P'(\rho)}{\rho},
\end{equation}
\begin{equation}
\label{ti8}
P(\rho)=\rho h(\rho)-u(\rho)=\rho u'(\rho)-u(\rho)=\rho^2 \left
(\frac{u}{\rho}\right )',
\end{equation}
\begin{equation}
\label{ti9}
P'(\rho)=\rho u''(\rho).
\end{equation}

The first principle of thermodynamics for a barotropic gas at $T=0$ [see Eq.
(\ref{ti2})]
provides a general relation between the density of internal energy $u(\rho)$ and
the
pressure $P(\rho)$. If we know the energy density $u=u(\rho)$, we can
obtain the pressure by
\begin{equation}
\label{ti10}
P=-\frac{d(u/\rho)}{d(1/\rho)}=\rho\frac{du}{d\rho}-u.
\end{equation}
Inversely, if we know the equation of state $P=P(\rho)$, we can obtain the
energy density by
\begin{equation}
\label{ti11}
u(\rho)=\rho\int^{\rho} \frac{P(\rho')}{{\rho'}^2}\, d\rho'.
\end{equation}

{\it Remark:} Comparing Eqs. (\ref{mad7}) and (\ref{mad8}) with Eqs. (\ref{ti8})
and (\ref{ti9}), we see that the potential $V(\rho)$ that occurs in the GP
equation (\ref{gpp1}) represents the density of
internal energy:
\begin{equation}
\label{ti11b}
u(\rho)=V(\rho).
\end{equation}
This justifies the expression of the internal energy in Eqs.
(\ref{gpp7}) and (\ref{mad12c}).

\subsection{Polytropic equation of state}

For a polytropic equation of state of the form
$P=K\rho^{\gamma}$ with $\gamma=1+1/n$, the density of internal energy [see
Eq. (\ref{ti11})] is explicitly given by
\begin{equation}
\label{ti12}
u=\frac{K}{\gamma-1}\rho^{\gamma}=\frac{P}{\gamma-1}=nP=nK\rho^{1+1/n},
\end{equation}
where we have set the constant of integration to zero. For the standard BEC
corresponding to $\gamma=2$ [see Eq. (\ref{gpp3})], we
have
\begin{equation}
\label{ti13}
u=P=\frac{2\pi a_s\hbar^2}{m^3}\rho^{2}\quad\Rightarrow\quad U=\frac{2\pi
a_s\hbar^2}{m^3}\int \rho^{2}\, d{\bf r}.
\end{equation}

\section{Derivation of the GPP equations in an expanding Universe}
\label{sec_eu}

In this Appendix, proceeding as in
Ref. \cite{aacosmo},  we derive the GPP equations in an expanding universe
starting from their expression in the inertial frame. Alternative derivations,
starting directly from the KGE equations written with the conformal Newtonian
gauge which is a perturbed form of the
Friedmann-Lema\^itre-Robertson-Walker (FLRW) metric accounting for the
expansion of the Universe, and
taking the nonrelativistic
limit $c\rightarrow +\infty$, can be found in Refs.
\cite{abrilph,playa,chavmatos,phi6}.

\subsection{Homogeneous solution}

In the inertial frame, the GPP equations are given by Eqs. (\ref{gpp1}) and
(\ref{gpp2}). The corresponding hydrodynamic equations, obtained from
the Madelung \cite{madelung} transformation, are given by
Eqs. (\ref{mad3})-(\ref{mad5}). Let us first show that these equations admit a
time-dependent spatially homogeneous solution describing an expanding
universe in a Newtonian cosmology.

We consider a
spatially homogeneous solution of Eqs. (\ref{mad3})-(\ref{mad5}) of the form
\begin{equation}
\label{da10a}
\rho_b({\bf r},t)=\rho_b(t),\qquad S_b({\bf r},t)=\frac{1}{2} H(t) m
r^2+S_0(t),
\end{equation}
\begin{equation}
\label{da10b}
{\bf u}_b({\bf r},t)=H(t){\bf r},\qquad \Phi_b({\bf
r},t)=\frac{2}{3}\pi G\rho_b(t) r^2,
\end{equation}
where $a(t)$ is the scale factor and $H=\dot a/a$ is the Hubble constant
(actually a function of time). The velocity is assumed to be proportional to the
distance (Hubble's law) and the gravitational potential has
been
determined from
the Poisson equation $\Delta\Phi_b=4\pi G\rho_b$. The corresponding wavefunction
is
\begin{equation}
\label{da10bg}
\psi_b({\bf r},t)=\sqrt{\rho_b(t)}e^{i\left\lbrack \frac{1}{2} H(t) m
r^2+S_0(t)\right\rbrack/\hbar}.
\end{equation}
The hydrodynamic equations (\ref{mad3})-(\ref{mad5}) 
then reduce to
\begin{equation}
\label{da11}
\frac{d\rho_b}{dt}+3H\rho_b=0 \qquad \Rightarrow\qquad  \rho_b\propto a^{-3},
\end{equation}
\begin{equation}
\label{da11b}
\frac{dS_0}{dt}=-mV'(\rho_b),
\end{equation}
\begin{equation}
\label{da12}
\dot H+H^2=-\frac{4}{3}\pi G \rho_b  \qquad \Rightarrow\qquad \ddot a
=-\frac{4}{3}\pi G\rho_b a.
\end{equation}
The first equation can be interpreted as the conservation of mass
\begin{equation}
\label{da13}
M=\frac{4}{3}\pi\rho_b a^3\qquad \Rightarrow \qquad \rho_b=\frac{3M}{4\pi a^3}
\end{equation}
and the third equation as the Newtonian equation of dynamics
\begin{equation}
\label{da14}
\ddot a =-\frac{GM}{a^2}=-\frac{\frac{4}{3}\pi G\rho_b a^3}{a^2}
\end{equation}
for a particle submitted to a gravitational field $-{GM}/{a^2}$ created by a
mass $M$. These equations can be justified in a Newtonian cosmology if
we view the Universe as a homogeneous sphere of mass $M$, radius $a(t)$ and
density $\rho_b(t)$ evolving under its 
own gravitation. Eq. (\ref{da14}) is then obtained by considering the force
experienced by a particle of arbitrary  mass $m$ on the surface of this sphere
and using Newton's law. The first integral of motion is
\begin{equation}
\label{da15z}
\frac{1}{2}\left (\frac{da}{dt}\right )^2-\frac{GM}{a}=E  
\end{equation}
implying
\begin{equation}
\label{da15}
\left (\frac{da}{dt}\right )^2=\frac{2GM}{a}+2E=\frac{8}{3}\pi
G\rho_b a^2+2E.
\end{equation}
We can check that the foregoing equations coincide with the Friedmann equations
in the nonrelativistic limit (or for pressureless matter). In the context of
general relativity, the
term $-2E$ represents the curvature of space $\kappa$, where $\kappa=-1,0,+1$
depending whether the Universe is open, critical, or closed. The
theory of inflation and the observations of the CMB favor a flat universe
($\kappa=0$) so we
shall take $E=0$. In that 
case, Eq. (\ref{da15}) reduces to
\begin{equation}
\label{da16}
\left (\frac{da}{dt}\right )^2=\frac{8}{3}\pi G\rho_b a^2 \qquad
\Rightarrow\qquad H^2=\frac{8}{3}\pi G\rho_b.
\end{equation}
Combining Eq. (\ref{da16}) with Eq. (\ref{da11}) we obtain the solution  
\begin{equation}
\label{da17}
a\propto t^{2/3},\qquad H=\frac{\dot a}{a}=\frac{2}{3t},\qquad
\rho_b=\frac{1}{6\pi Gt^2},
\end{equation}
which corresponds to the Einstein-de Sitter (EdS)
universe (we have assumed a vanishing cosmological
constant $\Lambda=0$).

\subsection{Comoving frame}

We now write the GPP equations in the comoving frame. To that purpose, we
make
the change of variables
\begin{equation}
\label{da18}
{\bf r}=a(t){\bf x},\qquad \psi({\bf r},t)=\Psi({\bf
x},t)e^{i\frac{1}{2}mHr^2/\hbar},
\end{equation}
where ${\bf r}$ is the proper distance. Eq. (\ref{da18}) is a change of
variables from proper locally Minkowski coordinates ${\bf r}$ to expanding
coordinates ${\bf x}$ comoving in the background model \cite{peeblesbook}.
The density is given by $\rho=|\Psi|^2$. Defining the
gravitational potential
$\phi({\bf x},t)$ by
\begin{equation}
\label{da19}
\Phi({\bf r},t)=\Phi_b({\bf r},t)+\phi({\bf x},t),
\end{equation}
we find that the Poisson equation (\ref{mad5}) becomes
\begin{equation}
\label{da20}
\Delta\phi=4\pi G a^2 (\rho-\rho_b),
\end{equation}
where the derivatives are with respect to ${\bf x}$ (the same is true for the
following equations unless explicitly stated).

In order to transform the GPP equations (\ref{gpp1}) and (\ref{gpp2}) to
the comoving frame we first  compute
\begin{eqnarray}
\label{da21}
\left (\frac{\partial\psi}{\partial t}\right )_{\bf r}&=&\left
(\frac{\partial}{\partial t}\right )_{\bf r}\Psi\left (\frac{{\bf
r}}{a(t)},t\right )e^{i\frac{1}{2}mHr^2/\hbar}\nonumber\\
&=&\left (\frac{\partial\Psi}{\partial t}-H {\bf
x}\cdot\nabla\Psi+\frac{i}{2\hbar}m{\dot H} a^2x^2\Psi\right
)e^{i\frac{1}{2}mHr^2/\hbar},\nonumber\\
\end{eqnarray}
and
\begin{eqnarray}
\label{da22}
\Delta_{\bf r}\psi=\biggl
(\frac{1}{a^2}\Delta\Psi+3\frac{i}{\hbar}mH\Psi+2\frac{i}{\hbar}mH {\bf
x}\cdot\nabla\Psi\nonumber\\
-\frac{m^2H^2}{\hbar^2}a^2x^2\Psi\biggr
)e^{i\frac{1}{2}mHr^2/\hbar}.
\end{eqnarray}
Substituting the foregoing relations into Eq. (\ref{gpp1}) we find after
simplification (using Eq. (\ref{da12})) that
\begin{equation}
\label{da24}
i\hbar \frac{\partial\Psi}{\partial
t}+\frac{3}{2}i\hbar
H\Psi=-\frac{\hbar^2}{2ma^2}\Delta\Psi+m\frac{dV}{d|\Psi|^2}
\Psi+m\phi\Psi.
\end{equation}
On the other hand, using Eq. (\ref{da16}), the Poisson equation (\ref{da20}) can
be written as
\begin{equation}
\label{da25}
\frac{\Delta\phi}{4\pi G a^2}= |\Psi|^2-\frac{3H^2}{8\pi G}.
\end{equation}

We can similarly transform the hydrodynamic equations (\ref{mad3})-(\ref{mad5}) 
to the comoving frame. The wave function can be written
as
\begin{equation}
\label{da26}
\Psi({\bf x},t)=\sqrt{{\rho({\bf x},t)}} e^{i{\cal S}({\bf x},t)/\hbar},
\end{equation}
where $\rho({\bf x},t)$ is the mass density  and ${\cal S}({\bf x},t)$ is the
action in the comoving frame. Making the Madelung \cite{madelung} transformation
\begin{equation}
\label{da27}
\rho({\bf
x},t)=|\Psi|^2\qquad {\rm and} \qquad {\bf v}=\frac{\nabla {\cal S}}{ma},
\end{equation}
where ${\bf v}({\bf x},t)$ is the velocity field in the comoving frame, and
comparing Eqs. (\ref{mad1}), (\ref{da18}) and  (\ref{da26}), we get
\begin{equation}
\label{da28}
S({\bf r},t)={\cal S}({\bf x},t)+\frac{1}{2}mHr^2\quad  \Rightarrow \quad {\bf
u}({\bf r},t)={\bf v}({\bf x},t)+H{\bf r},
\end{equation}
where ${\bf u}$ is the velocity field in the inertial frame and $H{\bf r}$ is
the Hubble flow.\footnote{This result can also be obtained as follows. Taking
the derivative with respect to time of the relation ${\bf r}=a(t){\bf x}$, we
get $d{\bf r}/dt=\dot a{\bf x}+a d{\bf x}/dt$. This can be written as ${\bf
u}=H{\bf r}+{\bf v}$ with ${\bf u}= d{\bf r}/dt$ and ${\bf v}= a d{\bf x}/dt$,
where ${\bf u}$ is the proper velocity  and ${\bf v}$ is the peculiar velocity.}
Then, we compute
\begin{equation}
\label{da29}
\left (\frac{\partial\rho}{\partial t}\right )_{\bf r}=\left
(\frac{\partial}{\partial t}\right )_{\bf r}\rho\left (\frac{{\bf
r}}{a(t)},t\right )=\frac{\partial\rho}{\partial t}-H{\bf x}\cdot\nabla\rho
\end{equation}
and
\begin{equation}
\label{da30}
\nabla_{\bf r}(\rho {\bf u})=\frac{1}{a}\nabla\cdot(\rho{\bf v})+H{\bf
x}\cdot\nabla\rho+3H\rho.
\end{equation}
With these relations, the continuity equation (\ref{mad3}) becomes
\begin{equation}
\label{da31}
\frac{\partial\rho}{\partial t}+3H\rho+\frac{1}{a}\nabla\cdot (\rho {\bf v})=0.
\end{equation}
Similarly, using
\begin{equation}
\label{da32}
\left (\frac{\partial {\bf u}}{\partial t}\right )_{\bf r}=\left
(\frac{\partial}{\partial t}\right )_{\bf r}{\bf v}\left (\frac{{\bf
r}}{a(t)},t\right )+\dot H {\bf r}=\frac{\partial {\bf v}}{\partial t}-H({\bf
x}\cdot\nabla){\bf v}+\dot H a {\bf x},
\end{equation}
and
\begin{eqnarray}
\label{da33}
({\bf u}\cdot\nabla_{\bf r}){\bf u}&=&\left\lbrack (H{\bf r}+{\bf
v})\cdot\nabla_{\bf r}\right\rbrack (H{\bf r}+{\bf v})\nonumber\\
&=&H^2a{\bf
x}+H({\bf x}\cdot\nabla){\bf v}+H{\bf v}+\frac{1}{a}({\bf v}\cdot \nabla){\bf
v},\nonumber\\
\end{eqnarray}
the quantum Euler equation (\ref{mad4}) becomes
\begin{equation}
\label{da34}
\frac{\partial {\bf v}}{\partial t}+\frac{1}{a}({\bf v}\cdot
\nabla){\bf v}+H{\bf v}=-\frac{1}{\rho a}\nabla
P-\frac{1}{a}\nabla\phi-\frac{1}{ma}\nabla
Q
\end{equation}
with the quantum potential
\begin{equation}
\label{da35}
Q=-\frac{\hbar^2}{2ma^2}\frac{\Delta
\sqrt{\rho}}{\sqrt{\rho}}=-\frac{\hbar^2}{4ma^2}\left\lbrack
\frac{\Delta\rho}{\rho}-\frac{1}{2}\frac{(\nabla\rho)^2}{\rho^2}\right\rbrack,
\end{equation}
where we have used Eq. (\ref{da12}) to simplify some
terms. These transformations can also be made at the level
of the action. Using
\begin{eqnarray}
\label{fin1}
\left (\frac{\partial S}{\partial t}\right )_{\bf r}&=&\left
(\frac{\partial}{\partial t}\right )_{\bf r}{\cal S}\left (\frac{{\bf
r}}{a(t)},t\right )+\frac{1}{2}m\dot H r^2\nonumber\\
&=&\frac{\partial {\cal
S}}{\partial
t}-H{\bf x}\cdot\nabla {\cal S}+\frac{1}{2}m\dot H r^2
\end{eqnarray}
and
\begin{equation}
\label{fin2}
\nabla_{\bf r}S=\frac{1}{a}\nabla{\cal S}+mH{\bf r},
\end{equation}
the Hamilton-Jacobi equation (\ref{mad3b}) becomes after simplification
\begin{equation}
\label{fin3}
\frac{\partial {\cal S}}{\partial t}+\frac{(\nabla
{\cal S})^2}{2ma^2}=-Q-m\phi-mV'(\rho).
\end{equation}
We can check that the above results return the equations
of Refs. \cite{aacosmo,abrilph,playa,chavmatos,phi6}
up to an obvious change of notations.

\end{document}